\documentclass[12pt]{article}
\usepackage[margin=2 cm]{geometry}
\usepackage{comment}
\usepackage{graphicx}
\usepackage{dsfont}
\usepackage[font={small,it}]{caption}
\usepackage{mwe}
\usepackage[nottoc]{tocbibind}
\usepackage{amsmath,amssymb,extarrows,mathtools,graphicx,subfigure,setspace}
\usepackage{cite}
\usepackage{stackengine}
\usepackage{tikz}\usetikzlibrary{calc}
\usepackage{braket}
\usepackage{epsfig}
\usepackage[section]{placeins}
\usepackage{slashed}
\usepackage{color}
\usepackage{caption}
\usepackage{amsmath}
\usepackage{hyperref}
\makeatother
\newmuskip\pFqmuskip

\newcommand{\be}{\begin{equation}}
\newcommand{\bea}{\begin{eqnarray}}
\newcommand{\eea}{\end{eqnarray}}
\newcommand{\ba}{\begin{array}}
\newcommand{\ea}{\end{array}}
\newcommand{\ee}{\end{equation}}
\newcommand{\bes}{\begin{equation*}}
\newcommand{\beas}{\begin{eqnarray*}}
\newcommand{\eeas}{\end{eqnarray*}}
\newcommand{\bas}{\begin{array*}}
\newcommand{\eas}{\end{array*}}
\newcommand{\ees}{\end{equation*}}

\newcommand*\pFq[6][8]{%
  \begingroup % only local assignments
  \pFqmuskip=#1mu\relax
  \mathcode`\,=\string"8000
  \begingroup\lccode`\~=`\,
  \lowercase{\endgroup\let~}\pFqcomma
  {}_{#2}F_{#3}{\left[\genfrac..{0pt}{}{#4}{#5};#6\right]}%
  \endgroup
}
\newcommand{\pFqcomma}{\mskip\pFqmuskip}

\numberwithin{equation}{section}

\begin{document}

\onehalfspacing
\vfill
\begin{titlepage}
\vspace{10mm}

\begin{center}

\vspace*{10mm}
\vspace*{1mm}
{\Large  \textbf{Black holes with quantum corrections in $3d$: The case of Page curve in Lindblad, greybody factor, and Lyapunov exponent}} 
 \vspace*{1cm}

{$\text{Mahdis Ghodrati}$}

{$\text{}$}

\vspace*{8mm}
{ \textsl{ $ $ International Centre for Theoretical Physics Asia-Pacific,
University of Chinese Academy of Sciences, 100190 Beijing, China}}

 \vspace*{0.45cm}

\textsl{e-mails: {\href{mahdisghodrati@ucas.ac.cn}{mahdisghodrati@ucas.ac.cn}}}
 \vspace*{2mm}

\vspace*{1.7cm}

\end{center}

\begin{abstract}
In this work, first we discuss black holes with quantum corrections as an open quantum system and apply Lindblad formalism to explain the ``zig-zag" behavior in the Hawking- Page curve and radiation process, specially by considering the effects of exceptional points. Then, we calculate quantum corrections for various parameters of $3d$ Cotler-Jensen theory which is $\text{AdS}_3$ with reparametrized modes. At each step of the calculations, we compare the results with the case of JT. We then calculate quantum-corrected greybody factor for various black hole solutions, and specially for the case of Cotler-Jensen theory. Finally, we study the effects of quantum corrections on Lyapunov exponent, potential, and quantum information structure of black holes.
 \end{abstract}

\end{titlepage}

\tableofcontents

\section{Introduction}

Recently, quantum corrections of black hole radiation for various situations, solution and quantities have been calculated, see \cite{Iliesiu:2020qvm,Heydeman:2020hhw, Iliesiu:2022onk, Kapec:2023ruw,Maulik:2024dwq,Nian:2025oei, PandoZayas:2025snm, Maulik:2025hax}, among other works. These quantum corrections become important specially at low-temperature and dominate the dynamics of the black hole.

In \cite{Brown:2024ajk}, it has been demonstrated that these quantum effects at low-temperature and in microcanonical ensemble would actually \textit{slow down} the emission of quantum radiation near the horizon of the black hole, by a factor of $10^{700}$. There, it has been shown that the black hole alternates between exponentially long epochs of integer and half-integer spin which have very different cooling rates. Here, in our first section, we argue that if we actually consider near the black hole region as an open quantum system, passing through the exceptional points (EPs) would actually cause the same effect, similar to the roles of  EPs in the dynamics of SYK model as discussed in \cite{Zheng:2025apa}.

In the canonical ensemble,  \cite{Brown:2024ajk} shows that the energy flux in the exact quantum-corrected solution would be larger than the semiclassical flux at fixed temperature $T$. This is in contradiction with the microcanonical result, where it would be a reduced flux relative to the semiclassical case at very low energies. So, in the canonical ensemble, the increase in typical energy leads to an increase in the expected flux relative to the semiclassical prediction. The resolution is that the relation $M-Q= \frac{2\pi^2}{E_{\text{brk}}}T^2$ would break at low temperatures. This also can be explained by considering the presence of EPs near the black hole horizon, as we show in our first section here.

Previously, in the semiclassical regime, the greybody factor and Hawking radiation have been calculated in \cite{Cvetic:1997uw, Lee:1998pd, Setare:2006hq, Jiang:2007pn, Cvetic:2009jn}. Recently, quantum corrections to these quantities, and many other thermodynamical quantities have also been calculated, mainly based on the new methods of \cite{Brown:2024ajk}.

In addition, in various recent works, the effects of quantum fluctuations at low temperature near extremal and near horizon geometries have been examined using the one dimensional Shwarzian effective action. Note that the Schwarzian modes are the zero modes in JT. One could then consider these fluctuations more precisely and go to one dimension higher and instead of near $\text{AdS}_2$ geometry, consider the full $\text{AdS}_3$ case or its reparameterized version \cite{Cotler:2018zff}, which is the main goal of this work.

In \cite{Cotler:2018zff},  the $\text{PSL}(2;\mathbb{R})$-invariant bilocal operators have been studied extensively, and the authors showed that the connected two-point function of bilocals is the Virasoro identity block.  Then, they argued for the gravity dual of a bilocal probe. They noted that, as bilocal operators in the Schwarzian theory and in nearly-$\text{AdS}_2$ gravity correspond to boundary insertions of sources for operators dual to bulk matter field, the extended picture in $3d$ could be envisioned for the reparametrized $\text{AdS}_3$ case of \cite{Cotler:2018zff} as well. The probe in this case would be dual to a bulk Wilson like which ends on the AdS boundary at the two points of the bilocal. This picture could help us to understand Schwarzian quantum corrections to various parameters of the black hole from one dimension higher point of view of the nearly $\text{AdS}_2$ case. We also use, the results of \cite{Fitzpatrick:2016mtp, Blommaert:2018oro}, along the way.

So the plan of this work is as follows. First, in section \ref{sec:Lindblad}, we study black holes in the Lindblad formalism and as an open quantum system. We compare each quantity, such as Green’s function, action, transition amplitude, bilocal operators and couplings, for the open quantum system and the theory of AdS$_3$ with reparametrization modes, as discussed in Cotler and Jensen work \cite{Cotler:2018zff}.
Then, in section \ref{sec:rep}, we study the theory of reparamtrized AdS$_3$ further. We compare the structure of this theory with JT, and find the quantum corrected free propagator and actions. Then, we find the one-loop corrected greybody factor for this theory, along with other black hole solution in section \ref{sec:greybody}. We also find the quantum corrected Lyapunov exponent in section \ref{sec:Lyapunov}. Next, we discuss quantum corrections to some other black hole quantities such as potential, and information structures such as information inequalities in section \ref{sec:others}. We conclude with a brief discussion in section \ref{sec:con}.

\section{Black hole in the Lindblad formalism}\label{sec:Lindblad}

In \cite{PandoZayas:2025snm}, the authors studied the effects of quantum fluctuations for near-extremal asymptotically $\text{AdS}_4$ black branes, where they found that at one-loop level, there is a coupling between the shear gravitational fluctuations to ``one of the would-be zero modes", and this coupling would affect the retarded Green's function which leads to low temperature violation of the shear viscosity to entropy density bound. On the other hand, in \cite{Zheng:2025apa}, the dynamics of Lindblad SYK model coupled to a bath has also been studied, where the authors found a nontrivial, non-monotonic feature, which is due to the presence of exceptional points and its proliferation. This non-monotonic behavior is actually in the behavior of dissipative gap versus the bath coupling $\mu$ of the system. Here, we aim to connect these two works and these two effects, and therefore explain the observed zig-zag behavior in the Page-cure of radiation with quantum corrections.

Note that in the extremal limit of black holes, and in its throat region, one has an effective theory which contains the Schwarzian geometry. One could imagine the quantum correction of this geometry as an open quantum system which contains a system coupled to a bath, and then apply the results and methods of \cite{Zheng:2025apa}, such as the vectorization of density matrix, and also apply the Lindblad formalism.

The gravitational path integral over the zero modes in this system, would affect the Green's function which were studied in \cite{PandoZayas:2025snm}. On the other hand, the connections between Green's function, partition function and dissipative gap were studied in \cite{Zheng:2025apa}.

For the SYK model, which has strong connection to Schwarzian model, the on-shell Green's function $G_{ab}$ in the large-N limit would be
\begin{gather}
G_{ab} (\tau) = \frac{1}{N} \sum_i \langle \psi_i^a (\tau) \psi_i^b (0) \rangle,
\end{gather}
where $a,b \in \{+,- \}$.

In the action, in addition to Green's function, one should consider another bilocal field $\Sigma_{ab}$ as well, which for the case of Lindblad SYK leads to \cite{Zheng:2025apa}
\begin{gather}
iS = \frac{1}{2} \log \det (\delta_{ab}\partial - \Sigma_{ab} )- \frac{i \mu}{2} \int_0^t d\tau (G_{+-}(\tau, \tau) - G_{-+} ( \tau, \tau))  \nonumber\\
- \frac{1}{2} \sum_{ab} \int_0^t \int_0^t d\tau d\tau' \lbrack \Sigma_{ab} ( \tau, \tau') G_{ab}( \tau, \tau') + \frac{J^2}{q} t_{ab} G_{ab} ( \tau, \tau')^q \rbrack.
\end{gather}

For the case of black hole, the Green's function has the form of \cite{PandoZayas:2025snm}
\begin{gather}
G^R_{xy, xy}(\omega, 0) = \frac{(2\pi)^6 \delta Z}{\delta \phi_0(- \omega, 0, 0) \delta \phi_0(\omega, 0, 0) } /Z \Big |_{\phi_0=0} \nonumber\\
= \frac{(2\pi)^6 e^{-S_{\text{on-shell}} }}{\delta \phi_0 (-\omega,0,0) \delta \phi_0(\omega,0,0) }/ e^{-S_{\text{on-shell}}} + \frac{(2\pi)^6 \delta \int \mathcal{D} h \ e^{-h_{\text{zero}} \Delta_L h_{\text{zero}}} }{\delta \phi_0 (-\omega,0,0) \delta\phi_0 (\omega,0,0)} /\left (\int \mathcal{D} h \ e^{-h_{\text{zero}} \Delta_L h_{\text{zero}} } \right) \Big |_{\phi_0=0}\nonumber\\
=G_{xy,xy}^{R(tree)} (\omega,0) + G_{xy,xy}^{R(one-loop)} (\omega,0).
\end{gather}

Then, the tree-level Green's function would be
\begin{gather}
G_{xy,xy}^{R(tree-level)} ( \omega, 0) = - \frac{1}{2 \kappa_4^2} \left ( \frac{r_0}{L} \right)^2 \left( i \omega+ \frac{2\pi i \omega L^2 T}{3 r_0} + \mathcal{O} (T^2) \right ).
\end{gather}

The one-loop Green's function here is
\begin{gather}
G_{xy,xy}^{R (\text{one-loop})} (\omega, 0) = (2\pi)^6 \frac{\delta \lbrack (2 \alpha T + \delta \Lambda_2 \lbrack \delta g \rbrack )^{-1}  \sqrt{\frac{2}{\pi} } (\alpha T)^{\frac{5}{2}} / (\frac{1}{\sqrt{2\pi}} (\alpha T)^{\frac{3}{2} } )  }{\delta c_1 (- \omega, 0, 0) \delta c_1 (\omega, 0, 0) } \Big |_{c_1 =0} = - \frac{9  \ i \pi \ \omega r_0}{L^2 T V}.
\end{gather}

One should then be able to define a coupling between these two sectors and show its non-monotonic behavior as in \cite{Zheng:2025apa}.

Now, one could imagine similar effects due to the presence of exceptional points in the Schwarzian modes of quantum fluctuations around the horizon of black holes. In addition, the behavior of such exceptional points from one dimension higher point of view could also be studied later.

The one-loop corrections can be considered by taking into account the gravitational fluctuations to quadratic order, as \cite{PandoZayas:2025snm}
\begin{gather}
Z= e^{-S_{on-shell} } \int \mathcal{D} h \ e^{-h \Delta_L h}.
\end{gather}

So, as discussed in \cite{PandoZayas:2025snm}, the bulk field $\phi$ couples to $h_{\text{zero}}$ via its action $h_{\text{zero}} \Delta_L h_{\text{zero}}$. The bulk field $\phi$ then could be considered as bath, and $h_{\text{zero}}$ with the action $h_{\text{zero}} \Delta_L h_{\text{zero}} $ as the system. If one takes $\mu$ as the coupling between them, one could see a non-monotonic behavior similar to the case of \cite{Zheng:2025apa}.

The Lichnerowicz operator here is
\begin{gather}
h^*_{\alpha\beta} \Delta_L^{\alpha\beta, \mu\nu} h_{\mu\nu} = - \frac{1}{2\kappa_4^2} h^*_{\alpha\beta} (\Delta_{EH}^{\alpha\beta, \mu\nu} - 2\Lambda \Delta_1^{\alpha\beta, \mu\nu}- 2L^2 \Delta_F^{\alpha\beta, \mu\nu}) h_{\mu\nu},
\end{gather}
where $\Delta_{EH}^{\alpha\beta, \mu\nu}$, $\Delta_1^{\alpha, \beta, \mu\nu}$, $\Delta_F^{\alpha\beta, \mu\nu}$ are the contribution from the Einstein-Hilbert, cosmological constant and the Maxwell action respectively. The change of saddle and the non-monotonic behavior of quantities such as dissipative gap $\Gamma_0$ versus bath coupling $\mu$ can be explained by the change in the contribution from each term. So, considering the sum of these operators as a matrix, the exponential points here could be detected by finding where the eigenvalue and eigenstates coalesces.  Note that, the theory of low energy Schwarzian modes \cite{Daguerre:2023cyx, Liu:2024gxr, Jensen:2016pah} is similar to SKY and Lindblad SYK model of \cite{Zheng:2025apa}.

In \cite{PandoZayas:2025snm}, the one-loop correction for the case of $n\ge 2$ has been calculated as
\begin{gather}
\int \mathcal{D} h \ e^{-h_{zero} \Delta_L h_{zero} } \sim ( \prod_{n \ge 2} \delta \Lambda_n)^{-1}\nonumber\\
= \left (\frac{4\pi l^2 T}{r_0} + \delta \Lambda_2 \lbrack \delta g \rbrack \right )^{-1} \prod_{n \ge 3} \delta \Lambda_n^{-1}\nonumber\\
=( 2 \alpha T +\delta \Lambda_2 \lbrack \delta g \rbrack )^{-1} \sqrt{\frac{2}{\pi} } (\alpha T)^{\frac{5}{2}},
\end{gather}
where $\alpha= \frac{2\pi l^2}{r_0}$.

The field $\delta g$ couples with the zero modes through the “lifted eigenvalue" $\delta \Lambda_n \lbrack \delta g\rbrack$ in the near-horizon “throat".

Also, one would expect that the dissipative gap $\Gamma_0$ versus the coupling between the shear gravitational fluctuations to the zero mode would be non-monotonic, similar to the case of \cite{Zheng:2025apa}. As times progress, this coupling changes and as the result, the dissipative gap changes, which affects all the thermodynamical quantities of the black holes and the quantum corrections effects, such as the ratio of shear viscosity to entropy density. 

At zero temperature, where $g=g^{(0)}$ and $A= A^{(0)}$, the form of graviton zero modes would be
\begin{gather}\label{eq:modes}
h_{\mu\nu}^{(n)} dx^\mu dx^\nu = C_n e^{in\tau} \left ( \frac{-1+Y}{1+Y} \right)^{\frac{n}{2}} \left ( - d\tau^2+ 2i \frac{d\tau dY}{Y^2-1}+\frac{dY^2}{(Y^2-1)^2}\right ).
\end{gather}
Note that the zero modes and Schwarzian modes are related with each other as discussed in \cite{Iliesiu:2022onk}. These are zero modes because they are pure gauge, and solutions of 
\begin{gather}
h_{\mu\nu}^{(n)} = \nabla_\mu \xi_\nu^{(n)} + \nabla_\nu \xi_\mu^{(n)}.
\end{gather}

Then, we can explain these connections and the map between the graviton modes $h^{(n)}_{\mu \nu}$ into Schwarzian modes $\varepsilon_n$, and then we can try to find the Lindbladian.
The zero modes of the extremal metric could be written as  \cite{Iliesiu:2022onk}
\begin{gather}\label{eq:zeros}
h^\varepsilon_{AB} dx^A dx^B  = \sum_{n \in \mathbb{Z}, | n | \ge 2 } 2n (n^2-1) i \varepsilon_n e^{i n \tau} \tanh^{|n|} \frac{\rho}{2} \left (- d\tau^2+ 2i \frac{d \tau d \rho}{\sinh \rho}  + \frac{d\rho^2}{\sinh^2 \rho}  \right ).
\end{gather}

By taking $Y= \cosh \rho$ and comparing the relations \ref{eq:modes} and \ref{eq:zeros}, we can find
\begin{gather}
C_n = 2 i n (n^2-1) \varepsilon_n, \ \ \  \text{or} \ \ \ 
\varepsilon_n = \frac{C_n}{ 2i n (n^2-1)}, \ \ \ \ \ |n| \ge 2.
\end{gather}
So we found the isomorphic map of graviton zero modes ${h_{\mu \nu}}^n$ to Schwarzian mode $\varepsilon_{n}$, ${h_{\mu \nu}}^n \leftrightarrow \varepsilon_n$. In \cite{PandoZayas:2025snm}, the normalization constant also has been found as 
\begin{gather}
C_n^2=\frac{l^2 L^4 n (n^2-1) \kappa_4^2}{6 \pi r_0^2 V}, \ \ \ \ \ \ V=L_x L_y.
\end{gather}
So, the Schwarzian modes can be written as
\begin{gather}
\varepsilon_n = \frac{\ell L^2 \kappa_4^2}{2i r_0  \sqrt{6\pi Vn(n^2-1)} }.
\end{gather}
So the graviton zero modes of \cite{PandoZayas:2025snm}, become the Schwarzian modes that enter in Lindblad equation of \cite{Zheng:2025apa}.

Also, at the level of gravitational path integral, we have the relation
\begin{gather}
\int \mathcal{D}h e^{- h_{\text{zero}} \Delta_L h_{\text{zero}} } = \int \mathcal{D} \varepsilon e^{-S_{\text{Schwarzian}} \lbrack \varepsilon \rbrack},
\end{gather}
where
\begin{gather}
S_{\text{Schwarzian}} \lbrack \varepsilon \rbrack = - \frac{1}{2g^2} \int d \tau \left ( (\varepsilon'')^2 - (\varepsilon')^2 \right).
\end{gather}
So, we also have the relation $C_n = \sqrt{\frac{2}{g^2}} \varepsilon$.

As explained in  \cite{Iliesiu:2022onk}, these modes do not mix with the rest of the fluctuations, so they can be considered as the one-loop determinant. They are exactly the Schwarzian sector coming from two dimensional JT gravity. They can also be written as pure diffeomorphism $x^A \to x^A + \zeta^A$ with
\begin{gather}
\zeta = \sum_{|n| \ge 2} \varepsilon_n e^{i n \tau} \tanh^{|n|} \frac{\rho}{2} \left ( \frac{i n (|n| + \cosh \rho) }{\sinh \rho} \partial_\rho - \frac{|n| ( |n| + \cosh \rho ) + \sinh ^2 \rho}{\sinh^2 \rho} \partial_\tau \right). 
\end{gather}

One could take the limit $\rho \gg 1$, and write the diffeomorphisms that act at infinity as 
\begin{gather}
\zeta \approx \sum_{|n| \ge 2} \varepsilon_n e^{i n \tau} ( in \partial_\rho - \partial_\tau)  
\approx \varepsilon'(\tau) \partial_\rho - \varepsilon (\tau) \partial_\tau, \ \ \ \ \ \ \varepsilon(\tau) \equiv \sum_{|n| \ge 2} \varepsilon_n e^{i n \tau}.
\end{gather}

This $\varepsilon(\tau)$ is precisely the boundary reparametrization variable. The $n=-1, 0, 1$ modes should be removed as the correspond to the $SL(2, \mathbb{R})$ isometries. 

This is the first step to find the Lindblad equation for the black hole evaporation. Then, we can treat the Schwarzian modes as the system, and the bulk fields as the bath.

Now the quantum corrected  ``cloud" around the naked quantity could be considered as the bath. If one turns on a small temperature, the lifted eigenvalues would be
\begin{gather}
\delta \Lambda_n = \int dx^4 \sqrt{g^{(0)}} h_{\alpha\beta}^{(n)*} \delta \Delta_L^{\alpha\beta,\mu \nu} h_{\mu\nu}^{(n)}.
\end{gather}
This lifted eigenvalue now left the system and lives in the bath. One then could consider the term $h_{\alpha\beta}^{(n)*} \delta \Delta_L^{\alpha\beta, \mu\nu} h_{\mu\nu}^{(n)}$, and find the couplings.

So, by considering SYK or the Schwarzian model and doubling the Hilbert state, one could get the following relationship
\begin{gather}
iS = \int_0^t d\tau \lbrack - \frac{1}{2} \sum_i ( \psi_i^+ \partial \psi_i^+ + \psi_i^- \partial \psi_i^-) - i \mu \sum_i \psi_i^+ \psi_i^- - \mu N \nonumber\\
- i ^{q/2+1} \sum J_{i_1 ...i_q} \psi^+_{i_1} \  ... \ \psi_{i_q}^+ + i^{q/2+1} \sum i^q J_{i_1 ... i_q} \psi^-_{i_1} \ ...\  \psi_{i_q}^- \rbrack.
\end{gather}

In the setup with quantum corrections, for a bosonic system, the transition amplitude would have the following path integral representation
\begin{gather}
\langle \phi_b | e^{- i t H} | \phi_a \rangle = \int \lbrack \mathcal{D} \pi \rbrack \int_{\phi(\mathbf{x}, 0)=\phi_a(\mathbf{x})}^{\phi(\mathbf{x},t)=\phi_b(\mathbf{x}) } \lbrack \mathcal{D \phi} \rbrack \nonumber\\
\times \ \text{exp} \left \lbrack i \int_0^t dt \int d^3x ( \pi ( \mathbf{x}, t) \frac{\partial \phi(\mathbf{x},t) }{\partial t} - \mathcal{H} \lbrack \pi ( \mathbf{x}, t ), \phi(\mathbf{x}, t ) \rbrack ) \right \rbrack.
\end{gather}

So the coupling between the system and the bath, (or the correlations between the bare system and the quantum corrected one) can move the system forward, leading to the following relation
\begin{gather}
\int d^3 x \left( \pi ( \mathbf{x}, t) \frac{\partial \phi(\mathbf{x}, t)}{\partial t} - \mathcal{H} \lbrack \pi ( \mathbf{x}, t), \phi(\mathbf{x}, t) \rbrack \right) \to - \frac{1}{2} \sum_i ( \psi_i^+ \partial \psi_i^+ + \psi_i^- \partial \psi_i^-)- i \mu \sum_i \psi_i^+ \psi_i^- - \mu N \nonumber\\
- i^{q/2+1} \sum J_{i_1 . . . i_q} \psi_{i_1}^+ ... \psi_{i_q}^+ + i^{q/2+1} + i^{q/2+1} \sum i^q J_{i_1 . . . i_q} \psi_{i_1}^- ... \psi_{i_q}^- .
\end{gather}

Then, one could propose the following relation between Lindblad SYK and Schwarzian mode of quantum corrections
\begin{gather}
\int d^3 x \left( \pi ( \mathbf{x}, t) \frac{\partial \phi(\mathbf{x}, t)}{\partial t} - \mathcal{H} \lbrack \pi ( \mathbf{x}, t), \phi(\mathbf{x}, t) \rbrack \right) \to  \frac{1}{2}  \log \det(\delta_{ab} \partial - \Sigma_{ab})- \frac{i\mu}{2} \int_0^t d\tau ( G_{+-} (\tau, \tau) - G_{-+}(\tau, \tau))\nonumber\\
- \frac{1}{2}\sum_{ab} \int_0^t \int_0^t d\tau d\tau' \lbrack \Sigma_{ab} (\tau, \tau')G_{ab}(\tau, \tau') + \frac{J^2}{q} t_{ab} G_{ab} (\tau, \tau')^q \rbrack,
\end{gather}
where $t_{++}=t_{--}=1$, $t_{+-}=t_{-+}=-(-1)^{q/2}$, $a,b \in \{+,-\}$, and in the large $N$, one has
\begin{gather}
G_{ab}(\tau) = \frac{1}{N} \sum_i \langle \psi_i^a (\tau) \psi_i^b (0) \rangle.
\end{gather}
Here $G_{ab}$ and $\sigma_{ab}$ are the Green's function and self-energy respectively, which by the above relation could be connected to the conjugate momentum $\pi (\mathbf{x})$.

Note that by a vectorization procedure, the Hilbert space would be doubled, where the Lagrangian becomes  \cite{Kulkarni_2022}
\begin{gather}
\mathcal{L}= - i H^+_{SYK} \otimes \mathcal{I}^- + i(-1)^{q/2} \mathcal{I}^+ \otimes H_{SYK}^- + i \mu \sum_i \psi_i^+ \psi_i^- -\mu \frac{N}{2} \mathcal{I}^+ \otimes \mathcal{I}^-.
\end{gather}

Therefore, one could imagine the quantum correction around the JT in the near horizon of black hole as a Markovian reservoir coupled to an SYK model, and the jump operator which are related to Hawking radiation, would be either linear or quadratic in the Majorana fermion operators. The distribution of linear jump operator or quadratic jump operators could be Gaussian, linear, or other statistical functions which needs to be specified for each case.

Next, we could write the fields as
\begin{gather}
i \partial_{t_1} G_{\alpha\beta} (t_1, t_2) = \int dt_3 \sum_{\gamma= +,-} \Sigma_{\alpha\gamma}(t_1,t_3) G_{\gamma\beta} (t_3,t_2) = \delta_{\alpha\beta}\delta (t_1 - t_2), \nonumber\\
\sigma_{\alpha\beta} ( t_1, t_2) = - i^{q_b} J^2 s_{\alpha \beta} G_{\alpha\beta}(t_1, t_2)^{q_b-1}+ \mu \epsilon_{\alpha\beta} \delta(t_1-t_2).
\end{gather}

In the large-$q_b$ limit, the expansion of the Green's function would be
\begin{gather}
G_{\alpha\beta}(t_1, t_2) = G_{\alpha\beta}^0 (t_1, t_2) \left(1+ \frac{1}{q_b} g_{\alpha\beta}(t_1, t_2) + ...\right),
\end{gather}
where $q_b$ is the q-body interaction in SYK model, which is related to Newton constant $g_N$, as
\begin{gather}
S= - \frac{\pi}{g_N^2}+ S^{(2)}+ g_N S^{(3)}+g_N^2 S^{(4)} +\mathcal{O}(g^3).
\end{gather}

The bilocal operator here is
\begin{gather}
\mathcal{B}(h; \theta_1, \theta_2) = \left ( \frac{\phi'(\theta_1)\phi'(\theta_2) }{4 \sin^2 (\frac{\phi(\theta_1)-\phi(\theta_2)}{2})^2} \right)^h.
\end{gather}

\begin{figure}[ht!]   
\begin{center}
\includegraphics[width=0.45\textwidth]{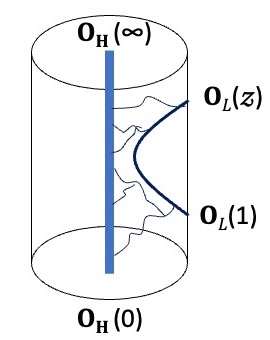} 
\caption{The schematic structure of graviton exchanges between heavy and light operators discussed in \cite{Fitzpatrick:2015foa}. The heavy operators could be considered as the system and the light operators as the bath and the graviton exchange between them could be considered as the coupling $\mu$ where its behavior were studied in \cite{Zheng:2025apa}.  }
\label{fig:gravitonExchange}
\end{center}
\end{figure}

The coupling of the above operator to the reparametrization field $\phi= \alpha_0 \theta + \frac{\epsilon}{\sqrt{C_{\mathcal{B}}}}$ can be found as \cite{Cotler:2018zff, Fitzpatrick:2015foa, Chen:2017dnl}
\begin{gather}
\mathcal{B} (h; \theta_1, \theta_2) = \left ( \frac{\alpha_0}{2\sin \left(\frac{\alpha_0 \theta_{12} }{2}\right ) }  \right)^{2h} \left(1+ \frac{h}{\sqrt{C_{\mathcal{B}}} } \mathcal{J}'^{(1)}_{12} \ . \ \epsilon + \frac{1}{C_{\mathcal{B}}} \left( \frac{h^2}{2} ( \mathcal{J}'^{(1)}_{12} \  . \ \epsilon)^2 + h {\mathcal{J}'_{12}}^{(2)} \ . \ \epsilon \right) + O\left(\frac{1}{C_{\mathcal{B}}^{3/2}} \right) \right).
\end{gather}
So, we expect that $q_b$ is related to $\frac{\sqrt{C_{\mathcal{B}}} }{h}$ or to the Newton's constant $g_N^2$. In fact, this picture that we construct here, is also related to the explicit exchange of Virasoro gravitons between heavy and light $\text{CFT}_2$ operators at large central charge, as explained in \cite{Fitzpatrick:2015foa,Fitzpatrick:2016mjq}, and shown in Fig \ref{fig:gravitonExchange}.

One then could use a vectorization procedure similar to \cite{Kulkarni:2021gtt,Zheng:2025apa}, and write the vectorized Liouvillian as
\begin{gather}
\mathcal{L} = - i H_{\text{Schwarzian}}^+ \otimes \mathcal{I}^- + i ( -1)^{\frac{\sqrt{C_{\mathcal{B}} }}{2h} } \mathcal{I}^+ \otimes H_{\text{Schwarzian}}^- + i \mu \sum_n \epsilon_n^+ \epsilon_n^- - \mu \frac{N}{2} \mathcal{I}^+ \otimes \mathcal{I}^-.
\end{gather}
Here we can see that the exponent $N$ is also related to Catalan number $C_k$ \cite{Fitzpatrick:2015foa}. By analytic continuation of the generating function of $C_k$, the Hawking temperature of BTZ black hole would arise \cite{Fitzpatrick:2015foa}.

The heavy-light OPE Virasoro blocks, and their connections to thermodynamics of black holes have also been studied in \cite{Fitzpatrick:2014vua, Fitzpatrick:2015zha}. Note that unlike the study of \cite{Qi:2019gny} which used the Feynman diagrams, in those works, the authors used the on-shell method which is simpler for using in various examples and problems.

The Catalan number $C_k$, then, is related to the number of distinct three diagrams with $k$ vertices which is related to the size of the bare system. Note that the bare system here is the BTZ black hole with Hawking temperature $T_H$ and with no quantum corrections.

The mode expansion of holomorphic stress tensor then can be written as
\begin{gather}
T(z)= \sum_n z^{-2-n} L_n,
\end{gather}
where these $L$-generators act on the vacuum and create non-trivial states, and they are related to the linear jump operators in the Lindblad formalism.

The exchange of graviton states between a heavy object (which is considered as the system here) and the light probe (which is considered as the bath in initial stages), could be written as
\begin{gather}
\mathcal{V} (z) = \left \langle \mathcal{O}_H(\infty) \mathcal{O}(1) \left( \sum_{ \{a_k\} , \{b_l\} } \frac{L_{-a_1} . . . L_{-a_k} \ket{0} \bra{0} L_{b_l} . . . L_{b_1}  }{\mathcal{N}_{ \{b_l \} , \{a_k \} } }\right ) \mathcal{O}_L(z) \mathcal{O}_L(0) \right \rangle.
\end{gather}

So the coupling between the two would be
\begin{gather}
\mu \to  \sum_{ \{a_k\} , \{b_l\} } \frac{L_{-a_1} . . . L_{-a_k} \ket{0} \bra{0} L_{b_l} . . . L_{b_1}  }{\mathcal{N}_{ \{b_l \} , \{a_k \} } } = \sum_{ \{a_i \}, \{b_j \}} \left( \ket{\{a_k\}} \mathcal{N}^{-1}_{ \{b_l \}, \{a_k \}} \bra{ \{b_l \}}   \right).
\end{gather}

The matrix $\mathcal{N}_{ \{b_l\}, \{ a_k\}}$ would vanish here unless $a_i=b_j$, and so we have
\begin{gather}
\mathcal{N}_{ \{ a_k\}, \{ a_k\}} = \prod_{i=1}^k \frac{a_i (a_i^2-1) }{12}.
\end{gather}

\begin{figure}[ht!]   
\begin{center}
\includegraphics[width=0.5\textwidth]{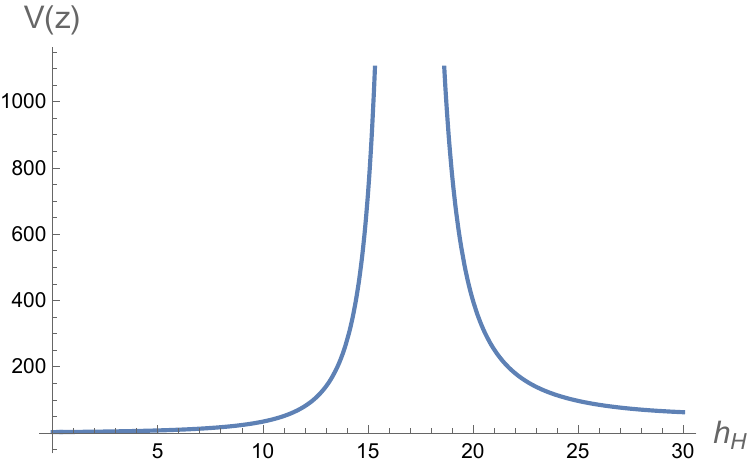} 
\caption{The behavior of $\mathcal{V} (z)$ versus $h_H$, which is non-monotonic and affects the non-monotonicity of energy gaps versus coupling.}
\label{fig:V(t)}
\end{center}
\end{figure}

The behavior of $\mathcal{V} (z)$ versus $h_H$ is shown in Figure \ref{fig:V(t)}, and the schematic behavior of this exchange of gravitons and the double-trace operators between heavy and light operators near the horizon of black holes is shown in Figure \ref{fig:gravitonExchange}.

The dissipative gap versus this coupling would have a non-monotonic behavior as in \cite{Zheng:2025apa}, which is related to the non-monotonicity of $\mathcal{V} (z)$.

Again one can see that the generating function of the Catalan number at $\frac{6h_H}{c}$ which builds the temperature of a BTZ black hole, is related to the size of SKY model $N$ as in
\begin{gather}
H_{L,R} = i^{q/2} \sum_{1\le j_1 < ... < j_q \le N} J_{j_1 ... j_q} \psi_{L,R}^{j_1} . . . \psi_{L,R}^{j_q},
\end{gather}
where for the case of thermofield double state (TFD), one can write
\begin{gather}
\ket{\text{TFD}} = \frac{1}{\sqrt{Z}} \sum_n e^{-\beta E_n/2} \ket{n}_L \otimes \ket{n}_R.
\end{gather}

With the Hamiltonian $H_{\text{tot}}=H_L+H_R$, the time evolution would be $U(t)= e^{-i H_{\text{tot} } t} $, and the fermionic bilinear coupling would be
\begin{gather}
e^{i \mu V}, \ \ \ \ \ \ V= \frac{1}{qN}H_{\text{int}}, \ \ \ \ \ H_{\text{int}} = i \sum_{i=1}^N \psi_L^i \psi_R^i.
\end{gather}

Note that an operator like $\langle \mathcal{O}_H (\infty) \mathcal{O}_H(1) \left(\sum_{\psi} \ket{\psi} \bra{\psi}\right) \mathcal{O}_L (z) \mathcal{O}_L (0) \rangle$ could also be written as partial wave decomposition which has all the dynamical data that we need to track the black hole and its quantum corrections. The interesting point is that, in the case of AdS, the vacuum would dynamically break the Virasoro symmetry, the stress energy tensor $T(z)$, and its derivatives, similar to the jump operators in Lindblad formalism. 

In \cite{Fitzpatrick:2015foa}, it has been shown that the operators of the order $\frac{1}{\sqrt{c}^{| k-l |}  }$ contribute in the heavy-light limit, and as this factor $| k-l |$ changes over the evolution of black hole radiation, the non-monotonic behaviors could be expected.

Also, in \cite{Fitzpatrick:2015zha}, using other methods, the heavy-light limit of vacuum Virasoro conformal block is found as
\begin{gather}
\mathcal{V}(z) = \text{exp} \left \lbrack 2h_L f \left ( \frac{h_H}{c}, z \right) \right \rbrack.
\end{gather}
Here $f$ is a function that will be shown only in the terms that are in single power of $h_L$, which would be also the case for the functionality of coupling $\mu$ between a SYK system coupled to a bath.

\begin{figure}[ht!]   
\begin{center}
\includegraphics[width=0.8\textwidth]{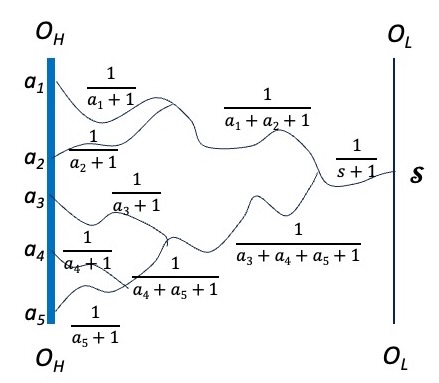} 
\caption{The propagations between a heavy and light operators and the diagrammatic rules discussed in \cite{Fitzpatrick:2015foa} are shown. The heavy operator can be considered as the system and the light operator as the bath and the propagation mechanism between them can create the coupling and its non-trivial behavior. The non-monotonic behavior of the dissipative gap versus coupling which has been found in \cite{Zheng:2025apa}, could be explained by the behavior of such propagations.}
\label{fig:prop4}
\end{center}
\end{figure}

\begin{figure}[ht!]   
\begin{center}
\includegraphics[width=0.55\textwidth]{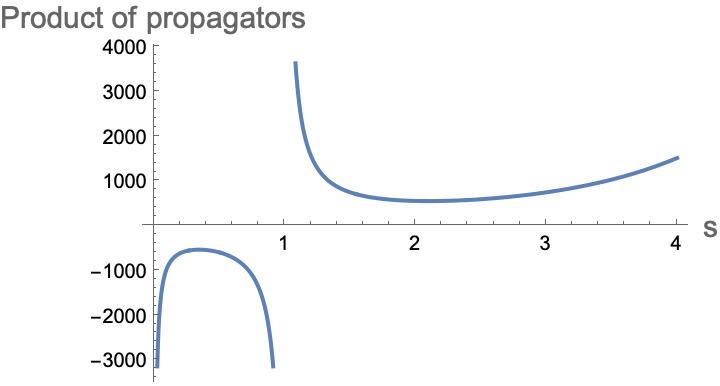} 
\caption{The behavior of the product of propagation $2 . \frac{6^k z^s}{s(s-1)} $ versus $s$ for $z=4$ and $k=2$. The non-monotonicity and the pole which correspond to EPs, as in \cite{Zheng:2025apa, Ghodrati:2025fah, Ghodrati:2023uef}, are evident here.}
\label{fig:product}
\end{center}
\end{figure}

Note that these $a_i$s which are depicted in Figure \ref{fig:prop4} can be considered as momenta flow between propagators. After taking the product of the propagators, based on \cite{Fitzpatrick:2015foa}, one should multiply the result by a factor of $2 . \frac{6^k z^s}{s(s-1)} $, where $s \equiv \sum_i a_i$, and as one could see from Figure \ref{fig:product}, this factor produce non-monotonicity in the correlation.

One can then consider the action in \cite{Fitzpatrick:2015foa} as
\begin{gather}
S= \int dt \int_0^{2\pi} d\theta \left(\phi^\dagger (\partial_t - \partial_\theta-1) \phi + g \phi^\dagger \phi( \phi^\dagger+\phi)\right),
\end{gather}
with the coupling $g\propto \frac{1}{\sqrt{c}}$. Note that the field $\phi(t, \theta)$ has a mode expansion of the form of
\begin{gather}
\phi_m= e^{i(m+1)t-im \theta},
\end{gather}
and with the propagator $\frac{i}{\omega-(m+1)}$.

Then, we could compare this mode expansion with the Loschmidt echo for dissipative quantum systems describing the overlap between the initial and time-evolved states of the vectorized density matrix as
\begin{gather}
iS(t) = \lim_{N\to\infty} \frac{\ln \text{Tr}_{\mathcal{H^+}\otimes \mathcal{H^-} } e^{t \mathcal{L}} }{N}.
\end{gather}

Thus, this Loschmidt echo can be considered as a relationship between two mode expansions, i.e., as a mode overlap.

The Hawking particle creation could then be modeled by a linear jump operator $L^i = \sqrt{\mu} \psi^i,  i=1,..., N$, or by quadratic jump operators $L^a = \sum_{1 \le i < j \le N} K_{ij}^a \psi_i \psi_j$, which would be related to one-loop or two-loops quantum corrections.

The Lindblad-SYK effective action is also given in \cite{Zheng:2025apa} as
\begin{gather}
iS = \frac{1}{2} \log \text{det} ( \delta_{ab} \partial - \Sigma_{ab} ) - \frac{i \mu}{2} \int_0^t d\tau ( G_{+-}(\tau, \tau) - G_{-+} ( \tau, \tau))\nonumber\\
- \frac{1}{2} \sum_{ab} \int_0^t \int_0^t d\tau d\tau' \lbrack \Sigma_{ab} ( \tau, \tau') G_{ab}(\tau, \tau') + \frac{J^2}{q} t_{ab} G_{ab} ( \tau, \tau')^q \rbrack.
\end{gather}

In black hole evaporation, the open dynamics can be considered as a sequence of Markov jump processes, producing entropy and determining the temperature. As in \cite{Chen:2025qdj}, one can choose a jump operator $L_k$ that satisfies the relation
\begin{gather}
L_k= e^{\phi^k/2} L_{k'}^\dagger,
\end{gather}
where $\phi^k = - \phi^{k'}$ is the entropy change in the environment after the action of the jump operator $L_{k}$.
On the other hand, the Virasoro vacuum conformal block has been found in \cite{Fitzpatrick:2015foa} to be 
\begin{gather}
\mathcal{V}(t) = e^{h_L t} \left ( \frac{\pi T_H}{\sin (\pi T_H t) } \right )^{2h_L},
\end{gather}
where 
\begin{gather}
T_H = \frac{\sqrt{\frac{24h_H}{c} -1 } }{2\pi},
\end{gather}
is the temperature, with $\beta=1/T_H$ the periodicity in Euclidean time $t$. Here $dE$ is proportional to $e^{h_L t}$. In Figure \ref{fig:prop4}, the terms proportional to $h_L$ in the correlator with light operators are
\begin{gather}
\langle X_{a_1, a_2} \mathcal{O}_L (z) \mathcal{O} (0) \rangle = \frac{h_L}{c} \frac{a_1 (a_1-1)a_2(a_2-1) }{s(s+1)} z^s+\mathcal{O}(h_L^2),
\end{gather}
which again indicates that the change of energy in the propagation of a graviton is proportional to $h_L$.
So we have
\begin{gather}
dS= \frac{dE}{T} \propto \frac{2\pi e^{h_L t} }{\sqrt{\frac{24 h_H}{c}-1} } \approx \phi^k.
\end{gather}

So, the best approximation for the jump operators would be
\begin{gather}
L_k \propto \frac{\pi e^{h_L t} }{\sqrt{\frac{24 h_H}{c} -1 }}.
\end{gather}

Also, for an Unruh-DeWitt (UDW) detector, the jump operators have been proposed as \cite{Chen:2025qdj},
\begin{gather}
L_1= \sqrt{\mathcal{C}(\omega)} \sigma_{-}, \ \ \ \ \ \   L_2= \sqrt{\mathcal{C}(-\omega)} \sigma^\dagger_{-},  \ \ \ \ \ \   L_3= \sqrt{\mathcal{C}(0)} \pi_{-},  \ \ \ \ \ \    L_4= \sqrt{\mathcal{C}(0)} \pi^\dagger_{-},
\end{gather}
where $\sigma_- = \ket{0} \bra{1}$ and $\pi_- = \ket{1} \bra{1}$, which are related to the light operator dimension $h_L$.

At first order of approximation,  we can assume, to first approximation, the map of $h_{\mu\nu}^{(n)} \to \varepsilon_{\mu\nu}$,  and by expanding the exponential functions,  we can write the simplified jump operators which are constructed from the Schwarzian modes as $L_n=\sqrt{\mu}\,\varepsilon_n$ and $L_{nm}=\sqrt{\mu}\,\varepsilon_n\varepsilon_m$, for linear and quadratic jump operators, respectively, corresponding to one-loop and two-loop quantum corrections. So, in the first approximate, we can assume that the structure of these jump operators would be similar to that of the Lindblad-SYK model of \cite{Zheng:2025apa}.

The general form of the Lindblad equation \cite{Lindblad:1975ef, Gielen:2009qya} can be written as
\begin{gather}
\frac{d\rho}{dt} = - i \lbrack H , \rho \rbrack - \frac{1}{2} \sum_{\alpha, \beta \ne 0} h_{\alpha, \beta} (L^\beta L^\alpha \rho + \rho L^\beta L^\alpha - 2 L^\alpha \rho L^\beta).
\end{gather}

For the SYK model coupled to a Markovian bath, the Liouvillian can be simplified to \cite{Zheng:2025apa}
\begin{gather}
\mathcal{L}(\rho)= \frac{d\rho}{dt} = - i \lbrack H_{\text{SYK}}, \rho \rbrack + \sum_k L_k \rho L^\dagger_k - \frac{1}{2} \lbrace L^\dagger_k  L_k, \rho \rbrace.
\end{gather}

Then, in a very simplified form and in the first approximation, and for the Schwarzian case, we could write a a Lindbladian of the form
\begin{gather}
\frac{d\rho}{dt} = - i \lbrack H_{\text{Schwarzian}} , \rho \rbrack+ \sum_k \left ( L_k \rho L^\dagger_k - \frac{1}{2} \lbrace L_k^\dagger L_k , \rho \rbrace \right ) + \sum_{n,m} \left ( L_{nm} \rho L^\dagger_{nm} - \frac{1}{2} \lbrace L^\dagger_{nm} L_{nm}, \rho \rbrace \right) \nonumber\\
= - i \lbrack H_{\text{Schwarzian}} , \rho \rbrack + \sum_k \left(\mu \varepsilon_k \rho  \varepsilon_k^\dagger - \frac{1}{2} \lbrace \mu \ \varepsilon_k^\dagger \varepsilon_k , \rho \rbrace \right)+ \sum_{n,m} \left ( \mu \varepsilon_n \varepsilon_m \rho \varepsilon_m^\dagger \varepsilon_n^\dagger - \frac{1}{2} \lbrace \mu \varepsilon_m^\dagger \varepsilon_n^\dagger \varepsilon_n \varepsilon_m , \rho \rbrace \right ).
\end{gather}

For this approximation to work, we should also have Markovianity and large central charge, so could be able to have perturbative expansion in $h_H/c$.

One would then expect that in these black hole systems, the decay rate toward steady states and the dissipative gap $\Gamma_0=\Gamma_0[\epsilon_n,\mu]$ would be non-monotonic functions of the bath coupling $\mu$ similar to the case of \cite{Zheng:2025apa, Jensen:2016pah} which exhibits the non-monotonic behavior observed in the Lindblad SYK model \cite{Zheng:2025apa}.

Also, in \cite{Chen:2025qdj}, the transition rates $\omega_k^{xy}(t) := | \langle x(t) | L_k | y(t) \rangle^2$, which for the case of large-$N$ is related to the Green's function $G_{ab}$ \cite{Zheng:2025apa},
\begin{gather}
G_{ab} ( \tau) = \frac{1}{N} \sum_i \langle \psi_i^a (\tau) \psi_i^b (0) \rangle.
\end{gather}

In \cite{Zheng:2025apa}, $G_{++}(\tau)$ has been found as
\begin{gather}
G_{++} (\tau) = A e^{- \Gamma_0 \tau} \sin (\Omega \tau+b),
\end{gather}
while in \cite{Chen:2025qdj}, the $++$ part of the transition rate has been found as
\begin{gather}
\omega_1^{++} = \omega_1^{--} = \frac{\mathcal{C}(\omega) \sin^2 \Theta }{2}.
\end{gather}

\begin{figure}[ht!]   
\begin{center}
\includegraphics[width=0.37\textwidth]{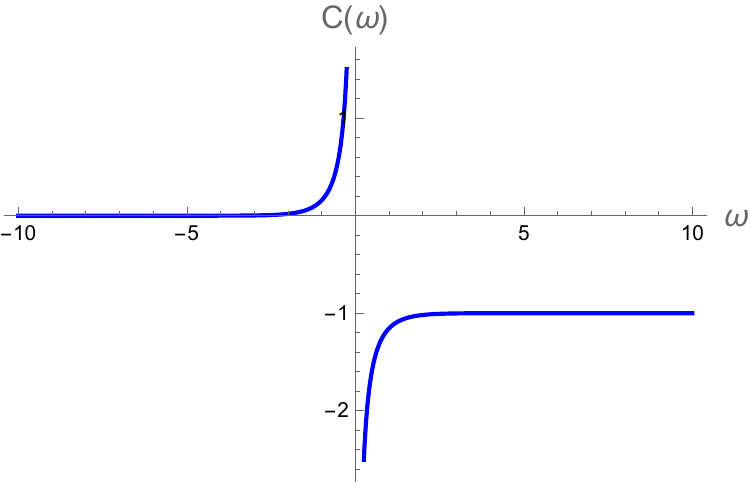}  \ \ \ \  \ \ \ \ \ \ \ \ \ \ \ 
\includegraphics[width=0.37\textwidth]{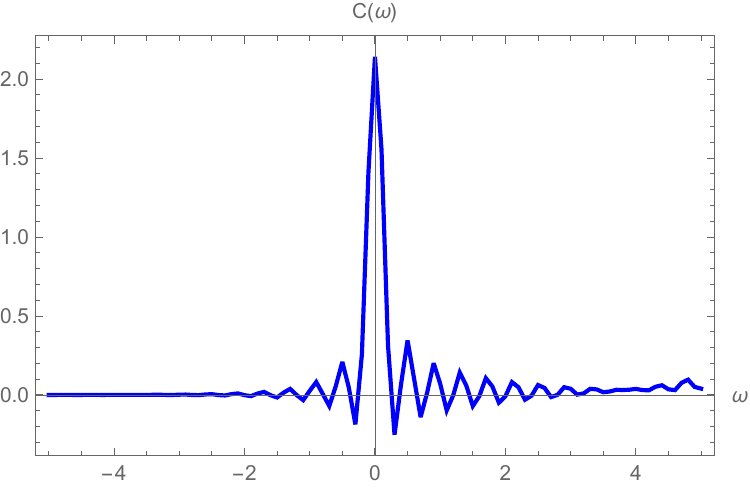} 
\caption{The behavior of the response function $\mathcal{C}(\omega)$ versus $\omega$ is shown, for the case of a Hawking-Unruh bath in Rindler, Schwarzschild, or de Sitter spacetime, on the left and for the case of Unruh-DeWitt detector outside a BTZ black hole, as calculated in \cite{Chen:2025qdj}, on the right.}
\label{fig:response}
\end{center}
\end{figure}

Therefore,  the response function $\mathcal{C}(\omega)$ is related to the dissipative gap $\Gamma_0$, and again based on the result of \cite{Zheng:2025apa}, one would expect a non-monotonic behavior. In fact, for the case of a Hawking-Unruh bath in Rindler, Schwarzschild or de Sitter spacetime, the transition rate is $\mathcal{C}(\omega)\sim ( e^{- \beta \omega}-1)^{-1} $, as shown in the left part of Fig. \ref{fig:response} and in the right part of Fig. \ref{fig:response}, the response for the case of rotating BTZ is shown. In both cases, the non-monotonicity due to the presence of EPs, similar to the case of dissipative gap in \cite{Zheng:2025apa}, can be noticed.

\section{Reparameterizations of AdS$_3$ gravity and boundary Liouville}\label{sec:rep}

In this section, we would like to consider the effects of quantum corrections, in $3d$, and mainly in the Cotler-Jensen theory of \cite{Cotler:2018zff}, which is a theory of reparameterizations for AdS$_3$ gravity, and also compare with the effects of quantum corrections in $2d$ JT. First, we review some aspect of Cotler-Jensen theory and then explain which parts of their methods and considerations are related to our studies of quantum corrections in $3d$ and $2d$.

As mentioned in \cite{Cotler:2018zff}, sometimes the dual of $\text{AdS}_3$ would be a boundary Liouville theory \cite{Coussaert:1995zp}. We would like to know, how this boundary Liouville theory could have an effect on the greybody factor and on Hawking radiation. The pure gravity in this case is on a Lorentzian cylinder. However, one should note that in the Liouville case, the spectrum is continuous and with no normalizable vacuum, while $\text{AdS}_3$ has discrete spectrum and normalizable vacuum. Finding how these features will play a role in our problem with quantum corrections, would be interesting.

As we try to find connections between $2d$ JT and $3d$ gravity, one should also note that a foundational paper is \cite{Ghosh:2019rcj}, where the authors by bootstrap methods showed that in most classes of $2d$ conformal field theories, there is a sector described by the Schwarzian theory. Specifically, they showed that at low-temperature and large angular momentum sourced by chemical potential, the dual of this Schwarzian sector is a near-extremal rotating BTZ black hole, with a nearly $\text{AdS}_2 \times S^1$ in their near horizon as its throat. They also could reproduced many CFT results by matching the physics in the throat to observables in $1d$ higher, and at $\text{AdS}_3$ boundary. Specially, they found the connections between the two-point function of CFT side and Schwarzian side 
\begin{gather}
G^{(\text{CFT})} (z, \bar{z} ; \beta_L, \beta_R) \sim  \left ( \frac{c}{6} \right)^{-2h} \left \lbrack \frac{1}{\pi T_R} \sinh ( \pi T_R \bar{z}) \right \rbrack^{-2 \bar{h}} G_h^{(\text{Schw})} (-z).
\end{gather}
This is somehow similar to our work here. However, in that work quantum corrections were not discussed, while here our emphasis is on how quantum corrections in this Schwarzian sector is related to quantum corrections in JT and also to quantum corrections in $\text{AdS}_3$ case. There, the connections to Liouville case also were discussed, which again is connected to our work here.

In a related interesting paper, the authors discussed the connection between the $3d$ and $2d$ points of view with emphasis on the important backreaction effects \cite{Castro:2025pst}. They discussed the effects of near-extremal gravitational backreaction on the dynamics of a scalar field. The effects of the backreaction on $\text{AdS}_2$ have been shown as the corrections in the two-point function of the operator dual to the scalar field at tree-level and the correlation function as
\begin{gather}
\frac{\langle \mathcal{O}_\Delta (\tau) \mathcal{O}_\Delta (0) \rangle } {\langle \mathcal{O}_\Delta (\tau) \mathcal{O}_\Delta (0) \rangle_{\text{free}}} = 1 + c_\Delta  \left( 2+ \Delta \pi \frac{1-2\tau/ \beta}{\tan (\frac{\pi \tau}{\beta})} + S_\Delta (\tau) \right) + O(\epsilon^2),
\end{gather}
where $S_{\Delta=1}(\tau)=0$, and the general form of $S_\Delta(\tau)$ is given in \cite{Castro:2025pst}. This relation also has been extended to the case where $\Delta$ is not an integer. These results can also be examined from the perspective of two coupled systems, where the corrections are related to the light operator and the backreaction and the free part is related to the heavy operator and the system, as we did in the previous section. Then, the exceptional point and its effects on the non-monotonicity of couplings, the dissipative gap and decay rate again could be studied using these results.

When one writes the pure $\text{AdS}_3$, on a cylinder, the boundary conditions which make the bulk asymptotically $\text{AdS}_3$ can constraint the gauge fields, and then the chiral Wess-Zumino-Witten (WZW) model would become a constrained $SO(2,1) \times SO(2,1)$ chiral WZW on the boundary. Then, by redefinition, it can become constrained $SO(2,1)$ WZW and then can be rewritten as a Liouville theory. So when considering the quantum fluctuation effects on the Hawking radiation and greybody factor, one should consider such constraints as well.

The field theory in this case, due to the boundary condition of $\text{AdS}_3$, keeps the $A_0$ part of $A=A_0 dt+ \tilde{A_i} dx^i$ non-zero, which then makes the boundary model to get quasi-local $SO(2,1) \times SO(2,1)$ quotient.

The main boundary theory that we consider is
\begin{gather}
S= - \frac{C}{24\pi} \int d^2x \left ( \frac{(\partial_+ \phi')\phi''}{\phi'^2} - (\partial_+ \phi) \phi' \right), \ \ \ \ \ \partial_+ = \frac{1}{2} (\partial_\theta+\partial_t),
\end{gather}
which is the quantum field theory of the (left-moving) boundary gravitons of pure $\text{AdS}_3$ gravity.

For the case of BTZ, the boundary action is
\begin{gather}
S= -S_- \lbrack \phi_+ \rbrack + S_+ \lbrack \phi_- \rbrack + S_+ \lbrack \bar{\phi}_+\rbrack - S_- \lbrack \bar{\phi}_-\rbrack.
\end{gather}

Similar to JT, the boundary description of the two-sided BTZ black hole has non-local constrains between the degrees of freedom and between the two asymptotic boundaries, and therefore has also a non-factorization property which would affect the greybody factor and the Hawking radiation.

An interesting feature of this theory is that at any fixed time, $\phi$ would be an element of $\text{Diff}(\mathbb{S}^1)/PSL(2; \mathbb{R})$. It would be interesting to check whether adding quantum corrections, which evolve during the Hawking radiation, this fact would remain true, and also to consider their effects on the fields in the bath and islands.

Another important point in the theory we consider here \cite{Cotler:2018zff}, is that the higher-order terms are expected to be computed from the higher loops in the $1/c$ expansion, and by comparing with \cite{Chen:2016cms}, the authors found that there would be non-perturbative contributions to the Virasoro blocks at large central charge, which would be suppressed as $\text{exp}(-c)$. This in fact, might be due to instanton configurations in the theory. The effects of these terms on greybody factor or Lyapanov exponent will be studied in or work here. One might think that these reparametrization fields and modes for the case of $\text{AdS}_3$, unlike the JT case and nearly-$\text{AdS}_2$ scenario, would ``not" correspond to matter fields in the gravity dual, and they just act as instantons, which only shift the central charge and move along the discrete spectrum. Also, note that this theory which we considered here, similar to JT, can support wormholes and has non-factorization behavior.

Now we can study the calculations of one-loop quantum corrections and their effects on various thermodynamical properties of black hole.

\subsection{Quantum correction, Schwarzian mode and Kaluza-Klein modes}

To understand the theory of \cite{Cotler:2018zff} better we need to track the Kaluza-Klein (KK) modes. The CFT dual of string theory on $\text{AdS}_3 \times \text{S}^3/\mathbb{Z}_k \times \mathbb{T}^4$ can be described by the subspace of the symmetric orbifold of $\mathbb{T}^4$. They contains the low-lying excitations on top of a certain reference state, where such ``non-perturbative” reference states are in the ``twisted sector”. This sector is associated to the conjugacy class, which only consist of k-cycles.

In \cite{Anninos:2010sq}, it has been noted that the one-loop quantum gravitational effects can change the tree-level results in holography. In three dimensions, we have infinite saddles and we could dimensionally reduce each of them to $2d$ and then sum them all and get the partition function of the $2d$ case. But the main issue is to find the weight of each saddle. The quantum correction around each saddle could also make one to reach to a new saddle as well, as
\begin{gather}
\bar{\phi} (x,y) \sim \bar{\phi} (x,y) + \bar{a} (y),
\end{gather}
which is a gauge symmetry.

The global $\text{AdS}_3$ has a winding number, as the boundary is torus, and therefore a winding number here could be defined. This is a winding around the $\text{PSL}(2,R)$. The Chern-Simon action also becomes a chiral $\text{SO}(2,2)$ WZW action. 

In $3d$, the modular sum develops a logarithmic divergence and the source of this divergence would be a zeta function. In recent work of \cite{Pathak:2026yyb}, the authors showed that the fluctuations in temperature is related to the near-horizon supertranslations of a black hole. Then, they could show that the temperature-dependent corrections to the near-horizon part of the Euclidean gravity action or free energy can be written in terms of the polynomial functionals of a near-horizon supertranslation for all orders. The shape and behavior of such polynomial functionals in higher dimensions would be similar. Note also that the near-horizon partition function could be written in terms of the sum over all possible near-horizon supertranslations.

Note that these quantum fluctuations would lead to a chaotic system, determined by the Lyapunov exponent $\lambda_L= 2\pi T (1- \alpha G)$ in $2d$, OTOC, and Virasoro vacuum blocks with block exponentiates of $\mathcal {F}(t) \sim \text{exp}\Big \lbrack \frac{2\pi}{\beta}t \left (1- \frac{12}{c} \right ) \Big \rbrack $ in $3d$. One could consider the intrinsic patterns and underlying laws of chaos and their relations between $2d$ and $3d$, based on the new studies of \cite{zhang2024brownianspinlockingeffect}, as there are some quantum effects that can become amplified by random chaotic interactions. We will do it in section \ref{sec:Lyapunov}, as we check how these effects can defuse from $2d$ to $3d$.

In $3d$, and for a torus boundary of complex structure $\tau$, the Maloney and Witten partition function would be a sum over fluctuations around real, non-singular, Euclidean saddles with boundary complex structure $\tau$. One could then consider how these Euclidean saddles would look like in a one-dimension lower where $1d$ is compactified, i.e, in $2d$. In fact, in $3d$, we have sum over topologies, so one might think that in one-dimension lower, we still would have sum over similar topologies, where one-dimension is compactified. In Maloney-Witten prescription, this sum would become a modular sum, so in $2d$, one could imagine a similar structure.

If one chooses a suitable modular transformation $\tau \to \tau'$, one could take this contractible circle to be the spatial one. This could be imagined in one dimension lower and it would become a contractible line, and still the sum over ``topologies" would become a modular sum. For any $\tau'$, there would be an Euclidean BTZ geometry, where its contribution to the bulk partition function, which is the saddle with all the fluctuations around it, would be the Virasoro character of the vacuum module at complex structure $\tau'$. This can be written as
\begin{gather}
\mathcal{Z}_{\text{vac}} (\tau', \bar{\tau}') = \Big | q'^{-k} \prod_{n=2}^\infty \frac{1}{1-q'^n} \Big |^2, \ \ \ \ q'= e^{2\pi i \tau'}, \ \ \ \ \ \ k=\frac{c}{24}.
\end{gather}

The Maloney-Witten partition function, which is a sum of modular image of vacuum character, would be
\begin{gather}
\mathcal{Z}_{\text{MW}} (\tau, \bar{\tau}) = \sum_{g \in PSL(2; \mathbb{Z})} \mathcal{Z}_{\text{vac}} (g\tau, g \bar{\tau}).
\end{gather}

This is divergent and needs a modular invariant regularization. After doing that, the result no longer would have a well-behaved small $q$ expansion, as there are logarithmic terms. Now, let's look at it from one dimension lower.

Writing the torus modulus as
\begin{gather}
\tau = \theta + i \frac{\beta}{2\pi \ell},
\end{gather}
in the limit of $\beta \gg \ell$, which is the case of low temperature and long circle, one could do a dimensional reduction and for large $\text{Im} \tau$, one gets
\begin{gather}
Z_{\text{vac}} \ \approx \ \text{exp} \left ( \frac{\pi^2 c}{3 \beta} \right ) \prod_{n=2}^\infty \frac{1}{\left ( 1- e^{-\frac{2\pi^2 n}{\beta} } \right )^2},
\end{gather}
where the exponential piece is the classical extremal entropy and the product is the quantum boundary gravitons, and so in this regime, the $\text{AdS}_3$ pinches to $\text{AdS}_2 \times S^1$. Also,
\begin{gather}
\prod_{n=2}^\infty \frac{1}{1-q^n} \longrightarrow \int \mathcal{D} f \text{exp} \left \lbrack -C \int dt \{ f(t),t \} \right \rbrack,
\end{gather}
where the central charge maps as $C= \frac{c}{24\pi}$.

If one considers quantized $k=1,2,...$ and includes complex saddles, one could get a sensible CFT partition function which is equal to Witten extremal CFT partition function. The picture of such complex saddles and complex BTZ in one dimension lower, and specially for the case of $k=1$ which corresponds to Monster CFT would be interesting. We  will further study them in section \ref{sec:complexBTZ}.

As the bulk Chern-Simons gauge fields are constrained by the boundary conditions,  the bulk spacetime should be asymptotically $\text{AdS}_3$ and the dual theory would be a constrained $SO(2,1) \times SO(2,1)$ chiral WZW model on the boundary which by field redefinition can be rewritten as a Liouville theory. The effects of this boundary condition in $3d$ on the one dimension lower of $2d$ JT gravity or $2d$ BF theory and then in the holographic dual side, on $1d$ Schwarzian mode and $1d$ particle on group could also be considered, as in the case of \cite{Ghodrati:2022hbb, Ghodrati:2026xky}.

\subsection{Comparison of Schwarzian and Coadjoint Orbits}

in this section, we compare the the Schwarzian action with the reparametrized theory of Cotler-Jensen \cite{Cotler:2018zff}.

Consider the Schwarzian action
\begin{gather}
S= -\frac{1}{2g^2} \int_0^{2\pi} d\tau \left (  \left(\frac{\phi''}{\phi'}\right)^2 - ( \phi')^2 \right), 
\end{gather}
and then compare it with the Cotler-Jensen theory
\begin{gather}
S= -\frac{C}{24\pi} \int d\theta dt \left ( \frac{(\partial_+ \phi') \phi''}{\phi'^2} - (\partial_+ \phi) \phi' \right ),
\end{gather}
where $\partial_+ = \frac{1}{2} ( \partial_\theta + \partial_t)$.

So one could find the following relation here
\begin{gather}
g \to \sqrt{\frac{12\pi}{C} }.
\end{gather} 

As $C= 24k$, one could also write
\begin{gather}
g \to \sqrt{\frac{2k}{\pi}}.
\end{gather}

In the Schwarzian theory, we have
\begin{gather}
\phi(\tau) \in \text{Diff} (S^1),
\end{gather}
which is a diffeomorphism of a circle, while in the Cotler-Jensen at fixed $\theta$, we have the following relation
\begin{gather}
\phi_{\pm} \in \frac{\text{Diff} (\mathbb{R} ) \times \text{Diff} (\mathbb{R}) }{PSL(2;R)}.
\end{gather}

The case of $g \ll 1$ is actually related to large central charge $C$. The diagrammatic expansion in $\frac{1}{C}$ which computes the $\frac{1}{C}$ expansion for the block at large central charges is related to the semi-classical analysis of the Schwarzian action, by expanding the diffeomorphism $\phi(\tau)$ around the saddle point $\phi(\tau) = \tau$ \cite{Qi:2019gny} as
\begin{gather}
\phi(\tau) = \tau + g \ \epsilon (\tau),
\end{gather}
while in the Cotler-Jensen we have the following relation
\begin{gather}
\phi = \phi_0 +\sum_{m,n} \frac{\epsilon_{m,n} }{(2\pi)^2} e^{\frac{imy}{\text{Im} (\tau) } + i n \left (\theta-\frac{\text{Re} (\tau) }{\text{Im} (\tau) } y \right )},\ \  \ \ \ \ \epsilon^*_{m,n} = \epsilon_{-m, -n}.
\end{gather}

The Schwarzian theory has $SL(2, \mathbb{R})$ symmetry as
\begin{gather}
\tan \frac{\phi}{2}  \ \  \longrightarrow  \ \ \frac{a \tan \frac{\phi}{2} + b }{c\tan \frac{\phi}{2} +d} \ \ \ \  \ \ \ (a,b,c,d \in \mathbb{R} \ \ \text{and} \ \ ad-bc=1), 
\end{gather}
while the Cotler-Jensen theory has quasi-local $SO(2,1) = PSL (2; \mathbb{R})$ redundancy
\begin{gather}
\tan \left ( \frac{\phi(\theta,t) }{2} \right ) \ \  \longrightarrow  \ \ \frac{a(t) \tan\left ( \frac{\phi(\theta,t) }{2} \right) + b(t)  }{c(t) \tan \left ( \frac{\phi(\theta,t) }{2} \right ) + d(t)}, \ \ \ \ 
\begin{pmatrix}
a(t) & b(t) \\
c(t) & d(t) 
\end{pmatrix} \in PSL(2; \mathbb{R}).
\end{gather}

In the Schwarzian, the physical degree of freedom lives on the quotient space $\text{Diff}(S^1) / SL(2)$ while in Cotler-Jensen theory, the $\phi$ field is an element of the quotient space $\text{Diff}(S^1) / PSL(2; \mathbb{R})$. In the Schwarzian case, the partition function is
\begin{gather}
Z\lbrack g \rbrack = \int \frac{\mu \lbrack \phi \rbrack }{SL(2, \mathbb{R})} \text{exp} \left \lbrack - \frac{1}{2g^2} \int_0^{2\pi} d\tau \left ( \left (\frac{\phi''}{\phi'} \right )^2 - (\phi')^2   \right ) \right \rbrack,
\end{gather}
which after $SL(2, \mathbb{R})$ gauge-fixing, the measure becomes
\begin{gather}
\mu \lbrack \phi \rbrack \equiv \frac{\mathcal{D} \phi}{\prod_\tau \phi'(\tau)} .
\end{gather}

For the case of Cotler-Jensen, then, the measure would be either $\lbrack d\phi \rbrack \text{Pf}(\omega)$ which can be written as 
\begin{gather}
\prod_t \left ( \text{Pf} ( \omega (\phi)) \prod_\theta d\phi \right),
\end{gather}
which is invariant under local $PSL(2; \mathbb{R})$ transformation, or we could write
\begin{gather}
\prod_{\theta, y} \frac{dF}{F'}, \ \ \ \ \ \ \ F= \tan \frac{\phi}{2},
\end{gather}
which are actually equivalent. So the measure in Cotler-Jensen can be written as
\begin{gather}
\mu \lbrack \phi(\theta,t) \rbrack \equiv \frac{\mathcal{D} \phi (\theta, t)}{\prod_t \left ( \text{Pf} ( \omega (\phi)) \prod_\theta d\phi \right)}.
\end{gather}

For the case of Schwarzian, then the measure $\mu \lbrack \phi \rbrack$ can be exponentiated by introducing a fermion $\psi(\tau)$ which makes the partition function 
\begin{gather}
Z \lbrack g \rbrack = \int \frac{\mathcal{D}\phi \mathcal{D}\psi}{SL(2, \mathbb{R}) } e^{-S},
\end{gather}
where the total action $S'$ would be
\begin{gather}
S= \int_0^{2\pi} d\tau \left \lbrack \frac{1}{g^2} \left ( \frac{\phi''}{\phi'} \right)^2- \frac{1}{g^2} ( \phi')^2 + \frac{\psi'' \psi'}{(\phi')^2}- \psi' \psi \right \rbrack.
\end{gather}

For the case of Cotler-Jensen, the path integral measure for $\phi$ can be found by considering that from the bulk measure for $A$. One can find the Haar measure for the boundary degree of freedom $g$, which can be written in terms of Gauss parameterization as
\begin{gather}
\lbrack dg \rbrack = \prod_{\theta, t} d\lambda d\Psi dF \lambda.
\end{gather}

Here the Gauss parameterization of $SL(2; \mathbb{R})$ group elements is
\begin{gather}
g= \begin{pmatrix}
1 & 0\\
F & 1 
\end{pmatrix}
\begin{pmatrix}
\lambda & 0\\
0 & \lambda^{-1} 
\end{pmatrix}
\begin{pmatrix}
1 & \Psi\\
0 & 1 
\end{pmatrix},
\end{gather}
and for the barred fields, it would be
\begin{gather}
\bar{g}= \begin{pmatrix}
1 & -\bar{F} \\
0 & 1 
\end{pmatrix}
\begin{pmatrix}
\bar{\lambda}^{-1} & 0\\
0 & \bar{\lambda}
\end{pmatrix}
\begin{pmatrix}
1 & 0\\
\bar{\Psi} & 1 
\end{pmatrix}.
\end{gather}

For the global $\text{AdS}_3$ with the metric
\begin{gather}
g=-(r^2+1) dt^2 + r^2 d\theta^2 + \frac{dr^2}{r^2+1},
\end{gather}
they would be
\begin{gather}
g = \begin{pmatrix}
\rho \cos \left ( \frac{x^+}{2} \right) & -\rho^{-1} \sin \left ( \frac{x^+}{2} \right)\\
\rho \sin \left ( \frac{x^+}{2}\right)  & \rho^{-1} \cos \left (\frac{x^+}{2}  \right)
\end{pmatrix}, \ \ \ \ \ \ \ \ 
\bar{g} = \begin{pmatrix}
\rho^{-1} \cos \left ( \frac{x^-}{2} \right) & -\rho \sin \left (\frac{x^-}{2} \right)\\
\rho^{-1} \sin \left ( \frac{x^-}{2}\right)  & \rho \cos \left (\frac{x^-}{2}  \right)
\end{pmatrix},
\end{gather}
where $x^{\pm}= \pm t + \theta$ and
\begin{gather}
\rho = \sqrt{\sqrt{r^2+1} + r}.
\end{gather}

The Chern-Simons gauge fields of $\text{AdS}_3$, due to the Einstein’s equations and torsion-free constraint, would follow the relations
\begin{gather}
A= g^{-1} dg, \ \ \ \ \ \ \ \ \bar{A} = \bar{g}^{-1} d\bar{g}^{-1}.
\end{gather}

By considering the constraints on $\lambda$ and $\Psi$, one can find the Haar measure as 
\begin{gather}
\prod_{\theta, t} dF \int d\lambda d\Psi \ \lambda \ \delta(\lambda^2 F'-r) \delta \left( \Psi + \frac{F''}{2rF'} \right)= \prod_{\theta, t} \frac{dF}{F'} = \prod_{\theta,t} \frac{d\phi}{\phi'}.
\end{gather}

Then, if we wish to add  ``fermions" or ghost, and similar fields to the Schwarzian case and exponentiate the action, we could consider the hidden supersymmetry of the phase space path integral.  These are actually being used to localize the path integral, which are based on the Stanford and Witten approach \cite{Stanford:2017thb}.

If one considers the presymplectic potential $\alpha = \alpha_i dx^i$ which satisfies $\omega= d\alpha$, the $\text{AdS}_3$ action can be written as
\begin{gather}
S=- \int dt (\dot{x}^i \alpha_i + H),
\end{gather}
which after Wick-rotation $t=-i y$ can be written as
\begin{gather}
S_E= \int_0^\beta dy \left ( i \frac{\partial x^i}{\partial y} \alpha_i + H \right ),
\end{gather}
and the partition function is
\begin{gather}
\mathcal{Z} = \int \lbrack dx^i \rbrack \text{Pf}(\omega) e^{-S_E}.
\end{gather}

In this theory, $Pf(\omega)$ is the measure on the phase space at each time, and similar to the Schwarzian case, can be exponentiated by introducing the Grassmann-odd fields $\psi^i$, which obeys the same boundary conditions as the $x^i$, i.e, $\psi^i(y)= \psi^i (y+\beta)$. In fact, $\psi^i$ are better to be considered as ghosts rather than fermions.

The partition function and the action can then be written as
\begin{gather}
\mathcal{Z} = \int \lbrack dx^i \rbrack \lbrack d\psi^j \rbrack e^{- S'_E},  \ \ \ \ \ \ S'_E = \int_0^\beta dy \left ( i \frac{\partial x^i}{\partial y} \alpha_i+H- \frac{1}{2}\omega_{ij} \psi^i \psi^j \right).
\end{gather}

The canonical symplectic form is 
\begin{gather}
\omega= \frac{1}{2} ( \partial_i \alpha_j - \partial_j \alpha_i ) dx^i \wedge dx^j = -dx^i \wedge dx^j \partial_j \alpha_i,
\end{gather}
which as found in \cite{Cotler:2018zff} is
\begin{gather}
\omega= - \int_0^{2\pi} d\theta \left \lbrace \frac{C}{48\pi}  \frac{d\phi' \wedge d\phi''}{\phi'^2} + b_0 d\phi \wedge d\phi' \right \rbrace.
\end{gather}

So the total action would be
\begin{gather}
S'=  \int d\theta dt \left \lbrack -\frac{C}{24\pi} \left(  \frac{(\partial_+ \phi') \phi''}{\phi'^2} - (\partial_+ \phi) \phi' \right )-\frac{1}{2} \left \lbrace \frac{C}{48 \pi} \frac{d\phi' \wedge d\phi''}{\phi'^2} + b_0 d\phi \wedge d\phi' \right \rbrace \psi^i \psi^j \right \rbrack.
\end{gather}

Note that the coadjoint vector is a pair shown by $(b(\theta),t) $ which is being paired by $(f,a) $ as
\begin{gather}
\langle (b, t), (f,a) \rangle = \int_0^{2\pi} d\theta b f + t a.
\end{gather}

The orbits of the Virasoro group are two types, one which coincides with the orbit of a constant coadjoint vector $(b_0, C)$ for some $b_0$, and those which do not. Also, depending on $b_0$, there are three types of orbits. If $b_0 \ne - \frac{C n^2}{48 \pi}$ the orbits are ordinary. If $b_0 = - \frac{C}{48 \pi}$, the orbit is the  ``first exceptional orbit'', and if $b_0 = - \frac{Cn^2}{48 \pi} $, $n>1$, the orbits are the ``higher exceptional orbits''.

Then, one can do the field redefinition as
\begin{gather}
f= \text{ln} \left ( e^{\sqrt{B} \phi } \right)', \ \ \ \ \ \ \ \ B=\frac{48 \pi b_0}{C}.
\end{gather}

So, the Euclidean action and the stress tensor become
\begin{gather}
S_E = \frac{C}{24 \pi} \int d^2 x ( \bar{\partial} f) f', \ \ \ \ \ \ \ \ \ T= - \frac{C}{12} \left ( f'' - \frac{f'^2}{2} \right).
\end{gather}

The symplectic form also is 
\begin{gather}
\omega = \frac{C}{48 \pi} \int_0^{2\pi} df \wedge df'.
\end{gather}

So the total action would be 
\begin{gather}
S'_E = \frac{C}{24 \pi} \left \lbrack \int d^2 x ( \bar{\partial} f) f' - \frac{1}{4} \int_0^{2\pi} df \wedge df' \psi^i \psi^j\right \rbrack,
\end{gather}
where $\bar{\partial} = \frac{1}{2} (\partial_\theta + i \partial_y)$.

The saddle here is $f= \sqrt{B} \theta$, and expanding around it would lead to
\begin{gather}
f= \sqrt{B} \theta + \epsilon,
\end{gather}
which then gives the $\epsilon$-propagator as
\begin{gather}
\langle \epsilon(w) \epsilon(0) \rangle = - \frac{12}{C} \ln(1-u), \ \ \ \ \ u=e^{i \text{sgn} (y) w}, \ \ \ \ \ w=\theta+i y.
\end{gather}

The stress tensor then would be
\begin{gather}
T = 2\pi b_0 - \frac{C}{12} \left ( \epsilon'' - \sqrt{B} \epsilon' - \frac{\epsilon'^2}{2} \right).
\end{gather}

The two point function of $T$ is given by the sum of an exchange diagram and a one-loop diagram which reads as
\begin{gather}
\langle\langle T(w) T(0) \rangle \rangle = \frac{C+1}{32 \sin (\frac{w}{2} )^4} - \frac{2\pi (b_0+ \frac{C}{48\pi}) }{2\sin^2 (\frac{w}.{2} ) }.
\end{gather}

This then leads to 
\begin{gather}
c=C+1, \ \ \ \ \ \ \ \ h= 2\pi b_0 + \frac{C}{24}.
\end{gather}

In \cite{Qi:2019gny}, the expansion of the action around the saddle $\phi(\tau) = \tau + g \epsilon (\tau)$ has been found as
\begin{gather}
S= - \frac{\pi}{g^2} + S^{(2)}+g S^{(3)}+g^2 S^{(4)}+ \mathcal{O}(g^3),
\end{gather}
which leads to 
\begin{flalign}\label{eq:S234}
S^{(2)} &= \frac{1}{2} \int_0^{2\pi} d\tau \lbrack (\epsilon'')^2 -(\epsilon')^2 + \psi'' \psi' - \psi' \psi\rbrack,\nonumber\\
S^{(3)} &= \frac{1}{2} \int_0^{2\pi} d\tau \epsilon' \lbrack -2 (\epsilon'')^2 -2 \psi'' \psi'\rbrack, \nonumber\\
S^{(4)} &= \frac{1}{2} d\tau (\epsilon')^2 \lbrack 3 (\epsilon'')^2 + 3\psi'' \psi'\rbrack.
\end{flalign}

For the case of Cotler-Jensen, taking $f= \sqrt{B} \theta+ g \epsilon(\theta)$, we get
\begin{gather}
\bar \partial f = \frac{1}{2} (\partial_\theta +i \partial_y) \left( \sqrt{B} \theta + g \epsilon(\theta)\right)= \frac{1}{2} \left ( \sqrt{B} + g \epsilon' \right).
\end{gather}

So the Euclidean action would be
\begin{gather}
S_E= \frac{C}{48 \pi} \int d^2x \left ( B+ 2g \sqrt{B} \epsilon' + g^2 \epsilon'^2\right),
\end{gather}
which is one-loop exact. So in this case we have
\begin{flalign}
S^{(2)} &= \frac{C}{48 \pi} \int d^2 x \left ( B - \frac{1}{2} \psi \psi' \right), \nonumber\\
S^{(3)} &= \frac{C}{48 \pi } \int d^2x \left(2g \sqrt{B} \right ) \epsilon' , \nonumber\\
S^{(4)} &= \frac{C}{48 \pi} \int d^2 x \ g^2 \epsilon^2.
\end{flalign}

We could also try another method, which has been employed in \cite{Simon:2024dwm}. Here the total action that we consider would be
\begin{gather}
S'_E = \int dy \left ( i \frac{\partial x^i}{\partial y} a_i + H \right)- \frac{1}{2} \int dy \ \omega_{ij} \psi^i \psi^j.
\end{gather}

The locally $\text{AdS}_3$ metrics can be written as 
\begin{gather}
ds^2 = \frac{dr^2}{r^2}+ ( r^2 + G^2 \mathcal{L} \bar{\mathcal{L}}) dx d\bar{x} + G \mathcal{L} dx^2 + G \bar{\mathcal{L} } d\bar{x}^2,
\end{gather}
where $x= t + \varphi$, $\bar{x} = \varphi - t$, $G$ is the Newton's constant and $\mathcal{L}(t, \varphi)$, $\mathcal{\bar{L}}(t, \varphi)$ describe the phase space configurations.

The set of infinitesimal transformations which preserve the black hole boundary conditions are generated by
\begin{gather}
\xi = \sigma r \partial_r + \left ( \epsilon + \frac{\bar{\partial} \sigma }{r^2} + O (r^{-4} ) \right) \partial + \left ( \bar{\epsilon} + \frac{\partial \sigma}{r^2} + O(r^{-4}) \right) \bar{\partial}, \ \ \ \ \ \ \sigma= - \frac{\epsilon' + \bar{\epsilon}'}{2},
\end{gather}
where $\partial = \partial_x$, $\bar{\partial} = \partial_{\bar{x}}$, and $\epsilon = \epsilon(x)$, $\bar{\epsilon} = \bar{\epsilon} (x)$.

The action of the metric would cause the change in the phase space $L$ and $\bar{L}$ as
\begin{gather}
\delta L = \epsilon L' + 2 \epsilon' L - \frac{c}{3} \epsilon''', \ \ \ \ \ \delta \bar{L} = \bar{\epsilon} \bar{L'}+2\bar{\epsilon'} \bar{L} - \frac{c}{3} \bar{\epsilon}'''.
\end{gather}

Their finite versions are
\begin{gather}
\tilde{L} = f'^2 L(f) - \frac{c}{3} \lbrace f , x \rbrace, \ \ \ \ \ \ \ \ \tilde{\bar{L}} = \bar{f'^2} \bar{L} ( \bar{f}) - \frac{c}{3} \lbrace \bar{f} , \bar{x} \rbrace,
\end{gather}
which are parameterized by two diffeomorphisms $f(x)$ and $\bar{f} ( \bar{x})$.

The left-moving part of $3d$ Chern-Simons action can be written as
\begin{gather}
S_{\text{CS}} \lbrack f, j_0 \rbrack = \int dt \int_0^{2\pi} d\varphi \left ( j_0 f' ( f'+ \dot{f}) + \frac{c}{48\pi} \frac{f'' (f''+ \dot{f}') }{f'^2} \right),
\end{gather}
which is an example of 
\begin{gather}
S_{\text{CS}} \lbrack f, j_0\rbrack = \int_\gamma ( a+ H_v) dt.
\end{gather}

The one-form $a$ is
\begin{gather}
a = \int_0^{2\pi} d\varphi \left (j_0 f' \delta f+ \frac{c}{48\pi} \frac{f'' \delta f'}{f'^2} \right),
\end{gather}
and the symplectic form $\omega$ is
\begin{gather}\label{eq:sympl}
\omega = \int_0^{2\pi} d\varphi \left ( j_0 \delta f' \wedge \delta f + \frac{c}{48\pi} \frac{\delta f''}{f'^2} \wedge \delta f' \right ).
\end{gather}

The Wick rotation $t \to - i y$ of
\begin{gather}
S_{\text{CS}} \lbrack f, j_0 \rbrack = \int dt \int_0^{2\pi} d\varphi \left ( j_0 f' ( f'+ \dot{f}) + \frac{c}{48\pi} \frac{f'' (f''+ \dot{f}') }{f'^2} \right),
\end{gather}
would lead to
\begin{gather}\label{eq:SE}
S_{\text{E}} = \int dy d\varphi  \left ( j_0 f' ( f'+ i \partial_y f) + \frac{c}{48\pi} f'' (f''+ i \partial_y f')   \right).
\end{gather}

Note that there is a small mistake in the work of \cite{Castro:2021csm} as the action should actually be written as,
\begin{gather}\label{eq:SE}
S_{\text{E}} = \int dy d\varphi  \left ( j_0 f' ( f'+ i \partial_y f) + \frac{c}{48\pi} \frac{f'' (f''+ i \partial_y f')}{ f'^2}  \right).
\end{gather}

The ghost action then can be derived by inserting the symplectic form \ref{eq:sympl} into $S_\omega= -\frac{1}{2} \int dy  \omega_{ij} \psi^i \psi^j $, leading to
\begin{gather}\label{eq:Somega}
S_\omega= \int dy d\varphi \left ( j_0 \psi \psi' + \frac{c}{48 \pi} \frac{\psi' \psi''}{f'^2} \right).
\end{gather}

Note that for deriving this, one uses the relations
\begin{gather}
- \frac{1}{2} \delta f' \wedge \delta f \psi^i \psi^j = \psi \psi' \nonumber\\
- \frac{1}{2} \delta f'' \wedge \delta f' \psi^i \psi^j = \psi' \psi''.
\end{gather}

The saddle solution of the action
\begin{gather}
S'_E= S_E + S_{\omega},
\end{gather}
is at \cite{Cotler:2018zff, Simon:2024dwm}
\begin{gather}
f_0 = \varphi - \Omega y, \ \ \ \ \ \ \psi=0.
\end{gather}

The expansion of $f$ and $\psi$ into Fourier modes around the saddle would lead to
\begin{flalign}\label{eq:fpsi}
f & = f_0 + \epsilon(\varphi, y) = f_0 + \sum_{m,n} \frac{\epsilon_{mn} }{(2\pi)^2} e^{-i n f_0 -  \frac{2\pi i m y}{\beta}},\nonumber\\
\psi & = \sum_{m,n} \frac{\psi_{mn} }{(2\pi)^2 \sqrt{\beta} } e^{-i n f_0 - \frac{2\pi i m y}{\beta} }. 
\end{flalign}

If we plug \ref{eq:fpsi} into \ref{eq:SE} and \ref{eq:Somega}, we get
\begin{flalign}\label{eq:allmodes}
S_E&= - 4\pi^2 i \tau j_0 + \frac{ic}{96 \pi^3} n(n^2+ \frac{48 \pi}{c} j_0) (m-n\tau) \big | \epsilon_{mn} \big|^2, \nonumber\\
S_{\omega} &= \frac{ic}{384\pi^4} \sum_{n,m} n (n^2 + \frac{48 \pi}{c} j_0) \psi_{mn} \wedge \psi^*_{mn},
\end{flalign}
where $\tau = \frac{\beta \Omega + i \beta}{2\pi}$.

The real and imaginary components of the modes satisfy the relations
\begin{flalign}
\epsilon_{mn}^R &= \epsilon_{-m,-n}^ R, \ \ \ \ \ \  \epsilon_{mn}^I = - \epsilon_{-m, -n}^I, \nonumber\\
\epsilon_{mn}^* &= \epsilon_{-m, -n}, \ \ \ \ \ \ \psi_{mn}^* = \psi_{-m, -n},
\end{flalign}
and the same could be written for $\psi_{mn}$.

The $\epsilon$-dependent part in the action determines the saddle contribution as
\begin{gather}
S_0= S_E (f_0) = - 4 \pi^2 i \tau j_0,
\end{gather}
and the Hamiltonian can be found by the real part of $S_E$ as 
\begin{gather}
\int_0^\beta dy H = \frac{\beta c}{192 \pi^4} \sum_{n,m} n^2 (n^2 + \frac{48 \pi}{c} j_0) \big | \epsilon_{mn} \big |^2.
\end{gather}

For $H$ to be bounded from below, so the partition function can converge, we should have the relation $j_0 \ge - \frac{c}{48\pi}$. The modes $n=0, \pm 1$ should also be excluded.

\subsection{Free propagators and quantum corrections}
Now we can find the free propagators and their quantum corrections in $3d$ and compare with the case of $2d$.
The soft mode propagator in \cite{Qi:2019gny} has been found from $S^{(2)}$ in \ref{eq:S234} as 
\begin{gather}
\langle \epsilon_{-n} \epsilon_n \rangle_{\text{free}} = \frac{1}{2\pi} \frac{1}{n^2 (n^2-1)}, \ \  \ \ \ \ \langle \psi_{-n} \psi_n \rangle_{\text{free}} = \frac{1}{2\pi i} \frac{1}{n(n^2-1)}.
\end{gather}

From the relation \ref{eq:allmodes}, we can read the $\epsilon$ mode propagators as
\begin{gather}
\langle \epsilon_m \epsilon_n \rangle_{\text{free}} = \frac{48 \pi^3}{c} \frac{1}{i n (n^2 + \frac{48 \pi}{c} J_0 )(m-n \tau)},
\end{gather}
and the $\psi$ propagator mode as
\begin{gather}
\langle \psi_{-n} \psi_n \rangle_{\text{free}} = \frac{192 \pi^4}{i c} \frac{1}{n \left(n^2 + \frac{48 \pi}{c} J_0 \right)}.
\end{gather}

On the other habd, in the Cotler-Jensen theory \cite{Cotler:2018zff}, the Fourier space $\epsilon$ propagator is found as 
\begin{gather}
\langle \tilde{\epsilon} (p_1) \tilde{\epsilon} (p_2) \rangle = \frac{24 \pi \alpha^2}{i n_1 (n_1^2 - \alpha^2) ( \omega - i n_1)} (2\pi)^2 \delta^{(2)} (p_1 + p_2),
\end{gather}
which implies
\begin{gather}
\int \frac{d \omega_1 d\omega_2}{(2\pi)^4} \sum_{n_1, n_2 \ne 0} e^{i n_1 \theta + i \omega_1 y} \langle \tilde{\epsilon} (p_1) \tilde{\epsilon} (p_2) \rangle, \nonumber\\
= \frac{6}{C} ( 2 \ln ( 1- \zeta)+ \Phi ( \zeta,1, \alpha) + \Phi( \zeta, 1, - \alpha)), \ \ \ \ \zeta= e^{i \text{sgn} (y) w },
\end{gather}
where $\Phi(w,s,a)$ is the Lerch transcendant
\begin{gather}
\Phi(w,s,a) = \sum_{n=0}^\infty \frac{w^n}{(n+a)^s}.
\end{gather}

For $s=1$, this is related to a certain incomplete Beta function as
\begin{gather}
B(w, a, 0) = w^a \Phi(w,1,a).
\end{gather}

Now one could discuss the supersymmetry and the localization term of the action $S'_E \equiv S_E + S_\omega = S_E - \frac{1}{2} \int dy \omega_{ij} \psi^i \psi^j $. It is convenient for the fermionic localization analysis to split the zero mode from $f(y, \varphi)$ as $f(y, \varphi) = f_0 + \epsilon(y, \phi)$,  and then one can work directly with the periodic functions $\epsilon(y, \phi)$ and $\psi(y, \varphi)$. 
After implementing the Pfaffian of the symplectic from in terms of the ghost fields $\psi^i$, the geometric action is the sum of
\begin{flalign}
S_E & = \int dy d\varphi \left( j_0 f' ( f' + i \partial_y f) + \frac{c f''(f''+ i \partial_y f') }{48 \pi f'^2} \right), \nonumber\\
S_\omega &= \int dy d\varphi \left ( j_0 \psi \psi' + \frac{c \psi' \psi'' }{48 \pi f'^2} \right ),
\end{flalign}
which is invariant under the supersymmetry transformations
\begin{gather}
Qf = \psi, \ \ \ \ \ Q\psi = - f' - i \frac{\partial f}{\partial y}.
\end{gather}

The full action then can be written as
\begin{gather}
S'_E = S_0 + \int dy d\varphi \left ( j_0 \epsilon ' ( i \partial_y \epsilon + \epsilon') + \frac{c}{48} \frac{\epsilon'' (\epsilon''+ i \partial_y \epsilon') }{(1+ \epsilon')^2} + j_0 \psi \psi' + \frac{c}{48 \pi} \frac{\psi' \psi''}{(1+ \epsilon')^2} \right),
\end{gather}
and the supersymmetry transformations would be
\begin{gather}
Q\epsilon = \psi, \ \ \ \ \ \ Q\psi = -\epsilon' - i \partial_y \epsilon.
\end{gather}

In the Cotler-Jensen reparametrization theory, the Hamiltonian also would be
\begin{gather}\label{eq:HH}
H= - \int_0^{2\pi} d\theta \left ( \frac{C}{24\pi} \lbrace \phi, \theta \rbrace-b_0 \phi'^2 \right ).
\end{gather}
Note that the Schwarzian derivative of $f$ with respect to $\theta$ is 
\begin{gather}
\lbrace f , \theta \rbrace = \frac{f'''}{f'} - \frac{3}{2} \left ( \frac{f''}{f'} \right )^2.
\end{gather}
Then, we can see that \ref{eq:HH} corresponds to the action
\begin{gather}
S_+ = - \frac{C}{24\pi} \int d^2 x \left (\frac{(\partial_+ \phi') \phi''}{\phi'^2} + B ( \partial_+ \phi) \phi' \right), \ \ \ \ \ B=\frac{b_0}{48\pi C},
\end{gather}
where $\partial_+ = \frac{1}{2} ( \partial_\theta + \partial_t)$, and the right-moving action is
\begin{gather}
S_-= - \frac{C}{24 \pi} \int d^2 x \left ( \frac{(\partial_- \phi') \phi''}{\phi'^2} + B (\partial_- \phi) \phi' \right ), \ \ \ \ \ \  \partial_- = \frac{1}{2} ( \partial_\theta - \partial_t).
\end{gather}

The expansion of $H$ near the critical point $\theta$ would be
\begin{gather}
H\lbrack \phi =\theta + \sum_n \phi_n e^{i n \theta} \rbrack = 2\pi b_0 + \frac{C}{12} \sum_n n^2(n^2+B) |\phi_n|^2 + O(\phi_n^3).
\end{gather}

The real part of the Euclidean action is $\text{Re} (S_E) = \int dy \ H(y)$ with
\begin{gather}
H = - \frac{C}{24 \pi} \int_0^{2\pi} d\theta \left ( \lbrace \phi, \theta \rbrace + \frac{\phi'^2}{2} \right ).
\end{gather}

As $H \ge - \frac{C}{24 \pi}$ then we have
\begin{gather}
\text{Re} (S_E) \ge - \frac{\pi C}{12} \text{Im} (\tau).
\end{gather}

For very large $C$, corresponding to small coupling $1/C$, the partition function has been calculated to one-loop order in \cite{Cotler:2018zff}. The field equation of the model is
\begin{gather}
\frac{\delta S_E}{\delta \phi} = \frac{C}{48 \pi \phi'} \bar{\partial} \left ( \lbrace \phi, \theta \rbrace + \frac{\phi'^2}{2} \right ) =0.
\end{gather}

The saddle points are at
\begin{gather}
\phi_0 = \theta - \frac{\text{Re} (\tau) }{\text{Im} (\tau) }y, \ \ \ \ \ \ S_0= \frac{\pi C}{12} i\tau,
\end{gather}
and the field $\phi$ can be expanded in fluctuations around the saddle as
\begin{gather}
\phi = \phi_0 +\sum_{m,n} \frac{\epsilon_{m,n} }{(2\pi)^2} e^{\frac{imy}{\text{Im} (\tau) } + i n \left (\theta-\frac{\text{Re} (\tau) }{\text{Im} (\tau) } y \right )},\ \  \ \ \ \ \epsilon^*_{m,n} = \epsilon_{-m, -n}.
\end{gather}

By using the local $PSL(2; \mathbb{R})$ redundancy, an infinite number of modes would vanish
\begin{gather}
\epsilon_{m,n} = {}_{-1,0,+1}=0.
\end{gather}
The Euclidean action then can be expanded as 
\begin{gather}
S_E= S_0 + \frac{iC}{96 \pi^3} \sum_{m= -\infty} \sum_{n \ne -1, 0, 1} n(n^2-1) (m- n\tau) |\epsilon_{m,n}|^2 + O(\epsilon^3_{m,n}).
\end{gather}

Using this expansion, the one-loop correction of partition function can be found as 
\begin{gather}
\mathcal{Z}_{1-\text{loop}} = N q^{- \frac{C}{24}} \prod_{m,n} (m-n\tau)^{-\frac{1}{2}},  \ \ \ \ \ \ \ \ \ q= e^{2\pi i \tau},
\end{gather}
while the vacuum partition function is
\begin{gather}
\mathcal{Z}_{\text{vac}} (\tau', \bar{\tau}') = \Bigg | q'^{-k} \prod_{n=2}^\infty \frac{1}{1-q'^n} \Bigg|^2, \ \ \ \ \ q'= e^{2\pi i \tau'}, \ \ \ \ \ k= \frac{c}{24}.
\end{gather}

Now, using these quantum corrected actions, partition functions, field equations, saddles, Hamiltonian, free-propagators, and quantum corrected propagators, we can study the quantum corrected greybody factor, Lyapunov exponent, potential, exceptional points in a quantum corrected Lindblad formalism, and many other quantities which we will study some of them in the next section.

\section{Greybody factor and its quantum corrections}\label{sec:greybody}
In this section we review the semiclassical calculation of greybody factor without quantum corrections first. Then, we move to adding the calculations of quantum corrections and the resulting plots in various examples. \
 
The gravitational anomaly method has been studied in \cite{Robinson:2005pd,Iso:2006wa,Iso:2006ut, Vagenas:2006qb, Murata:2006pt}. A scalar field theory in D-dimensional spacetime of 
\begin{gather}
ds^2= -f(r) dt^2+ \frac{1}{f(r)} dr^2 + r^2 d\Omega^2_{D-2},
\end{gather}
can be reduced to 2-dimension metric of
\begin{gather} \label{eq:twoDmetric}
ds^2=-f(r) dt^2 + \frac{1}{f(r)} dr^2,
\end{gather}
near the horizon, since the action of the scalar field would be
\begin{gather}
S\lbrack \varphi \rbrack = \frac{1}{2} \int d^D x \sqrt{-g} \varphi \nabla^2 \varphi \nonumber\\
= \frac{1}{2} \int d^D x \ r^{D-2} \sqrt{\gamma} \times \varphi \left(- \frac{1}{f} \partial_t^2 +\frac{1}{r^{D-2}} \partial_r r^{D-2} f \partial_r+\frac{1}{r^2} \Delta_{\Omega}\right )\varphi,
\end{gather}
where by taking the limit of $r \to r_H$ the action becomes
\begin{gather}
S\lbrack \varphi \rbrack = \frac{r_H^{D-2} }{2} \int d^D x \sqrt{\gamma} \ \varphi \left( -\frac{1}{f} \partial_t^2 + \partial_r f \partial_r \right) \varphi \nonumber\\
= \sum_n \frac{r_H^{D-2} }{2} \int dt \ dr \ \varphi_n  \left ( - \frac{1}{f} \partial_t^2 + \partial_r f \partial_r \right ) \varphi_n,
\end{gather}
which is infinite set of the scalar fields on the $2d$ metric \ref{eq:twoDmetric}.

The gravitational anomaly of the chiral scalar field in $1+1$ dimensions can be found by \cite{BERTLMANN2001137}
\begin{gather}
\nabla_\mu T^\mu_\nu = \frac{1}{96\pi \sqrt{-g} } \epsilon^{\beta \delta} \partial_\delta \partial_\alpha \Gamma_{\nu\beta}^\alpha.
\end{gather}

As this anomaly is purely timelike, it can be written as
\begin{gather}
\nabla_\mu T_\nu^\mu \equiv A_\nu \equiv \frac{1}{\sqrt{-g}}\partial_\mu, N^\mu_\nu,
\end{gather}
where the quantities $N_\nu^\mu$ are defined as
\begin{gather}
N_\nu^\mu= \frac{1}{96\pi} \epsilon^{\beta \mu} \partial_\alpha \Gamma^\alpha_{\nu\beta}.
\end{gather}

The components for Schwarzchild type black hole would be
\begin{gather}
N_t^t=N_r^r=0, \ \ \ \ N_t^r=\frac{1}{192\pi} (f'^2 + f'' f), \ \ \ \ N_r^t= - \frac{1}{192\pi f^2} ( f'^2 - f'' f).
\end{gather}

So the pure flux is 
\begin{gather} \label{eq:anomalyflow}
\Phi= N_t^r \Big |_{r_H}= \frac{1}{192\pi} f'^2 (r_H),
\end{gather}
and the surface gravity $\kappa$ is
\begin{gather}
\kappa= \frac{1}{2} \frac{\partial f}{\partial r} \Big |_{r=r_H} = \frac{1}{2} f'(r_H).
\end{gather}

A beam of massless black body radiation at this temperature $T_H$ has the flux
\begin{gather}\label{eq:beamflow}
\Phi=\frac{\pi}{12} T_H^2.
\end{gather}

So the flux of canceling the gravitational anomaly is precisely the thermal flux of black hole at Hawking temperature. Then,  the aim is to add the quantum correction to this anomaly and this specific flux, and compare the results. We could also compare these effects with the effects of exceptional points at each step.

Let's first check the $(2+1)$-dimension Banados, Teitelboim and Zanelli black hole \cite{Banados:1992wn,Banados:1992gq} which can be derived from a three dimensional gravity theory of 
\begin{gather}
S= \int d^3 x \sqrt{-g} (R^{(3)}+2 \Lambda) ,\ \ \ \ \ (\Lambda = \frac{1}{l^2}),
\end{gather}
with the line element of 
\begin{gather}
ds^2=- \left( -M + \frac{r^2}{l^2}+ \frac{J^2}{4r^2} \right) dt^2 + \frac{dr^2}{\bigg( -M + \frac{r^2}{l^2}+ \frac{J^2}{4r^2} \bigg) } + r^2 \left( d\theta - \frac{J}{2r^2}dt\right)^2,
\end{gather}
where $M$ is the Arnowitt-Deser-Misner (ADM) mass, $J$ is the angular momentum (spin) of the BTZ, $-\infty < t<+\infty$, $0 \le r <+\infty$, $0 \le \theta < 2\pi$.

By neglecting the classically irrelevant ingoing modes near the horizon, the effective $2d$ theory becomes chiral near the horizon and due to the gauge or gravitational anomalies, the gauge symmetry or general coordinate covariance becomes anomalous.

Outside of the horizon, the current is conserved, so we have $\partial_r J^r_{(o)}=0$. In the near horizon region, as there are only outgoing fields, the current satisfies the anomalous equation
\begin{gather}
\partial_r J_H^r = \frac{m^2}{4\pi} \partial_r A_t,
\end{gather}
where $m$ is the $U(1)$ charge of $2d$ massless field.

By solving these equations, one finds the two solutions of
\begin{gather}
J_{(o)}^r=C_0, \ \ \ \ \
J_H^r= C_H + \frac{m^2}{4\pi} (A_t(r) - A_t(r_+)),
\end{gather}
where $C_0$ and $C_H$ are integration constants. 

The total current then can be written as
\begin{gather}
J^\mu = J^\mu_{(o)} \theta_+(r) + J_H^\mu H(r),
\end{gather}
where $\theta_+(r) = \theta(r-r_+ - \epsilon)$ and $H(r)= 1- \theta_+(r)$.

Also, we have the relation
\begin{gather}
-\delta W = \int d^2 x \sqrt{-g_2} \lambda \nabla_\mu J^\mu,
\end{gather}
where after inserting $J^\mu$ in it, we get
\begin{gather}
-\delta W = \int d^2 x \lambda \lbrack \delta(r-r_+-\epsilon)(J^r_{(o)} -J^r_{(H)} + \frac{m^2}{4\pi} A_t)+ \partial_r(\frac{m^2}{4\pi} A_t H) \rbrack.
\end{gather}

The classically irrelevant ingoing modes have some ``quantum effects" being induced by the ingoing modes near the horizon, which can cause the last term to be cancelled out. Also, the coefficient of the delta-function should vanish. By connecting the coefficients of the current in the two regions we find
\begin{gather}
C_0= C_H- \frac{m^2}{4\pi} A_t(r_+),
\end{gather}
where $C_H$ is the value of consistent current at the horizon. As we have $J_H^{r^+}=0$,  we get $C_H=0$, and then for the value of the charge flux we get
\begin{gather}
C_0= - \frac{m^2}{2\pi} A_t(r_+)=-\frac{m^2}{2\pi} \Omega(r_+)=-\frac{m^2}{2\pi} \frac{J}{2r^2_+}=- \frac{m^2 r_-}{2\pi l r_+}.
\end{gather}

In a similar way, the flux of stress tensor being radiated from a rotating BTZ black hole has been calculated as
\begin{gather}
a_0= \frac{m^2}{4\pi} A^2_t(r_+) + N_t^r(r_+) = \frac{m^2}{4\pi} \Omega^2(r_+)+\frac{1}{192\pi} f'^2=\frac{m^2 r_-^2}{4\pi l^2 r^2_+}+\frac{1}{192\pi}f'^2.
\end{gather}

From the value on the horizon of the rotating BTZ metric, this quantity could be found as 
\begin{gather}
f' \big|_{r_+} = \frac{2r_+}{l^2}-\frac{J^2}{2r^3_+} = 4\pi T_H.
\end{gather}

So, $a_0$ would be
\begin{gather}
a_0= \frac{m^2 r^2_-}{4\pi l^2 r_+^2} + \frac{\pi}{12}T_H^2,
\end{gather}
which is the same as the Hawking flux from rotating BTZ black holes. Then, the greybody factor is 
\begin{gather}
\Gamma(\omega) = \frac{J_{\text{hor}}}{J_{\text{in}}},
\end{gather}
where for AdS case, we get
\begin{gather}
\hat{J}_{\text{asy}}= \frac{2\Omega_{2N+1} \omega^{2N+2}}{\pi \kappa^{4N+2}} \left( |\hat{C}_1|^2 - | \hat{C_2}|^2 \right). 
\end{gather}

As the total flux is being conserved for AdS case, we have $J_{\text{hor}}=\widehat{J}_{\text{asy}}$ so
\begin{gather}
\Gamma(\omega)= \frac{J_{\text{hor}}}{\widehat{J}_{\text{in}}}.
\end{gather}

So now various quantum correction types, for each term in the above derivation, could be considered which we will discuss next. The effects of these quantum corrections on the final thermodynamical quantities could be found and could be compared with the effects of EPs.

\subsection{Adding quantum corrections to greybody factor}

The greybody factor is actually the consequence of scattering a test field off the gravitational potential barrier surrounding the horizon. However, as recent studies showed Page curve is the result of transfer between two saddles, so greybody factor too would be affected by both the spacetime curvature and the quantum effects at the horizon. The interconnections of quantum effects, curvature and statistical mechanics would play interesting roles.

Now, one can consider the 1-loop quantum corrections to the gravitation anomaly of Robinson-Wilczek. First, one should note that due to Wess-Zumino consistency conditions, topological anomalies, and Adler-Bardeen-type non-renormalization theorems, the anomaly coefficient itself does not receive higher-loop corrections, but the quantum corrections enter indirectly, through the effective theory which is being used to compute the anomaly. These have three sources, which are the renormalization of background geometry, the change in effective central charge, and the change in the boundary and horizon degrees of freedom.

In the first case, $g_{\mu\nu}^{cl} \to g_{\mu\nu}^{\text{eff}}$, which changes the surface gravity $\kappa$, horizon location, and Hawking temperature $T_H$. The relation $\Phi= \frac{\pi}{12}T_H^2$ remains unchanged, but we have $T_H \to T_H^{\text{eff} }= T_H^{(0)}+ \delta T_H$. So the flux gets a 1-loop correction as $\delta \Phi = \frac{\pi}{6} T_H \delta T_H$, which in this case,  would be the dominant  source of quantum corrections.

Then, one should note that, as the anomaly coefficient would depend on the number of effective chiral degrees of freedom on the horizon and at one loop, some modes could decouple, some modes could acquire effective masses, and also ghosts and boundary modes can contribute. So, we have the relation $c \to c_{\text{eff} } = c+\delta c$, and then we get $\nabla_\mu {T^\mu}_\nu = \frac{c_{\text{eff}} }{96\pi} \epsilon_{\nu\mu} \partial^\mu R$. Also, $\Phi = \frac{\pi}{12} c_{\text{eff}} T_H^2$. Note that the Schwartzian or JT correction and also the heat kernel method is related to this correction.

Finally, the quantum gravity could add horizon edge modes, Schwarzian modes in the $\text{AdS}_2$ throat and soft hair contributions. These could modify the effective chiral action, and the matching condition between horizon and asymptotic regions leads to the relation $\Phi= \frac{\pi}{12} T_H^2 \times Z_{1-\text{loop}}$, where $Z_{1-\text{loop}}= \text{exp} (- \ln \text{det} ' \Delta)$.

The observed flux near the horizon according to Robinson-Wilczek is 
\begin{gather}
\Phi_\infty(\omega) = \Gamma(\omega) \Phi_{anomaly}(\omega),
\end{gather}
and at 1-loop we have
\begin{gather}
\Gamma(\omega) \to \Gamma_{cl}(\omega) \times \left | \frac{\text{det}(\omega- \omega_n^{\text{QNM}}) }{\text{det}(\omega- \omega_n^{\text{ref}}) } \right |^{-2}.
\end{gather}

These factors could affect several physical quantities of black holes and add quantum corrections to them. Here, we investigate several of these quantities and the solutions, in additions to the recent works of \cite{Chen:2025rcc, Cremonini:2025yqe, Nian:2025oei, Jiang:2025cyl, Liu:2024qnh, Liu:2024gxr, David:2021qaa}.

So, we could note that the gravitational anomaly derived by Robinson and Wilczek is one-loop exact and does not receive higher-loop corrections in its coefficient. But, the physical Hawking flux does receive quantum corrections at one loop through renormalization of the background geometry, effective central charge, and horizon boundary degrees of freedom. These corrections modify the temperature and multiplicative prefactors entering the anomaly-induced flux, and combine with quantum-corrected greybody factors to give the full corrected radiation spectrum.

Also, the connection between emission rate $\gamma(\omega)$ and greybody factor is
\begin{gather}
\gamma(\omega) = \left (  \frac{|A_{l,m} |^2 d^3k}{e^{\frac{\omega}{T_H} } (2\pi)^3}   \right),
\end{gather}
where the greybody factor would be a quantity that is dependent on frequency through $| A_{l,m}|^2$.

In \cite{Lv:2025nww},  the greybody factor of quantum Oppenheimer-Snyder-de Sitter spacetime has been studied, and the effective potential of this solution and the effects of physical parameters such as mass, rotational momentum and quantum parameters has been studied. There, they showed that by increasing the quantum parameter, the effective potential would decrease. Then, the absorption and scattering of massless scalar would be strongly affected by this. Here, we aim to do the same for several other black hole solutions, and check how quantum corrections affect the greybody factor, specifically in the $3d$ Cotler and Jensen theory.

Note that BTZ which is a locally AdS$_3$ with no curvature singularity, is an exact solution of string theory and therefore finding quantum corrections of Hawking radiation in such backgrounds could give more clear picture of how D-branes, strings and dynamics of this black hole would also behave under such fluctuations.
 
For a probe quantum field, at one loop, the quantum correction has the form
\begin{gather}
\Gamma(\omega) = \Gamma_{cl}\times \text{exp} \Big \{ - 2 \ \mathfrak{R} \Big \lbrack \Delta \ln Z_{1-\text{loop}} (\omega) \Big \rbrack \Big \},
\end{gather}
where $\Gamma_{cl}(\omega)$ is the classical tree-level greybody factor, and $\Delta \ln Z_{1-loop} (\omega)$ is the one loop correction to the scattering phase or density of states.

Up to local counterterms of the determinant, we have
\begin{gather}
\ln Z_{1-\text{loop}}^{(\text{BH})} = - \sum_{k \in \mathbb{Z}} \sum_{n  \ge 0} \ln (\omega- \omega_{n,k}) + (\text{counterterms/reference}).
\end{gather}

So we get
\begin{gather}
\Gamma(\gamma) = \Gamma_{cl}(\omega) \text{exp} \Big \{ - 2 \mathfrak{R} \Big \lbrack \sum_{k,n} \ln \frac{\omega-\omega_{n,k}^{\text{(BH)} }}{\omega- \omega_{n,k}^{\text{(ref)}} } \Big \rbrack \Big \}.
\end{gather}

The BTZ is special as the QNM spectrum of a scalar of (AdS) conformal weight $h$ and angular quantum number $k$ is known exactly. For a rotating BTZ with left and right temperatures $T_L$, $T_R$, the QNM frequencies take the form
\begin{gather}
\omega_{n,k}^{(L)} = k - 4\pi i T_L(n+h_L), \ \ \ \ \ \ \ \ \omega_{n,k}^{(R)} = - k -4\pi i T_R(n+h_R),
\end{gather}
with $n=0,1,2,...$ and $h_{L,R}$ fixed by the bulk mass $m$, and for non-rotating BTZ, $T_L=T_R=T_H$. As the QNM spectrum is exact in this case, the Denef-Hartnoll-Sachdev (DHS) product \cite{Denef:2009kn} can be constructed explicitly for BTZ \cite{Law:2022zdq, Cardoso:2001hn}.

 \begin{figure}[ht!]
 \centering
  \includegraphics[width=14cm] {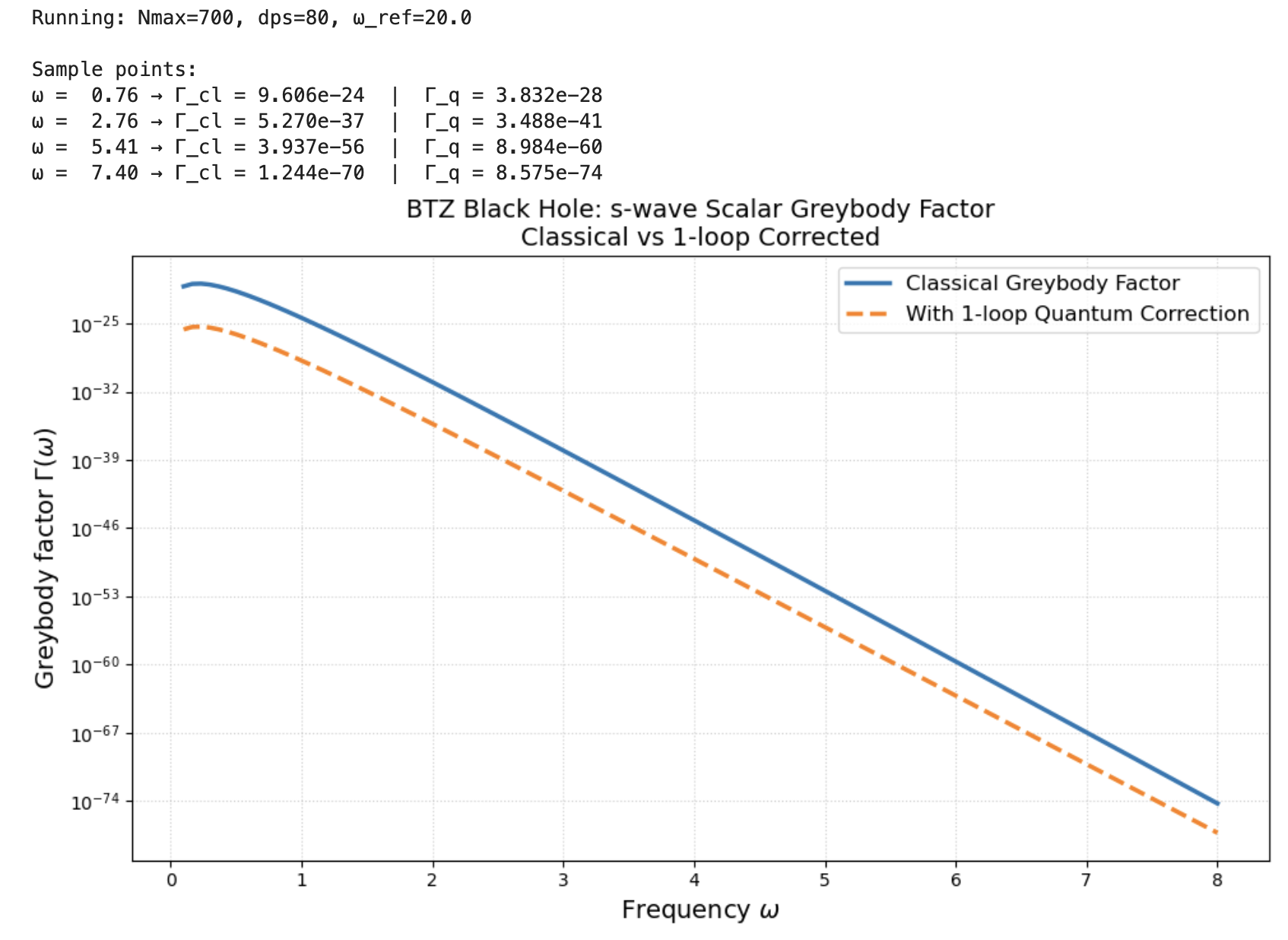} 
  \caption{The difference between classic and quantum corrected greybody factor for $2+1$-dimension BTZ black hole and for $r_+=10$, $r_-=3$, $\ell=1$, $h=2$. }
 \label{fig:QCBTZ}
\end{figure}

For the non-extremal rotating BTZ black hole, and for s-wave and $\ell=0$ sector, one gets
\begin{gather}
\Gamma(\omega) = \Big | \frac{\sinh \big(\pi \omega / (2\pi T_L) ) \sinh(\pi \omega/(2\pi T_R) \big) }{\sinh(2\pi \omega T_H) \sinh(2\pi h T_H) } \Big |^2 \times \text{exp} \lbrack -2 \text{Re} ( \Delta \log \text{det}) \rbrack,
\end{gather}
where the first factor is the classical $\Gamma_{\text{cl}}(\omega)$ and the second factor is the 1-loop correction. 

Here we have,
\begin{gather}
\Delta \log \text{det} = \lim_{N\to \infty} \sum_{n=-N}^N \Big \lbrack \log \left( 1- \frac{\omega_n^L}{\omega} \right) + \log \left( 1- \frac{\omega_n^R}{\omega} \right) - 2 \log \left( 1- \frac{\omega_n^{\text{ref}} }{\omega} \right ) \Big \rbrack \nonumber\\ - \left \lbrack \text{same sum at a larger reference}  \ \omega_{\text{ref}} \right \rbrack,
\end{gather}
with the highly overdamped s-wave BTZ quasinormal frequencies of 
\begin{gather}
\omega_n^L = - 2\pi i T_L(n+h), \ \ \ \ \ \ \omega_n^R = -2 \pi i T_R (n+h), \ \ \ \ n \in \mathbb{Z},
\end{gather}
and a simple regulator tower
\begin{gather}
\omega_n^{\text{ref}} = -2 \pi i (n+h).
\end{gather}

The final effect of quantum correction is shown in  Figure \ref{fig:QCBTZ}, which shows that quantum correction would \textit{suppress} the greybody factor in this case.

\subsection{Greybody factor with quantum corrections in higher dimensions}

Now, in this section, we would like to check the effects of quantum corrections in black holes in higher dimensions and specifically in $5d$.

In \cite{Lee:1998pd}, it has been shown that in the low-energy scattering, the greybody factor of $5d$ black hole (D1$+$D5 branes) in the dilute gas approximation is equal to the s-wave greybody factor of a minimally coupled scalar of $\text{AdS}_3 \  (\text{BTZ black hole}) \times S^3 \times T^4$ replacing the original geometry of $M_5 \times S^1 \times T^4$. Note that also the $5d$ black hole is U-dual of $\text{BTZ}\times S^3$ \cite{Hyun:1997jv, Sfetsos:1997xs}.
So the AdS theory contains the essential information about the bulk $5d$ black hole. Now, it would be interesting to check how quantum corrections to both geometries could change that.

The greybody factor could be calculated using a minimally coupled scalar, (or vector or matrix modes).
This calculation could be done just in $5d$ or the total, full $10d$ case. In fact, the extra dimensions $T^4$ would cause discrepancy.  The interconnections of the effects of these higher dimensions and quantum effects could be analyzed further.

One could picture the near-extremal $5d$ black hole as the bound states of $Q_1$ $1D$-branes and $Q_5$ $5d$-branes with momentum flowing along the 1D-brane. So one would expect the quantum fluctuations along this brane have bigger effects.

To find the greybody factor, one can consider the wave equation in the $10d$ background first, as
\begin{gather}\label{eq:box}
\square_{10} \Phi=0.
\end{gather}

The symmetries of the $10D$ metric force the following form of $\Phi$ as
\begin{gather}
\Phi= e^{-i \omega t} e^{i K_5 x_5} e^{i K_i x^i} Y_l (\theta_1, \theta_2, \theta_3) \phi(r),
\end{gather}
which for simplicity we could take $K_5=0$ and $r_1\simeq r_5$.

In the dilute gas limit, one could set $A_0=0$. However, one needs to check if this is also true when one considers the quantum corrections.

The solution of \ref{eq:box} would be a $5d$ wave equation for a free scalar as
\begin{gather}\label{eq:phi}
\left \lbrack \frac{d^2}{dr^2} + \left (\frac{h'}{h}+ \frac{3}{r} \right) \frac{d}{dr} \right \rbrack \phi+\frac{\omega^2 f}{h^2}\phi-\frac{l(l+2)}{hr^2}\phi-\frac{K^2 f_5^2}{h}\phi=0.
\end{gather}

In the near region, where the quantum fluctuations are stronger, and by taking $h=1-r_0^2/r^2$, the above wave equation can be written as
\begin{gather}
h(1-h)\frac{d^2 \phi}{dh^2}+ (1-h) \frac{d\phi}{dh}+ \left \lbrace -C + \frac{C+D+E}{h} + \frac{E}{1-h}\right \rbrace \phi=0,
\end{gather}
where
\begin{gather}
C= \left ( \frac{\omega r_1 r_5 r_n}{2 r_0^2} \right)^2,  \nonumber\\
D= \frac{l(l+2)}{4}+ \frac{\omega^2}{4r_0^2} \lbrack r_1^2 r_5^2 + r_n^2 (r_1^2+r_5^2) \rbrack +\frac{K^2 r_5^2}{4}, \nonumber\\
E= - \frac{l(l+2)}{4} + \frac{\omega^2 (r_1^2 + r_5^2 + r_n^2)}{4} - \frac{K^2 (r_5^2-r_0^2)}{4}.
\end{gather}

In \cite{Lee:1998pd}, only the low energy case of $r_1 \omega, r_5 \omega <1$ has been studied. For studying the interconnections of quantum fluctuations and curvature in terms of greybody factor, one needs to consider the higher energy, and maybe higher curvatures as well.

For solving the equation, the space is being divided into a far region $(r \gg r_1, r_5)$ and a near region $(r \ll r_1, r_5)$, and since in the dilute gas limit we have $\omega r_0 \ll \omega r_1, \omega r_5 <1$, the solution can be matched in the overlapping region ($\omega r_0 \ll \omega r \ll 1$).

In the \textit{far region}, one can rewrite $\phi=\tilde{\phi}/r$, so eq. \ref{eq:phi}  as
\begin{gather}
\frac{d^2 \tilde{\phi}} {du^2}+ \frac{1}{u} \frac{d\tilde{\phi}}{du}+\left \lbrack 1- \frac{\nu^2}{u^2} \right \rbrack \tilde{\phi}=0,
\end{gather}
where the new parameters are $u=\omega ' r$, $\nu^2 = (l+1)^2 - (r_1^2 + r_n^2) K^2 - ( \rho_1^2 + \rho_5^2 + \rho_n^2)$, and $\omega' = \sqrt{\omega^2 - K^2}$, $\rho_i = \omega' r_i$, and also $K_i r_i < \rho_i < \omega r_i$. In the far region one could guess that the quantum fluctuations would be suppressed. The solution of this equation would be given by the Bessel function when $\nu$ is not an integer. Here,
\begin{gather}
\phi_{\text{far}}= \frac{1}{u} \lbrack \alpha J_\nu(u) + \beta J_{-\nu} (u) \rbrack,
\end{gather}
where the main problem is to find the $\alpha$ and $\beta$ constants.

The incoming flux can be found from the large-$u$ behavior as 
\begin{gather}
\mathcal{F}_{\text{in}} (\infty)= \frac{2\pi}{i} \left \lbrack \phi_f^{\text{in}^*} r^3 \partial_r \phi_f^{\text{in}} - \phi_f^{\text{in}} r^3 \partial_r \phi_f^{\text{in}^*} \right \rbrack  \Big |_{r=\infty}\nonumber\\
= - \frac{2}{\omega'^2} \Big | \alpha e^{i (\nu+1/2) \pi/2} + \beta e^{i(-\nu+1/2) \pi/2} |^2,
\end{gather}
where $\phi_f^{\text{in}}$ is given by
\begin{gather}
\phi_f^{\text{in} } = \sqrt{\frac{1}{2\pi}} \frac{e^{-i u} }{u^{3/2}} \Big \{ \alpha e^{i(\nu+1/2) \pi/2} + \beta e^{i(-\nu+1/2) \pi/2} \Big \}.
\end{gather}

\begin{figure}[ht!]
 \centering
  \includegraphics[width=8cm] {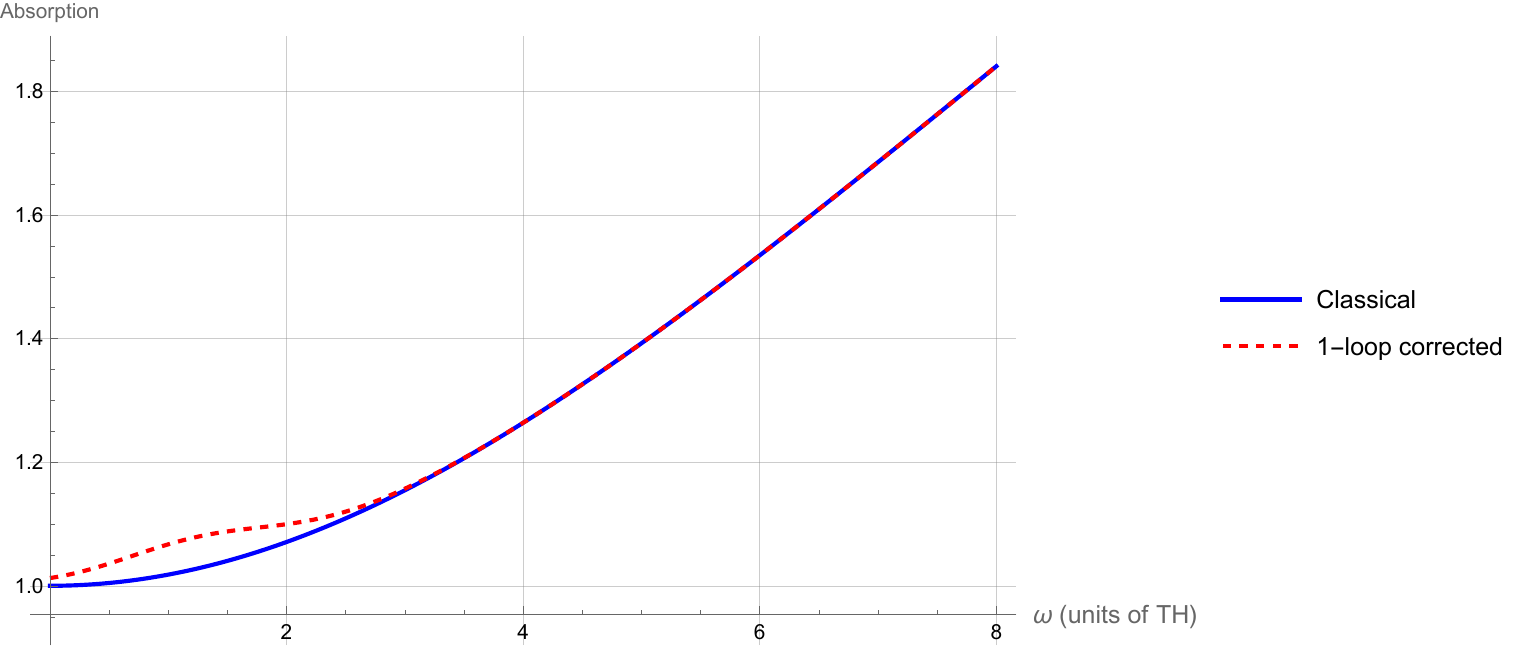} 
   \includegraphics[width=8cm] {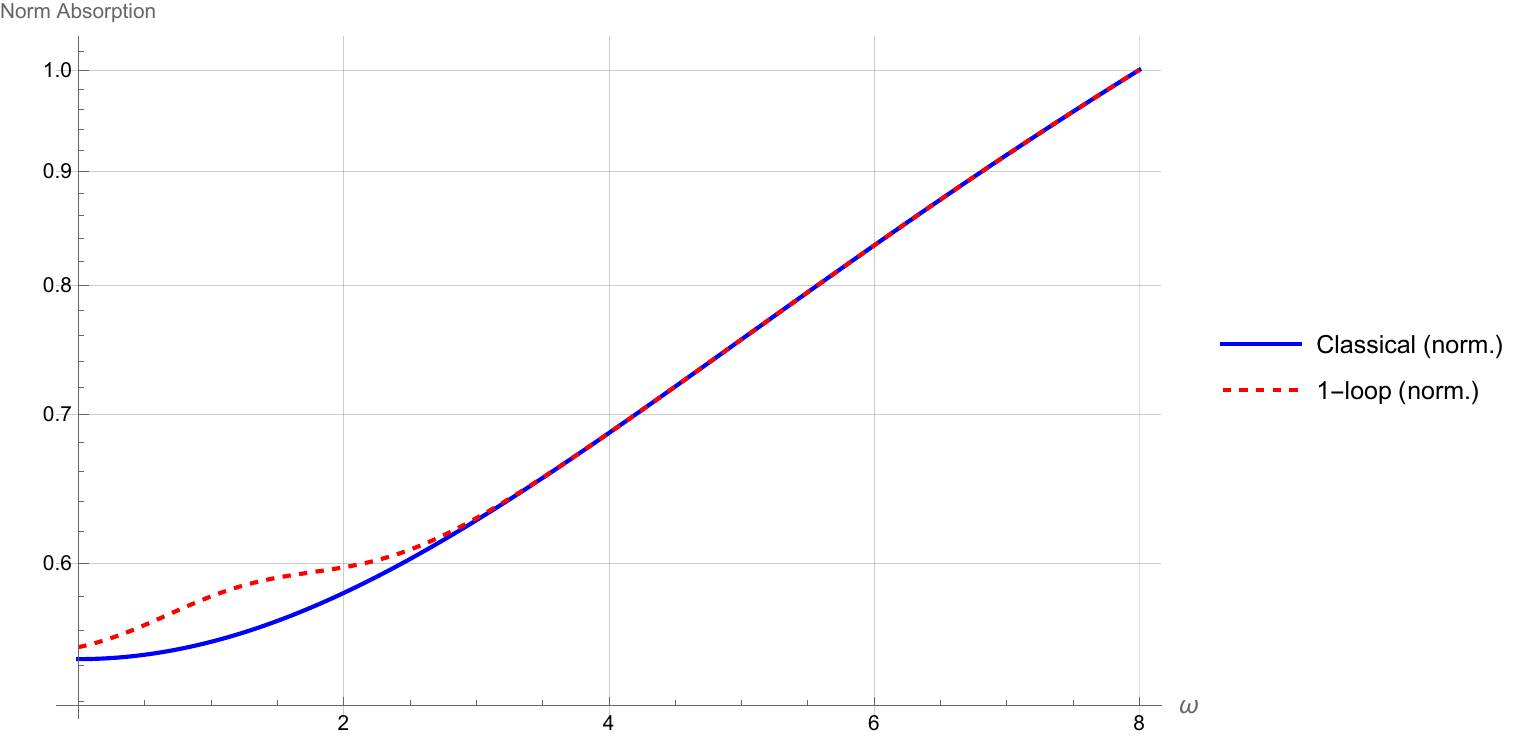} 
   \includegraphics[width=7cm] {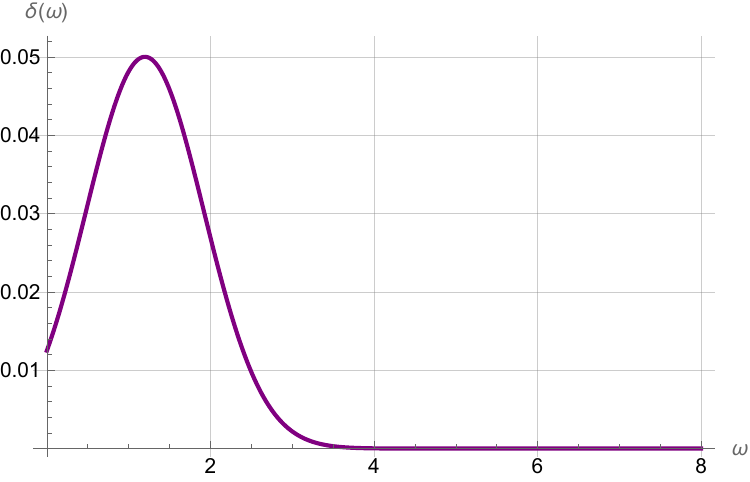}
  \caption{The classical $5d$ absorption factor with 1-loop correction, and a separate plot of $\delta(\omega)$.}
 \label{fig:5dOneloop}
\end{figure}

The classical and quantum corrected plots for greybody factor are shown in Figure \ref{fig:5dOneloop}, which shows at 
smaller $\omega$, quantum correction increases absorption, as it increases the potential as well.

One could consider the Schwarzian soft mode which are being coupled to the boundary insertion that creates or absorbs the bulk scalar, and then integrating out the soft mode dresses the boundary two-point, or the scattering kernel, producing loop corrections which are suppressed by inverse powers of $C$ i.e. powers of $1/S_0$. 

At leading nontrivial 1-loop order in the soft mode, the greybody is multiplicatively corrected as $G(\omega)= G_{cl}(\omega) \lbrack 1+\delta(\omega)\rbrack$, which for the $5d$ case would be
\begin{gather}
G(\omega) = G_{cl} (\omega) \left \lbrack 1+ \kappa \ C^{-2} \left ( \frac{\omega}{T_H}\right)^2 \ln \frac{\lambda}{\omega} + O(C^{-3}) \right \rbrack,
\end{gather}
where $G_{cl}(\omega)$ is the classical Lee-Myung greybody factor, $C$ is the Schwarzian coupling,  $C\sim S_0$, $T_H$ is the Hawking temperature, and $\Lambda$ is an effective UV scale where the Schwarzian description breaks down. Also, $\kappa$ is a dimensionless constant that depends on the precise coupling between the bulk scalar and the Schwarzian boundary mode, and on finite parts from the loop integral.

If we use the method of \cite{Qi:2019gny}, we get the following correction
\begin{gather}
\kappa_{\text{Qi}} = \frac{7 g^4 h^2}{576 \pi^2 \beta^2} \Big \lbrack 9 \pi^2 (\pi-16) + 168 \pi + 208 \Big \rbrack.
\end{gather}

\begin{figure}[ht!]
 \centering
  \includegraphics[width=9.5cm] {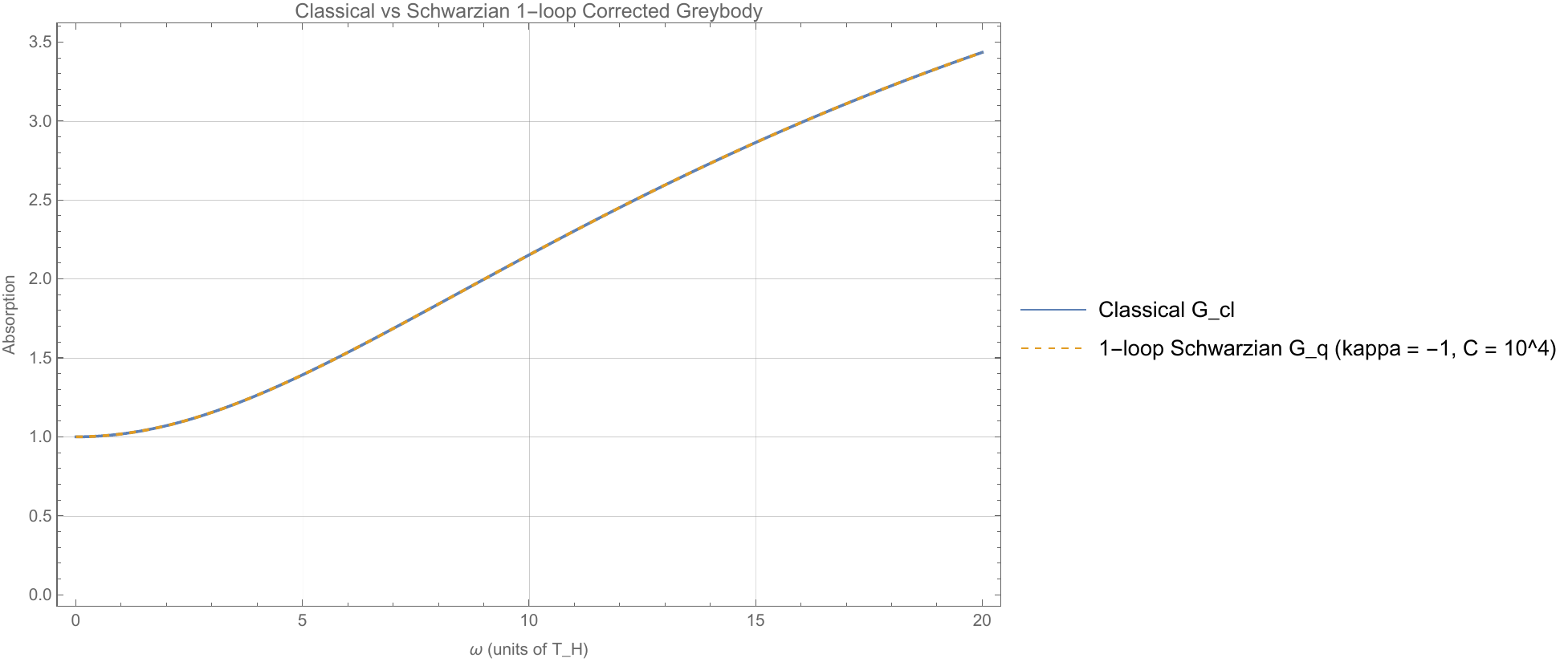} 
     \includegraphics[width=7.5cm] {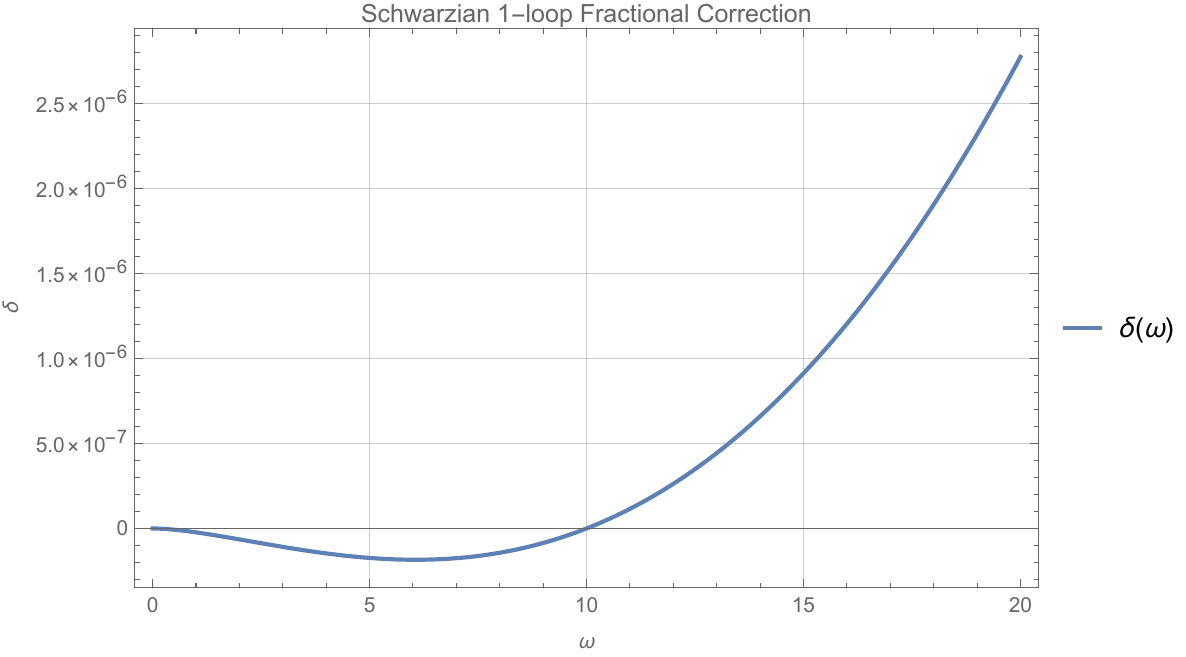} 
  \caption{The difference between classic and quantum corrected greybody factor for $5$-dimension black holes in the type IIB superstring theory compactified on $S^1 \times T^4$.}
 \label{fig:Qi-5D}
\end{figure}

So, in this case, one finds the Figure \ref{fig:Qi-5D}, which shows that quantum correction in this case lower the greybody factor.

\begin{figure}[ht!]
 \centering
  \includegraphics[width=18cm] {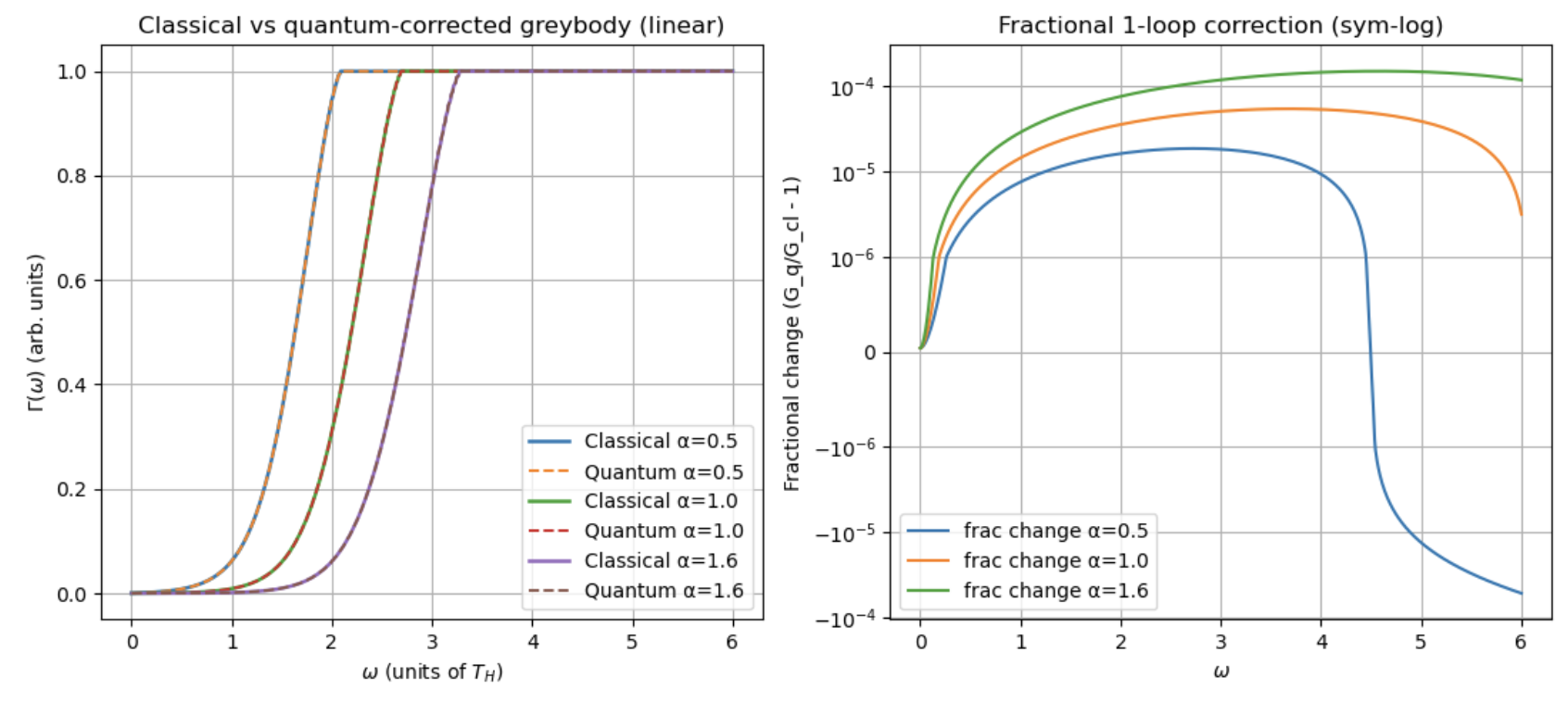} 
  \caption{The grey body factor with 1-loop correction for WBTZ geometry.}
 \label{fig:3dWBTZ}
\end{figure}

One could also find the greybody factor for the case of warped $\text{AdS}_3$ space as well, which is shown in Figure \ref{fig:3dWBTZ}. In this case, we could not track the change in the greybody factor and therefore further studies are needed.

Since the mechanism of non-local encoding information in the island has only be understood for the case of theories with ``massive gravitons" \cite{Geng:2025rov,Geng:2024xpj}, it would be interesting to check quantum corrections of greybody factor and Hawking radiation for these cases as well. In fact, for this scenario, one could either consider warped CFT case or BCFT models. So the results of here would be useful in studying islands as well.

For different $\alpha$ we have the plots shown in Figure. \ref{fig:3dWBTZ}.

\subsection{Quantum corrections and extremality of black holes}

It is expected that the quantum correction only changes the speed that black holes approach the extremality. Then, the black hole should go away from extremality as it has been shown in \cite{Brown:2024ajk}, producing a ``zig-zag" curve for black hole evaporation. As we explained in the first section of this work, this zigzag behavior could also be explained by considering the roles of exceptional points in the dynamics of SYK-like model and Swarzian modes around the horizon as in \cite{Zheng:2025apa}.

In \cite{David:2021qaa}, and also \cite{Nian:2019pxj, Nian:2019buz, Nian:2020bzf}, it has been shown that the Hawking radiation of an uncharged rotating black hole tends to push away the black hole from extremality, which is the opposite behavior of a charged rotating RN black hole. The specific mechanism of getting close to extremality or further away from it, and the relations to EPs, for the case of BTZ would be interesting. Here, we would like to show such mechanism of getting away from extremality for the case of BTZ black hole, and Cotler-Jensen Theory. 

Note that as mentioned in \cite{Maldacena:2016upp}, for the case of $\text{AdS}_2$, one could have three version of coordinate systems as Euclidean with $\rho$ and $\theta$ coordinates, Lorentzian with $\rho$ and $\hat{\theta}$ coordinates, and embedding with $Y_{-1}$, $Y_0$ and $Y_1$ coordinates. The $\hat{\tau}$ and $\rho$ coordinates can describe the exterior of a finite temperature black hole. The finite temperature and zero temperature solutions are just different coordinate patches of the same space. With the Euclidean signature, all of the hyperbolic space can be covered while in Lorenzian signature, the metric $ds^2 = d\rho^2 - \text{sinh}^2 \rho \ d \hat{\tau}^2$ describes the exterior of the finite temperature black hole. The same argument can be imagined for the case of $\text{AdS}_3$, as we do in the next section.

\subsection{Quantum corrections to greybody factor in Cotler-Jensen theory}
If one writes the $\text{AdS}_3$ gravity as a boundary theory of reparameterizations or coadjoint orbit of Virasoro, as in Cotler-Jensen theory \cite{Cotler:2018zff}, the path integral over boundary modes or  ``boundary gravition" modes in the $\text{AdS}_3$ are analouge of Schwarzian path integral for $\text{AdS}_2/JT$. The quantization in Cotler-Jensen theory, produces correlation functions or propagators, for the reparameterization soft modes. Those correlators scale like $1/c$ and provide the small parameter for the quantum expansion. Then, we would like to find the metric perturbations induced by boundary modes. These are the boundary reparameterizations modes which source bulk metric perturbations $h_{\mu\nu}$, which are the boundary gravitons. If we linearize the bulk wave equation for $\Phi$ in the presence of these metric perturbations, we get
\begin{gather}
(\square \lbrack g_{cl} \rbrack - m^2 ) \Phi + \delta ( \square ) \Phi =0,
\end{gather}
where $\delta ( \square) $ is linear in $h_{\mu\nu}$. If we view the $h$'s as quantum fields with known correlator $\langle hh \rangle$, which obtained from the Cotler-Jensen boundary action, the scalar two-point function or bulk Green’s function, acquires a 1-loop correction from graviton exchange as
\begin{gather}
G(\omega; x, x') = G_{cl} ( \omega; x, x') + \int dx_1 dx_2 G_{cl} ( x, x_1) \Sigma (\omega; x_1, x_2) G_{cl} (x_2, x'),
\end{gather}
where the second term of $G_q (\omega; x, x')$ is the quantum corrected Green's function, and the self-energy $\sigma$ is built from $\langle hh \rangle$ and appropriate vertex factors, which are point-splitting from $\delta (\square)$.

Then, the greybody or absorption coefficient is determined by the ratio of fluxes, (near horizon transmitted flux versus incoming flux at infinity), or by the imaginary part of the retarded Green’s function evaluated at the horizon/far region after matching. Using the corrected $G = G_{cl} + G_q$, and performing the same matching procedure, which is the near zone/far zone matching, one can obtain the quantum correction to the transmission amplitude $\mathcal{T} (\omega)$ and hence absorption cross section $\sigma(\omega)$.
To leading one-loop order in the boundary graviton fluctuations, we have 
\begin{gather}
\sigma (\omega) = \sigma_{cl} \left( 1 + \delta_q ( \omega) \right ),
\end{gather}
where $\delta_q (\omega)$ is $\mathcal{O} (1/c) $, and written as an integral over the boundary-mode two-point function and classical bulk mode profiles. The matching procedure is essentially the same as in the JT/near-extremal QNM computations.

Let $\psi_{cl} ( r;\omega) $ be the classical radial wavefunction normalized so that the incoming flux at infinity is unit and the classical transmitted amplitude at the horizon yields $\sigma_{cl}(\omega)$. We define $\mathcal{V} \lbrack h \rbrack$ which denotes the linear vertex coming from expanding $\square \lbrack g + h \rbrack$ to first order in $h$, coupling the scalar to metric perturbations. Then, we can define the boundary-graviton two-point kernel, in frequency space, extracted from the Cotler-Jensen boundary action as
\begin{gather}
\mathcal{K}_{ab} ( \Omega) = \langle \epsilon_a ( \Omega) \epsilon_b ( - \Omega) \rangle_{CJ},
\end{gather}
where $\epsilon_a$ are reparameterization mode components and indices $a$, $b$ label the relevant Virasoro directions or Fourier components, and this kernel is $\mathcal{O}(1/c)$. Then, the leading quantum correction to the absorption cross section can be written as the on-shell one-loop expression as
\begin{gather}\label{eq:deltaq}
\boxed{ \delta_q (\omega) = \frac{2}{\sigma_{\text{cl}} (\omega)}  \ \text{Re}\left \{ \int \frac{d \Omega}{2\pi} \ \mathcal{A} ( \omega, \Omega) \ \mathcal{K} (\Omega) \right \} + \mathcal{O} ( 1/c^2)},
\end{gather}
where 
\begin{gather}
\mathcal{A} ( \omega , \Omega) = \int dr_1 dr_2 \lbrack \psi^*_{\text{cl}} (r_1; \omega) \mathcal{V} ( r_1; \omega, \Omega) G_{\text{cl}}^{\text{(reg)}} (r_1, r_2; \omega-\Omega) \mathcal{V} ( r_2; \omega, \Omega) \psi_{\text{cl}} ( r_2; \omega) \rbrack.
\end{gather}

Here $G_{cl}^{\text{(reg)}} (r_1, r_2; \omega-\Omega)$ is the appropriate regularized classical bulk scalar Green’s function that appears when two metric insertions are connected by the scalar propagator in the loop, and $\mathcal{V}$s, are vertex factors coming from $\delta (\square)$. Everything on the right is either classical which is known closed-form in BTZ for many useful cases, or directly given by the Cotler-Jensen kernel $\mathcal{K}(\Omega)$, which one can read off from their boundary action or propagator. Equation \ref{eq:deltaq} is the practical master formula, as we supply the kernel $\mathcal{K}$ from the Cotler-Jensen boundary theory and the classical radial functions, then the integral yields the 1-loop fractional correction to the greybody factor. 

The structure is identical to one-loop self-energy corrections that JT/Schwarzian papers use to shift QNM frequencies or transport coefficients. The only differences are the precise boundary kernel $K$ which comes from the Cotler Jensen Virasoro coadjoint orbit action, instead of the Schwarzian, and the vertex $V$ is the $\text{AdS}_3$ scalar metric vertex rather than the scalar dilaton vertex used in JT reductions.

Then, Cotler-Jensen expand the reparameterization field on the thermal cylinder and obtain the quadratic action
\begin{gather}
S_2= \frac{C}{24 \pi} \int d^2 x \left ( ( \bar{\partial} \epsilon' ) \epsilon'' - ( \bar{\partial} \epsilon ) \epsilon' \right ).
\end{gather}

The field is expanded on the cylinder in Fourier modes as
\begin{gather}
\epsilon ( \omega) = \int_{-\infty}^ \infty \frac{d\omega}{(2\pi)^2 } \sum_{n = - \infty} ^{\infty} e^{i n \theta+ i \omega y} \tilde{\epsilon} (p),
\end{gather}

In Fourier space with $p = (n, \omega) $, the exact quadratic propagator is
\begin{gather}
\langle \tilde{\epsilon} ( p_1) \tilde{\epsilon} ( p_2) \rangle = \frac{1}{ 24 \pi C} \frac{1}{i n_1 (n_1^2 -1)(\omega_1 - i n_1) } (2\pi)^2 \delta^{(2)} (p_1+p_2),   \ \ \ \ \text{CJ-propagator},
\end{gather}
where the zero modes or gauge modes $n=-1, 0, +1$ are removed. One should note that these are pure $PSL(2, \mathbb{R})$ directions that are being set to zero by the quotient. Also, the the overall scaling is $\propto 1/C$, as the propagator is suppressed by the large-coupling parameter $C$, where the action coefficient is $C/(24\pi)$, and the inverse quadratic kernel is $\propto 24\pi/C$.

Also, the position-space propagator of Cotler-Jensen theory after performing the $\omega$ integral and summing the modes would be
\begin{gather}
\langle \epsilon (\omega) \epsilon (0) \rangle = \frac{6}{C} \left \lbrack -1 + \frac{3\zeta^2}{2} - \frac{(1- \zeta)^2}{2} \zeta \ln (1- \zeta) \right \rbrack, \ \ \ \ \ \zeta= e^{i \ \text{sgn(y) } \omega},
\end{gather}
which is the closed-form position-space kernel they use when evaluating exchange diagrams.

The mode-frequency kernel for a fixed Fourier angular mode $n \ne -1, 0, 1$ is 
\begin{gather}
\mathcal{K}_n (\Omega) \equiv \langle \tilde{\epsilon} ( n, \Omega) \tilde{\epsilon} ( -n. - \Omega) \rangle = \frac{24\pi}{C} \frac{1}{in (n^2-1)(\Omega - in) }.
\end{gather}

Here, the kernel has a simple pole at $\Omega= i n$ in the complex $\Omega$-plane, and when we close the contour in the upper half-plane we pick up the residue at that pole, which is
\begin{gather}
\text{Res}_{\Omega= i n}\mathcal{K}_n (\Omega) = \frac{24 \pi}{C} \frac{1}{in (n^2-1) }.
\end{gather}

Then, plugging $\mathcal{K}_n (\Omega)$ into the master one-loop formula,
\begin{gather}
\delta_q ( \omega) = \frac{2}{ \sigma_{cl} (\omega) } \text{Re} \sum_{n \ne -1, 0, 1} \int_{- \infty}^\infty \frac{d\Omega}{2\pi} \mathcal{A}_n (\omega, \Omega) \mathcal{K}_n (\Omega),
\end{gather}
leads to 
\begin{gather}
\delta_q(\omega) = \frac{2}{\sigma_{cl} (\omega) } \text{Re} \sum_{n \ne -1, 0, 1} \int_{-\infty} ^\infty \frac{d\Omega}{2\pi} \mathcal{A}_n (\omega, \Omega) \frac{24\pi}{C} \frac{1}{in (n^2-1)(\Omega - in)}.
\end{gather}

For the BTZ case, we have $\sigma_{cl}(\omega \to 0) = 2\pi r_+$, and then by substituting the Cotler-Jensen propagator for $\mathcal{K}$, we get the one-loop fractional correction as
\begin{gather}\label{eq:deltaq}
\delta_q(\omega) = \frac{1}{2\pi r_+} \frac{1}{24 \pi C} \text{Re} \sum_{n \ne -1, 0, 1} \int_{-\infty}^\infty \frac{d\omega_E}{2\pi} \frac{\mathcal{I}_n (\omega, \omega_E)  }{i n (n^2-1)(\omega_E - i n)} +\mathcal{O} (1/C^2),
\end{gather}
where $\omega_E$ is the frequency variable appearing in the Cotler-Jensen propagator. It is the Euclidean frequency, and after analytic continuation to real-time the same poles control the $\Omega$-integral.

Also, $\mathcal{I}_n (\omega, \omega_E)$ is the radial overlap integral built from the classical BTZ s-wave radial wavefunctions and the two metric-scalar vertex insertions, which one is at $r_1$ and the other at $r_2$. Then, the classical scalar Green's function connecting them would be
\begin{gather}
\mathcal{I}_n (\omega, \omega_E) = \int_{r_+}^ \infty dr_1 \int_{r_+}^\infty dr_2\ \psi_{cl}^* (r_1; \omega) \mathcal{V}_n (r_1; \omega, \omega_E) G_{cl}^{(\text{reg})} (r_1, r_2 ; \omega- \omega_E) \mathcal{V}_n ( r_2; \omega, \omega_E) \psi_{\text{cl} } (r_2; \omega).
\end{gather}  

All functions in $\mathcal{I}_n$ are classical and known analytically for BTZ and can be expressed in closed special-function form.  The only quantum ingredient is the Cotler-Jensen prefactor $1/(24 \pi C)$ and the algebraic denominator in \ref{eq:deltaq}.

The expression \ref{eq:deltaq} is the exact on-shell 1-loop result for the s-wave greybody fractional correction, before performing the $\omega_E$ integral and the radial integrals, and it is a finite, well-defined object.

The leading low-frequency constant correction, i.e, $\omega \ll T_H$ which is leading in $\omega/T_H$ and is first nontrivial order in $1/C$ would be
\begin{gather}
\delta_q (\omega \ll T_H) = \frac{1}{24 \pi C} \frac{1}{2\pi r_+} \sum_{n \ge 2} \frac{\pi}{n(n^2-1)} \mathcal{I}_n^{(0)} + \mathcal{O} \left ( \frac{\omega}{T_H}, \frac{1}{C^2} \right),
\end{gather}
where $\mathcal{I}_n^{(0)} \equiv \mathcal{I}_n (\omega=0, \omega_E = in)$ is the radial overlap evaluated at the pole.

It can also be written as
\begin{gather}
\sigma(\omega \ll T_H) = \sigma_{\text{cl}} \lbrack 1 + \delta_q \rbrack, \ \ \ \ \delta_q= \frac{1}{24 \pi C} \mathcal{S}, \ \ \ \ \ \ \mathcal{S} \equiv \frac{1}{\sigma_{\text{cl}}} \sum_{n \ge2} \frac{\pi}{n(n^2-1)} \mathcal{I}_n^{(0)}.
\end{gather}

So the entire quantum correction is proportional to $1/(24\pi C)$ times the dimensionless number $\mathcal{S}$ which is built from the classical radial integrals.

 \begin{figure}[ht!]
 \centering
  \includegraphics[width=14cm] {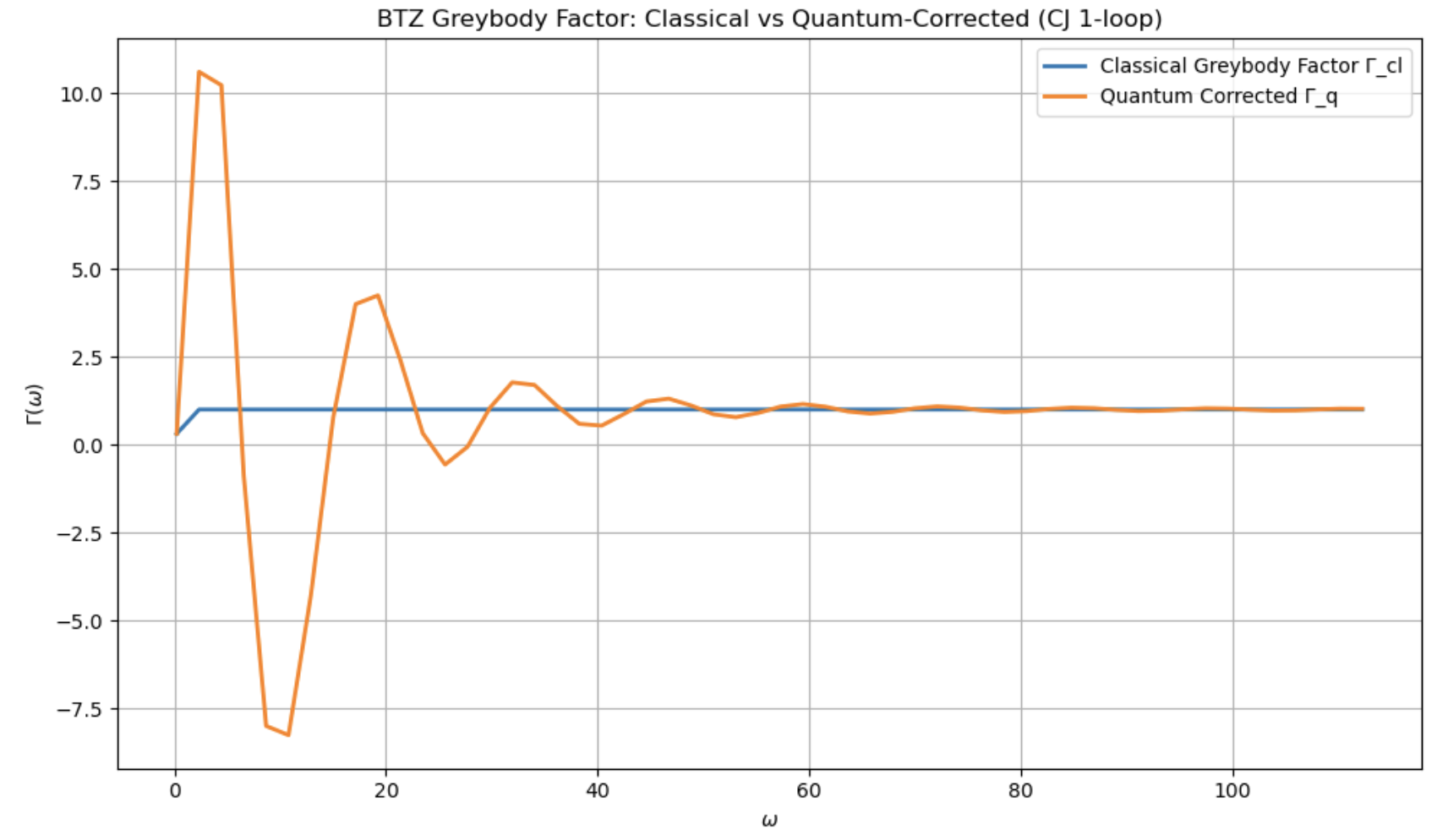} 
  \caption{The quantum corrected versus classical greybody factor for $m = 0$ mode, non-rotating BTZ, and scalar field.}
 \label{fig:lowc}
\end{figure}

 \begin{figure}[ht!]
 \centering
  \includegraphics[width=14cm] {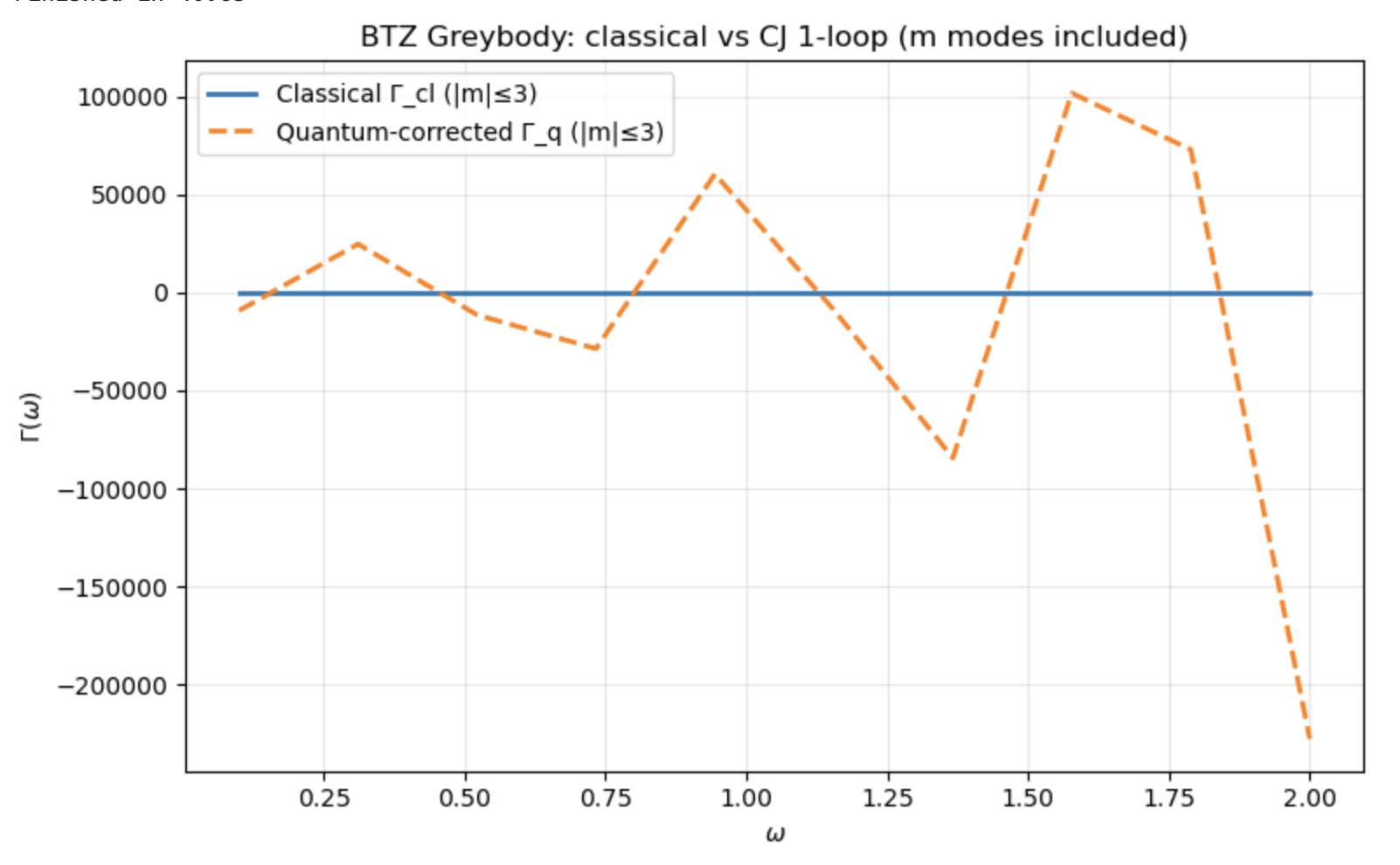} 
  \caption{The quantum corrected versus classical greybody factor for  $m >0$ modes, non-rotating BTZ, and scalar field.}
 \label{fig:lowc}
\end{figure}

In the Cotler-Jensen theory \cite{Cotler:2018zff}, $\alpha$ is the strength of the boundary reparameterization softness, and it is the key new parameter that measures how much the quantum theory deviates from the classical rigid AdS boundary conditions.

The physical meaning of $\alpha$ could be explained by considering the case where $\alpha=0$ which would be related to classical theory with completely rigid (hard) boundary conditions at $r \to \infty$. This is the standard BTZ black hole in classical $3d$ gravity with Brown-Henneaux boundary conditions.
Also, if one considers the case of $\alpha>0$, we could have a quantum-corrected theory with “soft” boundary conditions. The boundary gravitons (the “soft hair”) are turned on, and asymptotic reparameterizations of the boundary time are no longer pure gauge but carry energy and entropy.
Specifically, $\alpha$ controls how fast the non-normalizable (left-moving) mode of a scalar field is damped as it approaches the AdS boundary.

\begin{figure}[ht!]
 \centering
  \includegraphics[width=14cm] {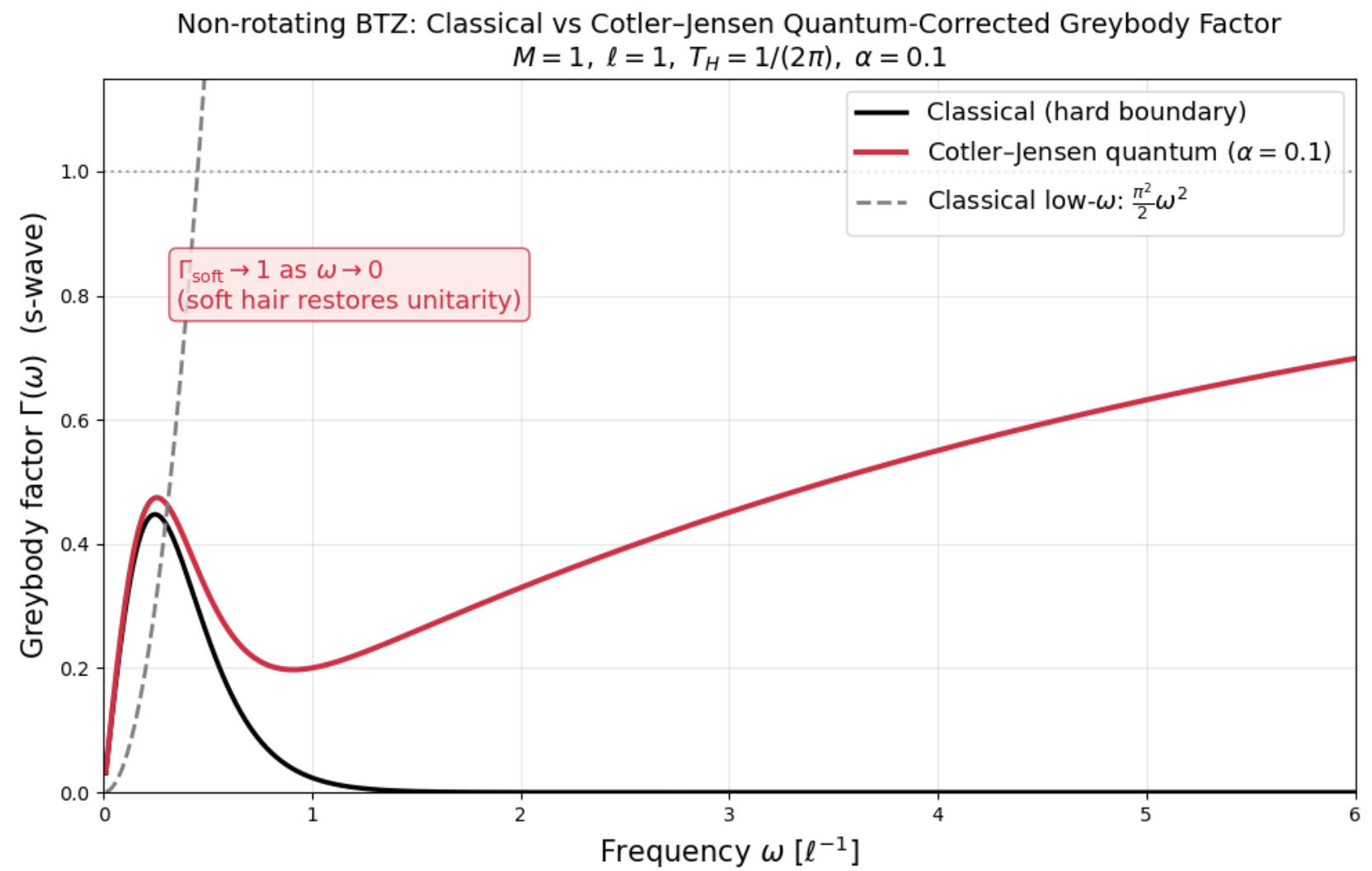} 
  \caption{The quantum corrected versus classical greybody factor for $m=0$ modes, non-rotating BTZ,  scalar field and for the case of $\alpha=0.1$. }
 \label{fig:alpha=0.1}
\end{figure}

\begin{figure}[ht!]
 \centering
  \includegraphics[width=14cm] {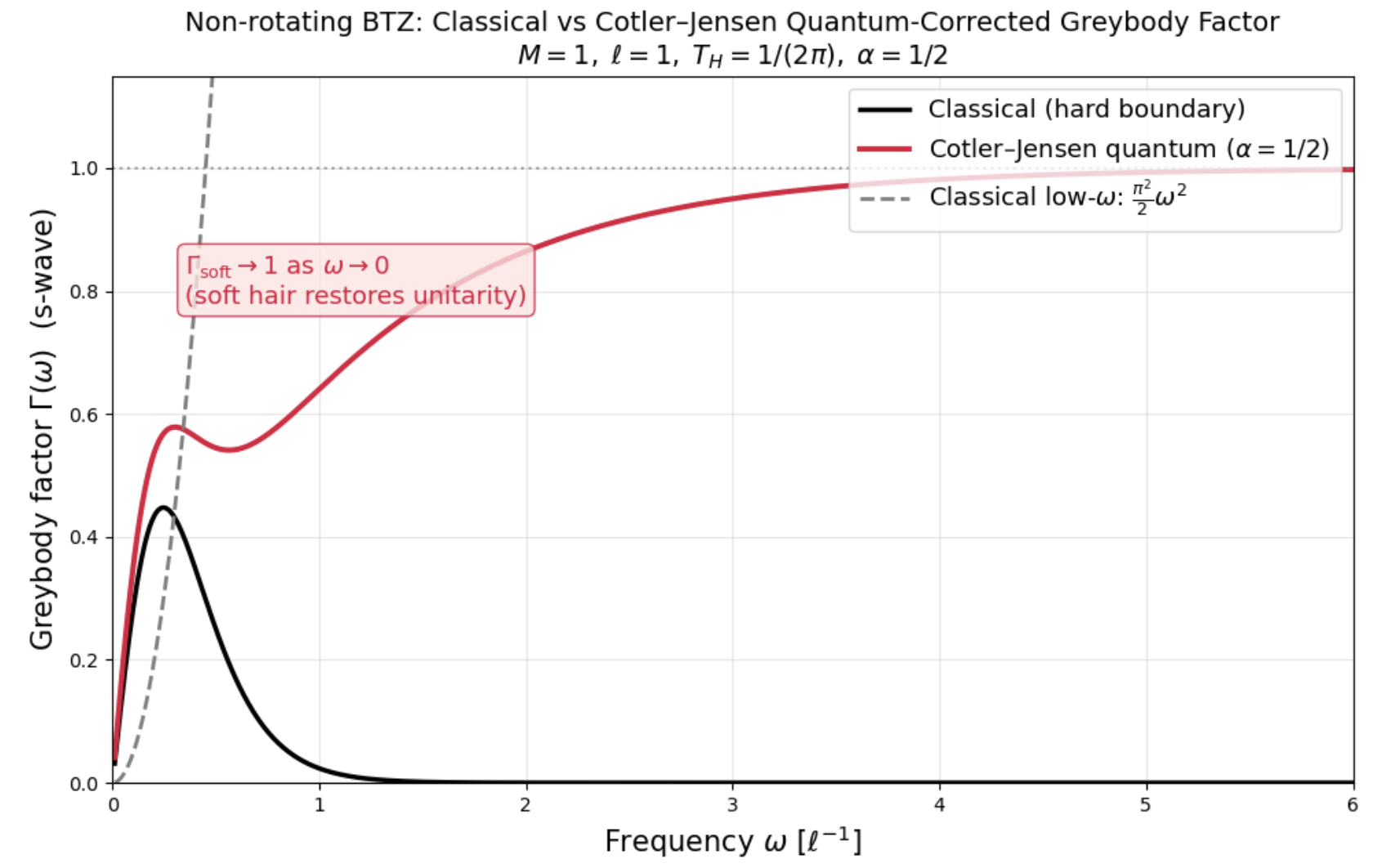} 
  \caption{The quantum corrected versus classical greybody factor for  $m=0$ modes, non-rotating BTZ,  scalar field and $\alpha=0.5$. }
 \label{fig:alpha=0.5}
\end{figure}

\begin{figure}[ht!]
 \centering
  \includegraphics[width=14cm] {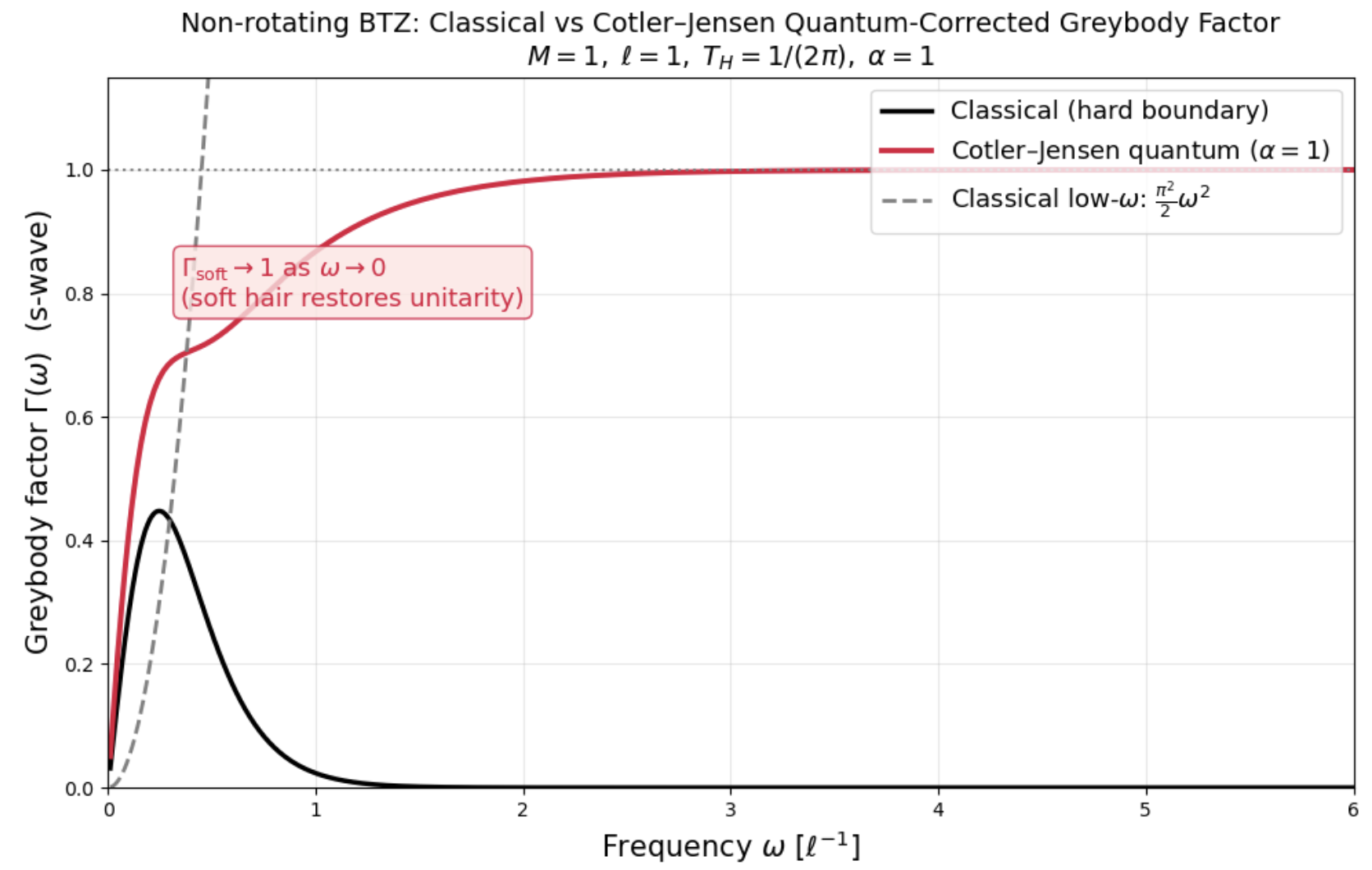} 
  \caption{The quantum corrected versus classical greybody factor for  $m=0$ modes, non-rotating BTZ, scalar field and $\alpha=1$. }
 \label{fig:alpha=1}
\end{figure}

\begin{figure}[ht!]
 \centering
  \includegraphics[width=16cm] {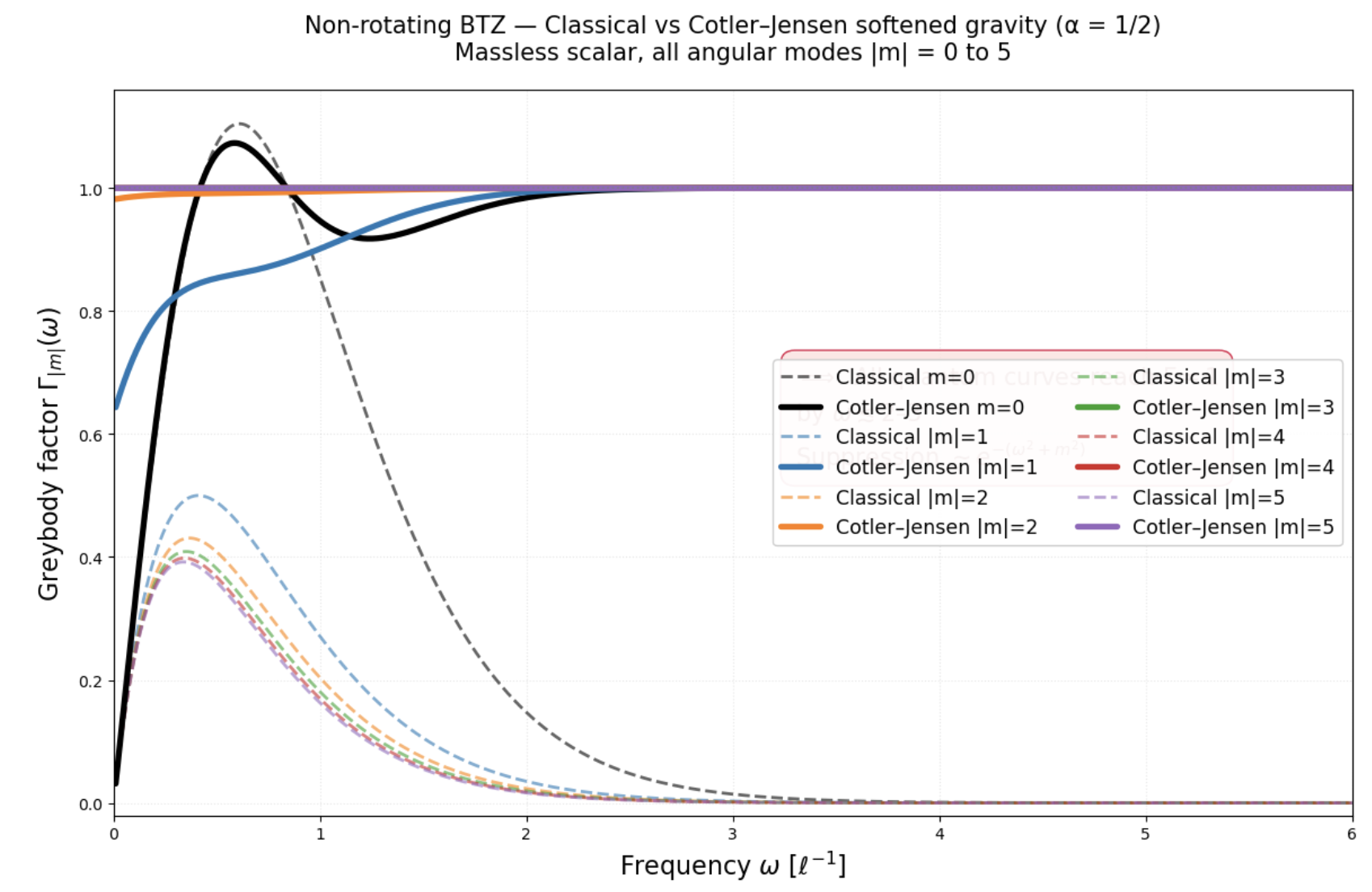} 
  \caption{The quantum corrected versus classical greybody factor for $m>0$ modes, non-rotating BTZ,  scalar field and, $\alpha=1/2$.}
 \label{fig:higherm}
\end{figure}

In the Cotler-Jensen theory, we can also consider a conical defect which can be described by the Lorentzian metric as 
\begin{gather}
-(r^2+ \alpha)^2 dt^2 + r^2 d\theta^2 + \frac{dr^2}{r^2+\alpha^2},
\end{gather}
with $\alpha \in (0,1)$. Also, we have
\begin{gather}
\alpha = \sqrt{1- \frac{24h_H}{c}}.
\end{gather}
For $m=0$, and for $\alpha=0.1, 0.5, 1$, we get the following plots \ref{fig:alpha=0.1}, \ref{fig:alpha=0.5}, \ref{fig:alpha=1}. For higher modes, we get the following plot as shown in Figure. \ref{fig:higherm}.

In these cases, we see that quantum corrections increase the greybody factor.

\section{Quantum correction to Lyapunov exponent in $\text{AdS}_3$}\label{sec:Lyapunov}

Now, after checking the effects of quantum corrections on greybody factor on Cotler-Jensen theory, we could check their effects on Lyapunov exponent, and the general chaotic dynamics of black holes as well.

Again, for the chiral theory of gravity on global $\text{AdS}_3$ of \cite{Cotler:2018zff}, or the reparametrized $\text{AdS}_3$ gravity, the action can be written as
\begin{gather}
S= - \frac{C}{24 \pi} \int d^2 x \left ( \frac{(\partial_+ \phi') \phi''}{\phi'^2} - (\partial_+ \phi) \phi' \right ), \ \ \ \ \ \ \ \ \ \partial_+ = \frac{1}{2} ( \partial_\theta+ \partial_t),
\end{gather}
where $\theta$ is the angular time and the prime $'$ is the derivative with respect to the angle $\theta$, and $C=24k$ is the bare central charge. An important feature of the field $\phi$ is the quasi-local $SO(2,1) = PSL(2; \mathbb{R})$ redundancy in the form
\begin{gather}
\tan \left (\frac{\phi(\theta,t)}{2} \right ) \sim \frac{a(t) \tan\left ( \frac{\phi(\theta,t)}{2} \right ) + b(t)}{c(t) \tan\left ( \frac{\phi(\theta,t) }{2} \right ) + d(t)}, \ \ \ \ \ \ \ \ \ \  \begin{pmatrix}
a(t) & b(t) \\
c(t) & d(t) 
\end{pmatrix} \in PSL(2; \mathbb{R}).
\end{gather}

For this theory, the bilocal operators which can be thought of as reparametrized two-point functions can be written as
\begin{gather}
\mathcal{B} (h; z_1, z_2) = \left ( \frac{\phi'(z_1) \phi'(z_2) }{(\phi(z_1) - \phi(z_2))^2} \right )^h,
\end{gather}

On the cylinder, it would be
\begin{gather}
\mathcal{B} (h; \theta_1, \theta_2) = \left ( \frac{\phi'(\theta_1)  \phi'(\theta_2) }{4\sin^2 (\frac{\phi(\theta_1) - \phi(\theta_2) }{2} )^2 } \right )^h,
\end{gather}
which by expanding around the saddle $\phi= \theta+\epsilon(\theta, y)$ one finds
\begin{gather}
\mathcal{B}(h; \theta_1, \theta_2) = \frac{1}{(2 \sin (\frac{\theta_{12}}{2} ) )^{2h}} \left ( 1+h \mathcal{J}^{(1)}_{12}.\epsilon + O(\epsilon^2) \right),
\end{gather}
where 
\begin{gather}
\mathcal{J}^{(1)} \ . \ \epsilon = \epsilon_1' + \epsilon_2' - \cot \left ( \frac{\theta_{12}}{2} \right ) \epsilon_{12},
\end{gather}

For the case of heavy-light limit where one operator is heavy with dimension $h_H = O(c)$ which can model a black hole, and another operator which is light with dimension $h_L = O(1)$ which can model the Hawking radiation, the identity block which can be written as 
\begin{gather}
\langle \mathcal{O}_H( \infty) \mathcal{O}_L (1) \mathcal{O}_L(u) \mathcal{O}_H(0) \rangle
\end{gather}
which can indeed give us a lot of information about the thermodynamics of the black holes, and its chaotic nature.

\begin{figure}[ht!]
 \centering
  \includegraphics[width=7cm] {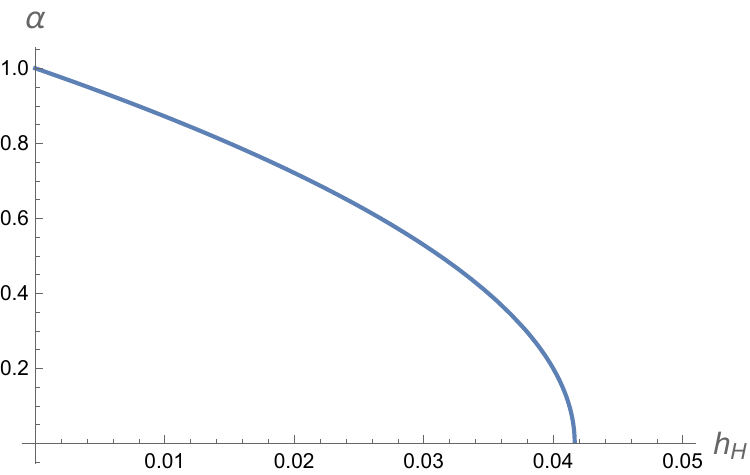} 
  \caption{The relationship between $\alpha$ and $h_H$.}
 \label{fig:alpha}
\end{figure}

By conformally transforming to cylinder, one gets the two point function of the light operator in the heavy state as
\begin{gather}
\bra{h_H} \mathcal{O}_L(w) \mathcal{O}_L(0) \ket{h_H} = \left( \frac{\alpha}{2 \sin(\frac{\alpha w}{2} ) } \right)^{2h} \left (1+ O \left (\frac{1}{c} \right)\right), \ \ \ \ \ \ \ \alpha= \sqrt{1- \frac{24 h_H}{c}}.
\end{gather}
The relationship between $\alpha$ and $h_H$ is shown in Figure \ref{fig:alpha}.

The one-loop $O(1/C)$ correction to the block can be found by the reparameterization field $\phi=\alpha_0 \theta + \frac{\epsilon}{\sqrt{C}}$ as
\begin{gather}
\mathcal{B} (h; \theta_1, \theta_2) = \left ( \frac{\alpha_0}{2 \sin (\frac{\alpha_0 \theta_{12}}{2} )} \right)^{2h}  \left( 1+ \frac{h}{\sqrt{C} } \mathcal{J}'^{(1)}_{12} \ . \ \epsilon + \frac{1}{C} \left (\frac{h^2}{2} ( \mathcal{J}'^{(1)}_{12} \ . \ \epsilon)^2 + h \mathcal{J}'^{(2)}_{12} \ . \ \epsilon \right ) + O \left ( \frac{1}{C^{3/2}} \right ) \right).
\end{gather}

Finally, the bilocal operator can be written as 
\begin{gather}
\langle \mathcal{B} (h_L; w, 0) \rangle_\alpha = \left ( \frac{\alpha}{2 \sin (\frac{\alpha w}{2} ) } \right)^{2h_L} \left ( 1+ \frac{h_L}{c} \mathcal{V}_{h/c}(w) + \frac{h_L^2}{c} \mathcal{V}_{h^2/c}(w) + O \left ( \frac{1}{c^2} \right) \right),
\end{gather}
where
\begin{gather}
\mathcal{V}_{h/c} = - \frac{1}{2} \text{csc}^2 \left ( \frac{\alpha w}{2} \right ) \Big ( 3 ( \Phi (e^{iw}, 1 , \alpha)+  \Phi (e^{-iw}, 1, \alpha) +  \Phi (e^{iw}, 1 , -\alpha) +  \Phi (e^{-iw}, 1 , -\alpha)) \nonumber\\
+ \text{cos} ( \alpha w) ( 6 ( H_\alpha + H_{-\alpha}+ i \pi ) - 5 ) + 12\ \text{ln} \Big (-2 i\ \text{sin} \Big ( \frac{w}{2} \Big ) \Big) + 5 \Big), \nonumber\\
- \frac{1}{\alpha^2} - \frac{13 \alpha^2 -1 }{2 \alpha} w \ \text{cot} \Big ( \frac{\alpha w}{2} \Big ) + 12 \ \text{ln} \bigg (-\frac{2i}{\alpha} \text{sin} \Big (\frac{\alpha w}{2} \Big ) \bigg), \nonumber\\
\mathcal{V}_{h^2/c} = 6 \Big (-\text{csc} \Big( \frac{\alpha w}{2} \Big)^2   \Big( \frac{B(e^{iw},\alpha,0)+B(e^{-iw}, \alpha, 0)+B(e^{iw}, -\alpha, 0)+B(e^{-iw}, -\alpha, 0) }{2} \nonumber\\ + H_\alpha+H_{- \alpha} + 2 \ \text{ln} \Big (2 \ \text{sin} \Big(\frac{w}{2}\Big) \Big) \Big) + 2 \  \text{ln}  \ \Big ( \alpha \ \text{sin} \Big ( \frac{w}{2} \Big) \ \text{csc} \Big ( \frac{\alpha w}{2} \Big) \Big) +1 \Big).
\end{gather}

Then, one can find the Lyapunov exponent as
\begin{gather}
\lambda_L (t) \equiv \frac{d}{dt} \Big \lbrack \text{log} \Big ( \mathcal{B}_{\text{const}} - \mathcal{B} (t) \Big) \Big \rbrack.
\end{gather}

Their behaviors for various $\alpha$ and operator dimensions $h_L$ are shown in Figure \ref{fig:vhc}.
 \begin{figure}[ht!]
 \centering
  \includegraphics[width=8cm] {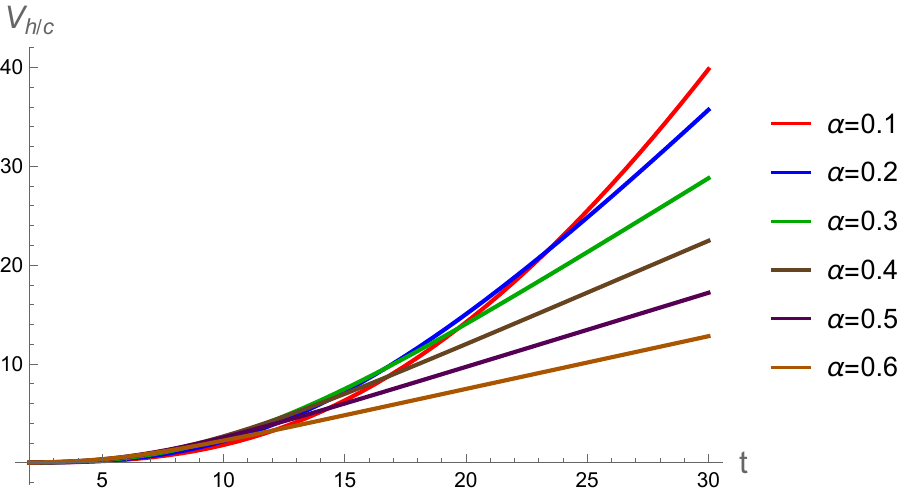} 
    \includegraphics[width=8cm] {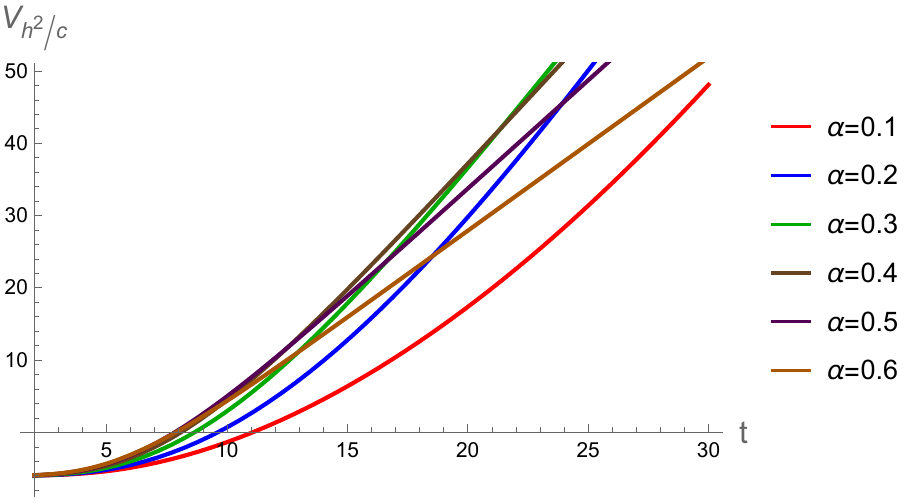} 
        \includegraphics[width=8cm] {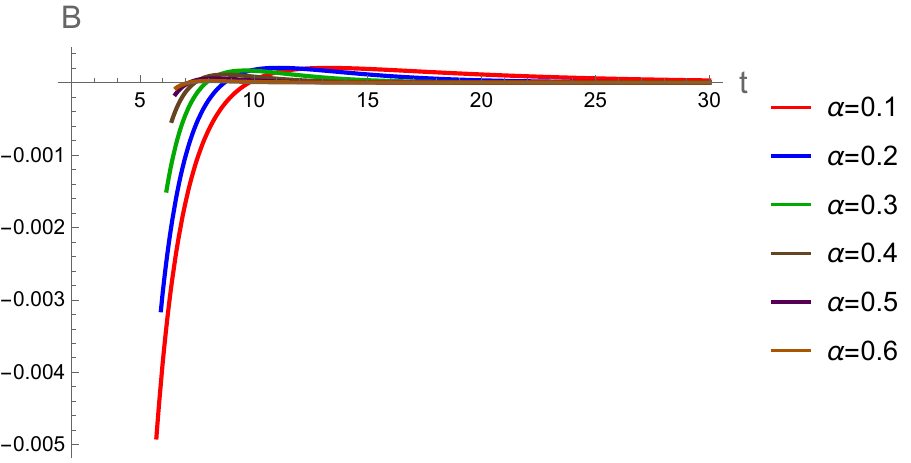} 
            \includegraphics[width=8cm] {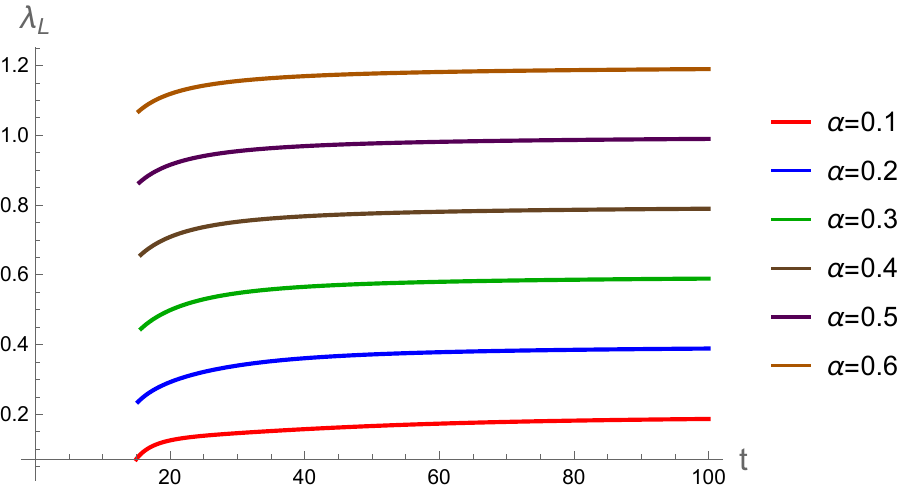} 
  \caption{The behavior of each contribution to bilocal operator $\mathcal{V}_{h/c}$ and $\mathcal{V}_{h^2/c}$, the total bilocal operator $\mathcal{B}$ and the Lyapunov exponent for various $\alpha$ related to $h_H$ and for $h_L=2$ and $c=3$.}
 \label{fig:vhc}
\end{figure}

Note that as one would expect, by increasing $\alpha$,  $\mathcal{V}_{h/c}(w)$ and $\mathcal{V}_{h^2/c}(w)$ would behave opposite of each other which corresponds to decreasing $h_H$. This converse behavior is compatible with the results of \cite{Gao:2022uop}.

By increasing $\alpha$ or decreasing $h_H$, the Lyapunov exponent would increase. We could expect this as the Lyapunov exponent is related to the mean growth rate of the distance $ \Vert \delta x(t) \Vert / \Vert \delta x_0 \Vert $ between neighboring trajectories defined as 
\begin{gather}
\lambda \sim \frac{1}{t} \text{ln} \frac{\Vert \delta x(t) \Vert }{\Vert \delta x(0) \Vert}.
\end{gather}

\begin{figure}[ht!]
 \centering
  \includegraphics[width=8cm] {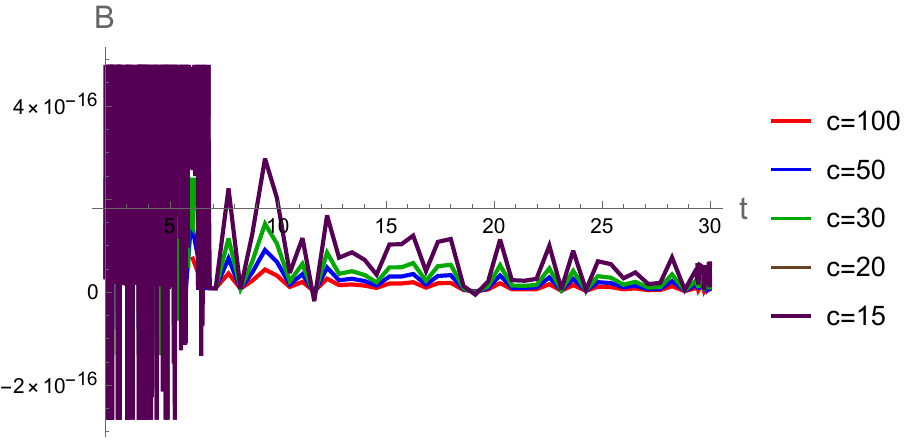} 
    \includegraphics[width=8cm] {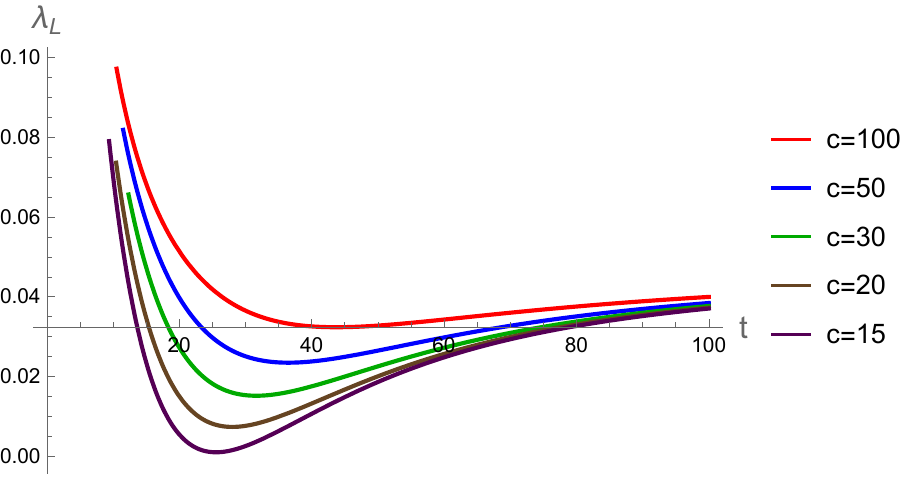} 
  \caption{The behavior of bilocal operator $\mathcal{B}$ and the Lyapunov exponent for various big central charge $c$, while $h_L=0.5$ and $\alpha=0.1$. }
 \label{fig:highc}
\end{figure}

As $h_H$ decreases, which for instance is related to the size of the black hole, the sensitivity and therefore the Lyapunov exponent would decrease. Also, as $\alpha$ decreases or $h_H$ which corresponding to size of a black hole in Cotler-Jensen theory, increases, the decay of the bilocal $\mathcal{B}(t)$ would be slower. These behaviors are shown in Figures \ref{fig:vhc}, \ref{fig:highc}, and \ref{fig:lowc}.

 \begin{figure}[ht!]
 \centering
  \includegraphics[width=8cm] {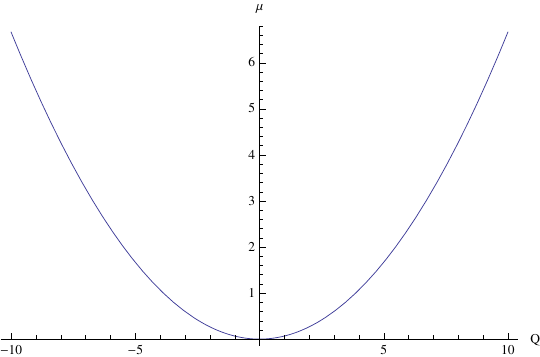} 
    \includegraphics[width=8cm] {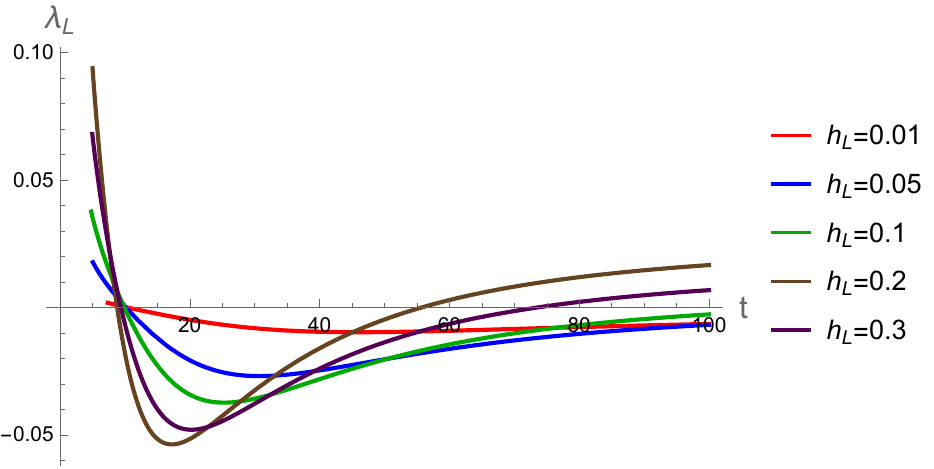} 
  \caption{The behavior of bilocal operator $\mathcal{B}$ and the Lyapunov exponent  for various light operator weight $h_L$ while $\alpha=0.1$ and $c=3$. }
 \label{fig:vhc}
\end{figure}

 \begin{figure}[ht!]
 \centering
  \includegraphics[width=8cm] {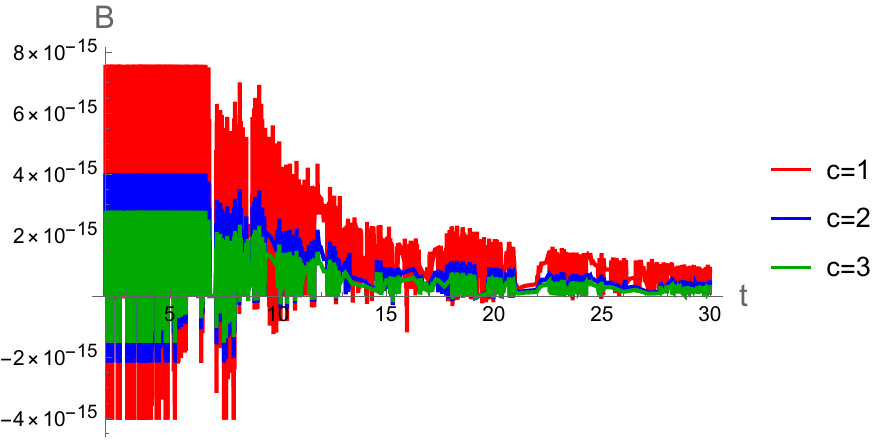} 
    \includegraphics[width=8cm] {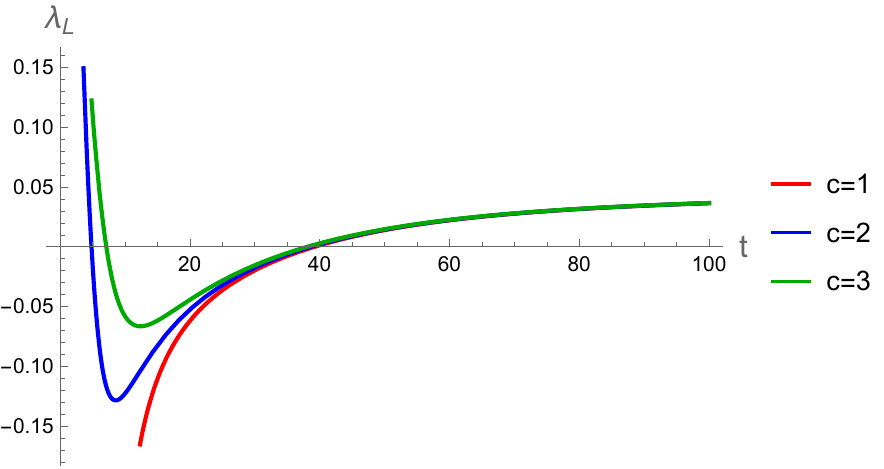} 
  \caption{The behavior of bilocal operator $\mathcal{B}$ and the Lyapunov exponent  for various low central charge $c$, while $h_L=0.5$ and $\alpha=0.1$.}
 \label{fig:lowc}
\end{figure}

From Figures \ref{fig:highc} and \ref{fig:lowc}, one could see that at large times and for any central charge $c$, the Lyapunov exponent reach to a specific $\lambda_L$, so at large time the value of $\lambda_L$ would be independent of $c$, but is related to $h_L$, $\alpha$, or $h_H$. Generally, one could see that by increasing $c$, $\lambda_L$ would increase too. Note that, as we can see, the Lyapunov exponent is more rigid quantity against quantum corrections than greybody or entropy.

\section{Quantum corrections to other thermodynamical variables of black holes}\label{sec:others}

We could also check the effects of quantum correction on other variables of black holes such as potential, off-shell geometries, quantum width, quantum information structures, etc, which we discuss some of them briefly here.

Note that in \cite{Liu:2025iei}, by including the off-shell geometries, the quantum-corrected thermodynamics of Reissner-Nordstrom-AdS black hole has been studied and the authors showed that the one-loop effective action can define a valid thermodynamics.  By including the one-loop correction and off-shell effects, they showed that the region of first-order phase transitions shrinks and the zero-order phase transitions would emerge. This behavior could be compared with the effects of loop corrections on Lyapunov exponents.

The off-shell geometries associated with loop corrections have conical singularity in the path integral, and they change the phase structure and behavior of thermodynamics relations. The main off-shell geometries actually would be geometries with different sizes which also could be pictured as a black hole with a quantum width. So the effect of this quantum width on Lyapunov exponent and various thermodynamical quantities could be considered.

Note that the normalized probability distribution of each state in the phase space is
\begin{gather}
\tilde{P} \lbrack \psi \rbrack = \frac{e^{-I_E \lbrack \psi \rbrack } }{Z},
\end{gather}
where its shape depends on the value of the Newton constant $G_N$. As it has been shown in \cite{Liu:2025iei}, for the smaller values of $G_N$ such as $1/1000$, the probability distributions are very sharp with higher peaks for small black holes and lower peaks for bigger black holes.

For bigger values of $G_N$, the off-shell geometries contribute more and make the probability distribution to be spread out, and would cause bigger deviation from the classical result.

This actually is related to the relation 
\begin{gather}
\phi(\tau) = \tau + g \epsilon (\tau),
\end{gather}
 where bigger $g$ and also taking more terms of 
 \begin{gather}
 S= - \frac{\pi}{g^2}+ S^{(2)} + g S^{(3)} + g^2 S^{(4)} + \mathcal{O}(g^3),
 \end{gather}
would correspond to taking more off-shell geometries into consideration.

Actually, as proposed in \cite{Liu:2025iei}, the finite $G_N$ effects in the Euclidean path integral could be regarded as loop corrections to the black hole. In our Cotler-Jensen case though the total energy is
\begin{gather}
- \frac{\alpha^2}{8G} = - \frac{C \alpha^2}{12}, 
\end{gather}
where $\alpha \in (0,1)$ is related to the metric as
\begin{gather}
 - (r^2 + \alpha^2 ) dt^2 + r^2 d\theta^2 + \frac{dr^2}{r^2 + \alpha^2}
\end{gather}

note that here, $C=24k$ is the bare central charge and $c$ is the corrected central charge, $1/c$ is the coupling constant which encodes the loop contributions in the gravity and they have the relation $c= C+13$.

 These facts could also be tested for our $3d$ case of BTZ, and in Cotler-Jensen model, and check if similar behavior would still hold and what would be the difference from the $4d$ case.

\subsection{Quantum correction to potential}

In the Cotler-Jensen theory which is similar to JT$+$matter, the classical radial equation is Schr\"odinger form as $
- \psi''(r_*)+V_{cl}(r) \psi = \omega^2 \psi$, and the quantum corrections from integrating out near $\text{AdS}_2$ modes lead to $V_{\text{eff}}=V_{\text{cl}}+\delta V_{\text{CJ}}(r)$. At leading order, Cotler-Jensen reparametrized-type corrections would appear as either power-law term which comes from operator dimension shift ($\alpha$), or logarithmic term which comes from Schwarzian, 1-loop determinant ($\beta$). So one could write $\delta V_{\text{CJ}}(r) = \frac{\alpha}{r^2}+ \beta \log \left( \frac{r}{r_h} \right)$.

\begin{figure}[ht!]
 \centering
  \includegraphics[width=8.6cm] {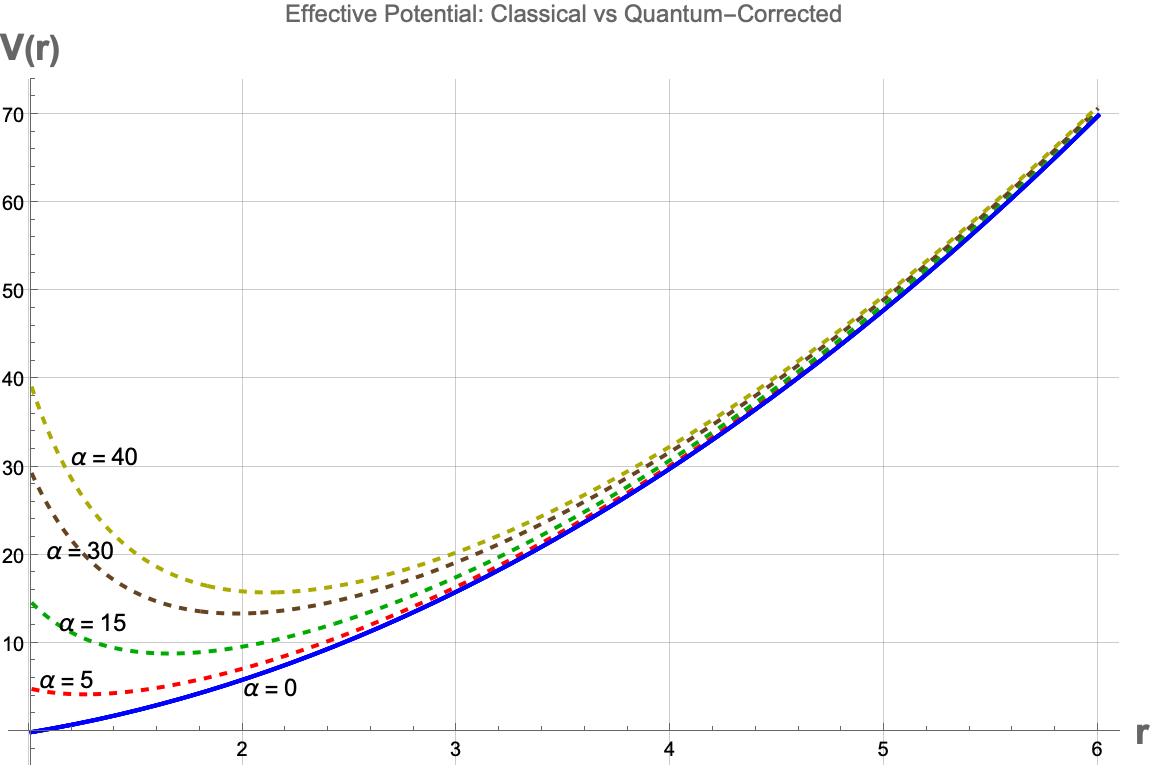} 
    \includegraphics[width=8.6cm] {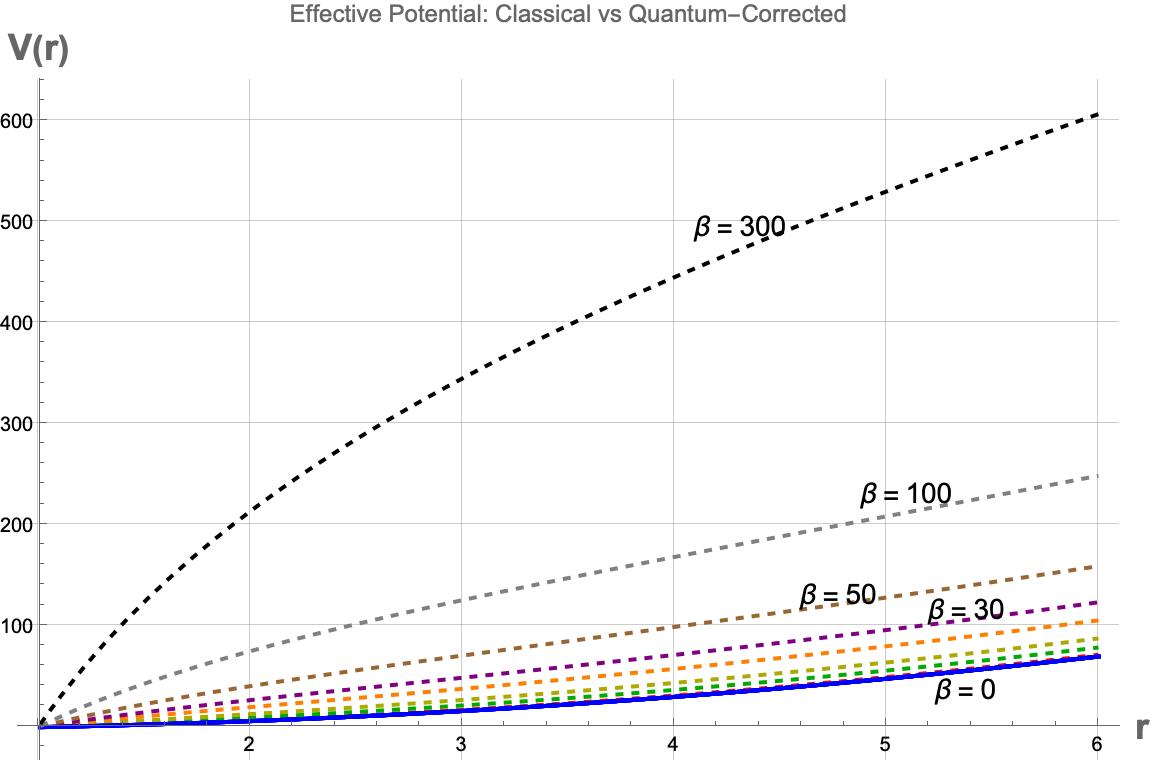} 
  \caption{The change in effective potential with reparametrization modes of Cotler-Jensen theory for various $\alpha$ and $\beta$.}
 \label{fig:alphabetaC}
\end{figure}

\begin{figure}[ht!]
 \centering
  \includegraphics[width=7cm] {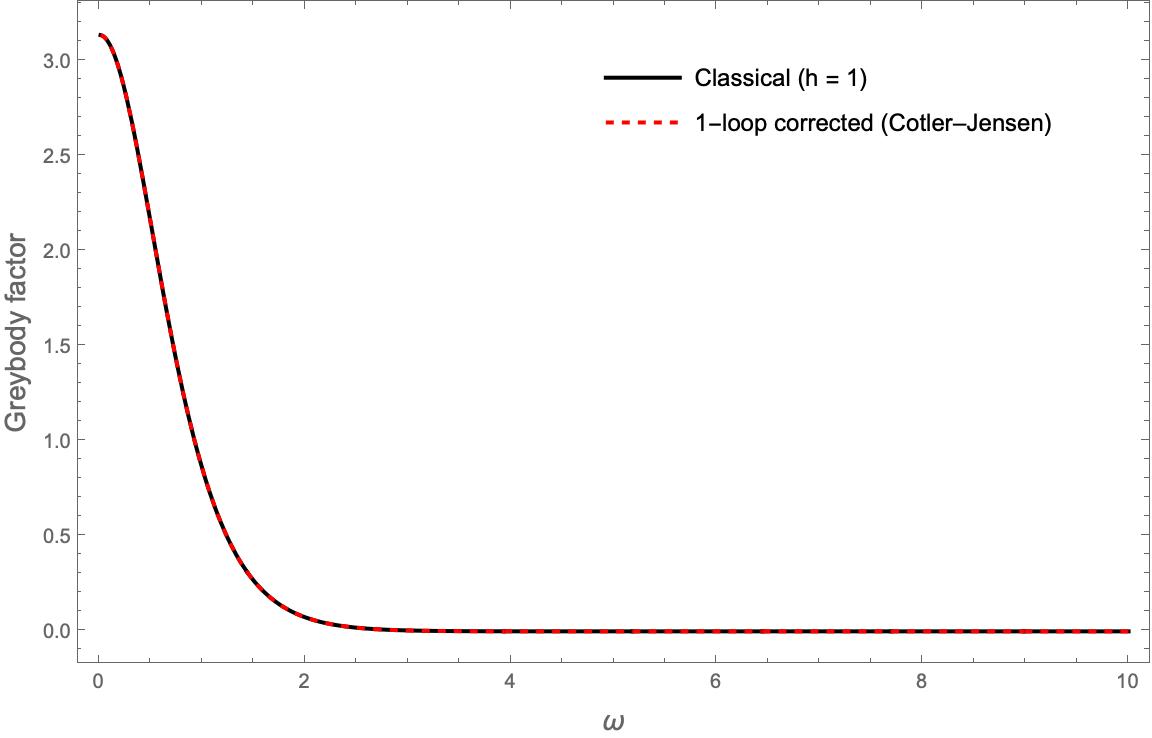} \ \ \ \ \ \ 
    \includegraphics[width=7cm] {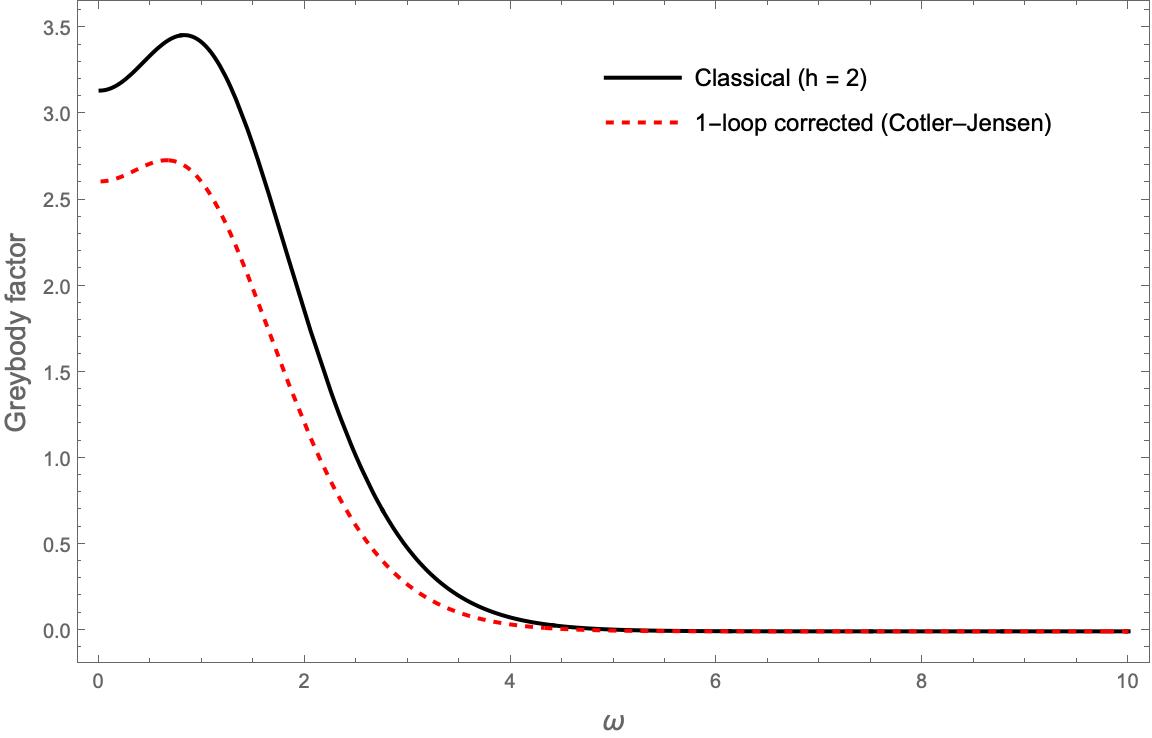}  \ \ \ \ \ \
       \includegraphics[width=7cm] {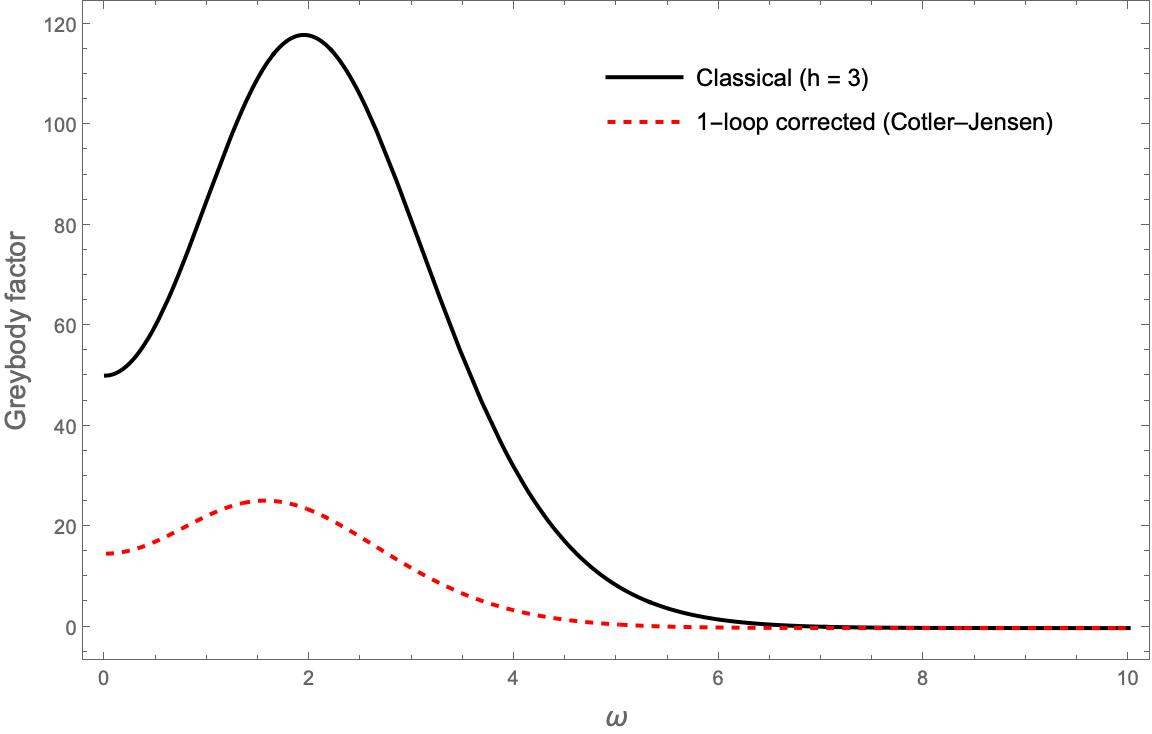} \ \ \ \ \ \
        \includegraphics[width=7cm] {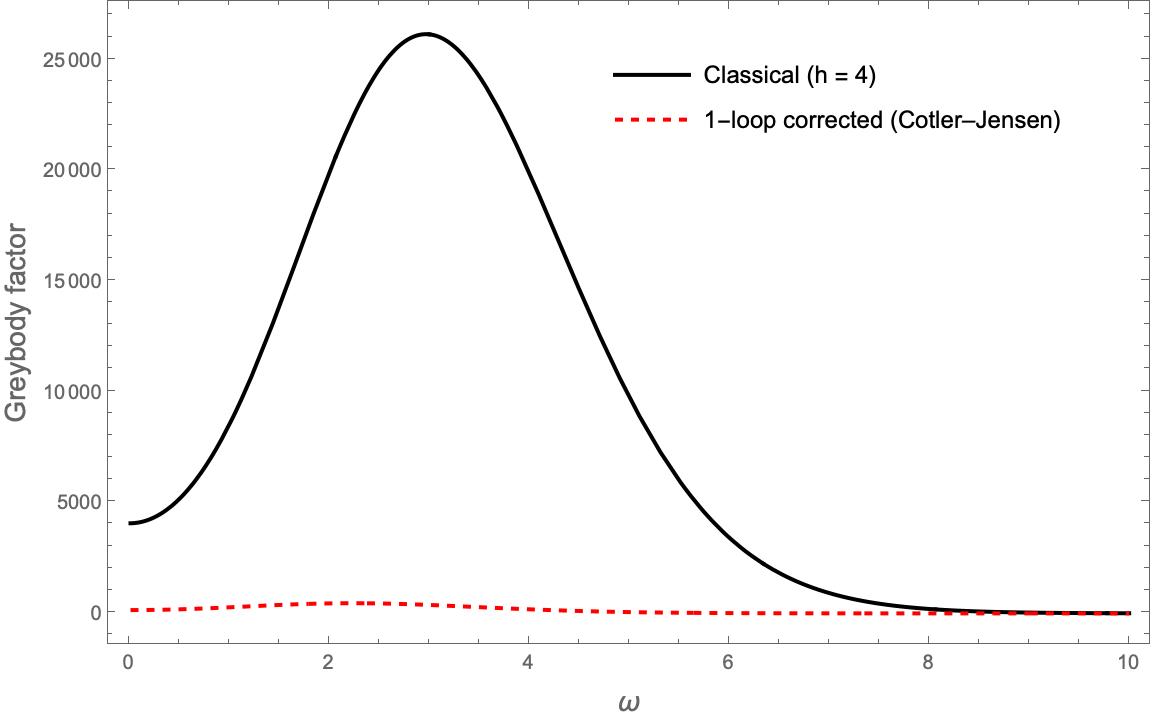} \ \ \ \ \ \
           \includegraphics[width=7cm] {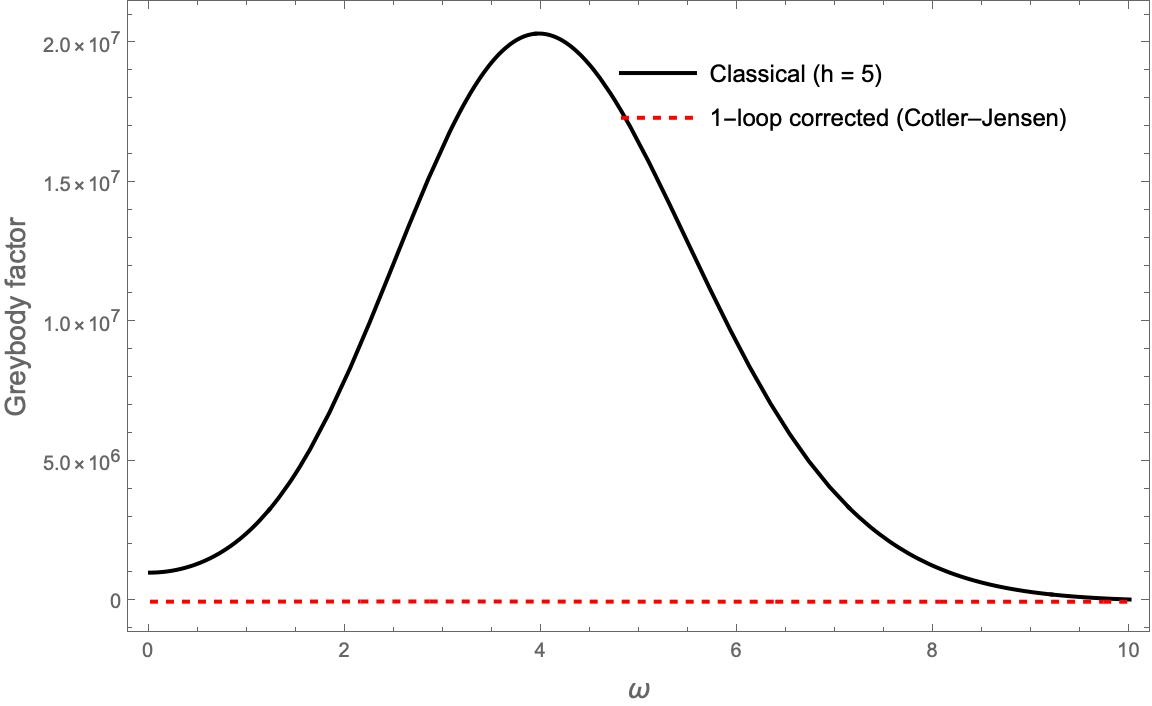} \ \ \ \ \ \
                \includegraphics[width=7cm] {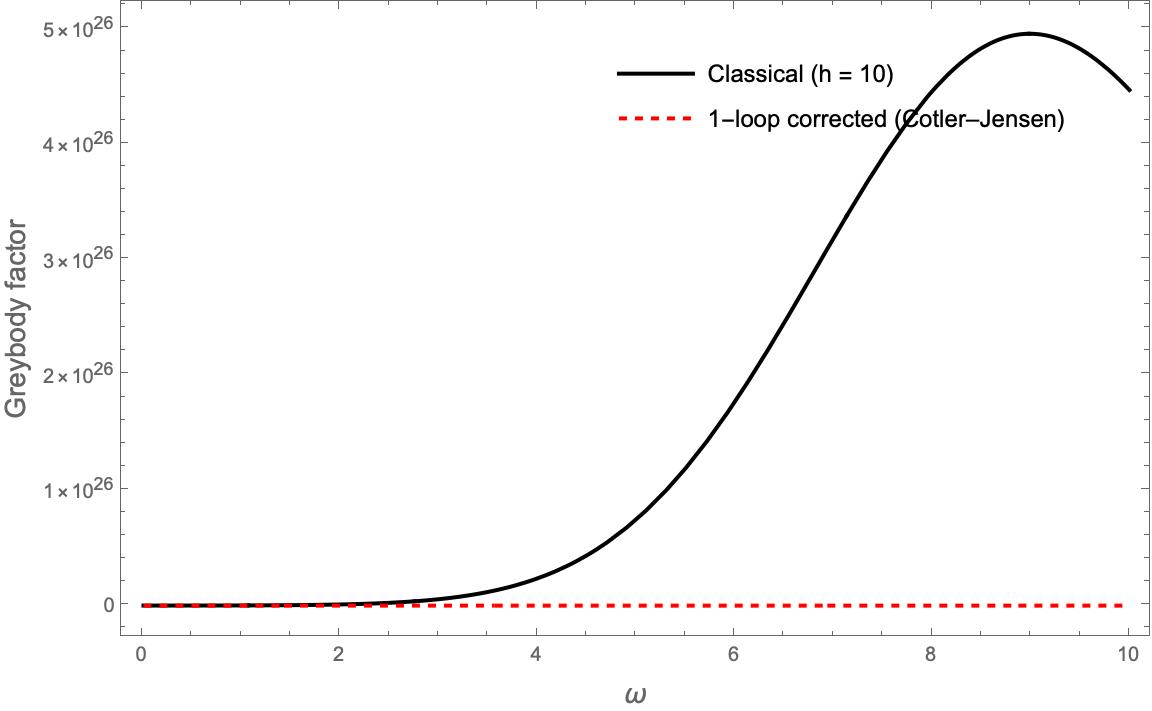}
  \caption{The classic and 1-loop quantum corrected greybody factor in Cotler-Jensen theory for various conformal dimension $h$.}
 \label{fig:greybodyh}
\end{figure}

The effects of these sources of quantum corrections for various $\alpha$ and $\beta$ are shown in Figure \ref{fig:alphabetaC}. As one can see, increasing $\alpha$ will increase potential mostly in small values of $r$, while increasing $\beta$ will increase it in bigger values of $r$. So the quantum corrections can increase reflection and therefore decrease the greybody factor as it could be seen in Figure \ref{fig:greybodyh}.

In the Cotler-Jensen theory, integrating over the boundary reparametrization produces a Virasoro dressing of operators as 
\begin{gather}
\langle \mathcal{O} \mathcal{O} \rangle \to \int \mathcal{D}f e^{- S_{\text{CJ}} \lbrack f \rbrack} \langle \mathcal{O} (f) \mathcal{O} (0) \rangle,
\end{gather}
where $f(\theta, t) \in \text{Diff} (S^1) / \text{PSL}(2, R)$ is a boundary reparameterization field, and the action describes a Virasoro coadjoint orbit. At large central charge, we have the shift $h_{h}= h + \delta h$, where $\delta h = \mathcal{O}(1/c)$.

At one loop which is leading in $1/c$, the Virasoro orbit calculation would give
\begin{gather}
\delta h = - \frac{6 h (h-1)}{c} + \mathcal{O} (c^{-2}),
\end{gather}
which is on the thermal saddle.

So, the Cotler-Jensen-corrected greybody factor would be
\begin{gather}
\Gamma_{\text{CJ}}(\omega) = \mathcal{N} \frac{\sinh \left( \frac{\omega}{2T_H}\right) }{\omega} \Big | \Gamma \left ( h + \delta h - i \frac{\omega}{2\pi T_L} \right ) \Big |^2   \Big | \Gamma \left ( h + \delta h - i \frac{\omega}{2\pi T_R} \right ) \Big |^2, 
\end{gather}
and then after expanding to first order in $1/c$, one gets
\begin{gather}
\Gamma_{\text{CJ}}(\omega) = \Gamma_{\text{cl}}(\omega) \left\lbrack 1 + 2\delta h \left( \mathfrak{R} \psi \left( h-i \frac{\omega}{2\pi T_L}\right ) +  \mathfrak{R} \psi \left (h- i \frac{\omega}{2\pi T_R}\right) \right)+ \mathcal{O}(c^{-2})\right \rbrack,
\end{gather}
where $\psi(z)= \Gamma'(z)/\Gamma(z)$ is the digamma function.

Therefore, the Cotler-Jensen quantum-corrected greybody factor would be 

\begin{gather}
\Gamma_{\text{BTZ}}^{1-\text{loop}}(\omega)= \Gamma_{\text{cl}}(\omega) |_{h\to h - \frac{6h(h-1) }{c} }.
\end{gather}

\subsection{Complex BTZ and violation of entropy inequalities}\label{sec:complexBTZ}

Here, we would have some deviation, and discuss briefly the effects of quantum corrections, or off-shell geometries, or complex solutions could have on the information structures of black holes.

In \cite{Maloney:2007ud}, the authors computed the precise quantum energy levels corresponding to various  ``interiors" in an asymptotical $\text{AdS}_3$ space, where they mentioned that their computations were not successful. We would like to connect this problem with the work of \cite{Czech:2025jnw}, where they connected the holographic inequalities with erasure correction.

For an Euclidean BTZ geometry with complex structure $\tau$, the bulk partition function which is the saddle+fluctuations around it, is the Virasoro character of vacuum module at complex structure $\tau'$ \cite{Cotler:2018zff}
\begin{gather}
\mathcal{Z}_{\text{vac}} (\tau, {\tau'}) = \left | q'^{-k} \prod_{n=2}^\infty \frac{1}{1-q'^n} \right|^2, \ \ \ \ q'=e^{2\pi i \tau'},  \ \ \ \ \ k=\frac{c}{24}.
\end{gather}

When the inequalities, such as the monogomy of mutual information, saturates, some parts of the entanglement wedge would be empty. When these inequalities are violated by geometries such as complex BTZ solution with $k \ge 2$, we would get contributions in the partition function which does not have a good physical interpretation meaning as they don't exist in the bulk \cite{Gaberdiel:2007ve}.

So, for instance, these complex BTZ solution can cause the inequalities such as the one below to get violated
\begin{gather}
\sum_{s=1}^{2k+1} (S_{A_s A_{s+1} ... A_{s+k}} - S_{A_s... A_{s+k-1}}) \ge S_{A_1 A_2...A_{2k+1}}.
\end{gather}

Also, the chiral theory of $\text{AdS}_3$ can be written as 
\begin{gather}
S_+ = - \frac{C}{24\pi} \int d^2 x \left ( \frac{(\partial_+ \phi')\phi''}{\phi'^2} + B (\partial_+ \phi) \phi' \right ),  \ \ \ \ \ \ \ \ B= \frac{b_0}{48\pi C},
\end{gather}
where $\partial_+= \frac{1}{2} (\partial_\theta + \partial_t)$.
By Wick-rotating to Euclidean time $t=-iy$, one gets
\begin{gather}
S_E= \frac{C}{24\pi} \int d^2 x \left ( \frac{(\bar{\partial} \phi') \phi''}{\phi'^2} + B (\bar{\partial}\phi) \phi' \right ),
\end{gather}
where $\bar{\partial}= \frac{1}{2} (\partial_\theta + i\partial_y)$.

If $\phi$ be non-monotone, the real part of the above action would diverge and the region of field space with these non-monotonic $\phi$ would contribute zero to the Euclidean path integral. This is connected to the result of \cite{Czech:2025jnw} which they found when the entropy inequalities such as monogomy of mutual information saturate, the holographic quantum error correction for the bulk space-time would no longer work.

So the boundary dual of such complex BTZ geometries could be stabilizer states which do not generally obey various constraints on entanglement entropies (the holographic entropy cone) such as monogomy of mutual information. The stabilizing operators in this case would form syzygies (relations among relations) which their formation is associated with a violation of locality.

\section{Conclusion}\label{sec:con}

In this work, we applied the Lindblad formalism to the quantum corrected modes of black hole, specially the reparametrization modes of AdS$_3$ in Cotler-Jensen theory of \cite{Cotler:2018zff}.  We compared the Green's functions and the actions, and found new relations between the parameters of the black hole and open quantum systems. We also showed how the exceptional points which affect the non-monotonicity of energy gaps versus coupling, as in the case of SYK model, is related to the non-monotonicity of Page curve.

Then, we reviewed the case of Cotler-Jensen theory \cite{Cotler:2018zff} as a $3d$ model of AdS$_3$ with reparametrization modes. In this theory, we studied quantum corrected partition function, reparamtrization modes versus Schwarzian and KK modes and quantum corrected propagators. Then, we discussed quantum corrections to greybody factor. For the study of greybody factor, we first review its calculations in Robinson-Wilczek gravitational theory with anomalies, and then its quantum corrections. We did the same for black holes in $5d$, warped BTZ black holes, and finally in Cotler-Jensen theory.

Next, we found the quantum corrections to bilocal operatos and Lyapunov exponent in $3d$.
We also examined quantum corrections to various other quantities, such as potential of black holes, and information structures of black holes.

In the future works, we aim to draw more exact connections between the effects of EPs on the structures of black holes near horizon, which is an open quantum system, and these quantum correction modes. These quantum corrections could also be found in various other models and solutions and for other quantities, in condensed matter, AdS/QCD and quantum gravity.

\section*{Acknowledgments}

I would like to thank Jun Nian for introducing the topic, and Jun Nian, Alessia Segati, and Federico Faedo for discussions and initial collaborations.

\appendix

\section{Higher order corrections in Kaplan's theory}

In \cite{Chen:2016cms}, higher order computation for Virasoro conformal blocks for the case of $\text{AdS}_3/ \text{CFT}_2$ and in the small $1/c \propto G_N$ expansion has been done. The Virasoro vacuum block has been calculated to order $1/c^2$ and $1/c^3$, which correspond to 2 and 3 loop calculation in $\text{AdS}_3$. These equations could also be found in the gravity side using the method of \cite{Qi:2019gny} as well, which is a topic for future works.

At order $1/c$, they claimed
\begin{gather}
\log \mathcal{V} \supset \frac{h_L}{c} (f_{10} + h_L f_{01} ),
\end{gather}
 and at order $1/c^2$, 
 \begin{gather}
 \log \mathcal{V} \supset \frac{h_L}{c^2} ( f_{20} + h_L f_{11} + h_L^2 f_{02})
 \end{gather}

The linear $\eta_H$ and $\eta_H^2$ terms at order $1/c^2$ are
\begin{gather}
\log \mathcal{V} \supset \frac{h_L}{c^2} \left( \eta_H ( f_{200} + h_L f_{110} + h^2_L f_{020}) + \eta^2_H (f_{201} + h_L f_{111} + h^2_L f_{021} ) \right)
\end{gather}

Here $\eta_H \equiv \frac{h_H}{c} <1$.
 
At order $1/c^3$, we have 
\begin{gather}
\log \mathcal{V} \supset \frac{h_L \eta_H}{c^3} (f_{300} + h_L f_{210} + h_L^2 f_{120} + h_L^3 f_{030}),
\end{gather}

Note that $f_{210} = f_{201}$, $f_{120} = f_{102}$, $f_{030} = f_{003}$, $f_{110} = f_{101}$ and $f_{020} = f_{002} $.

The coefficients are 
\begin{flalign}
f_{10} &= \frac{{\text{csch}}^2 (\frac{\alpha t}{2} ) }{2} \Big \lbrack 3 ( e^{-\alpha t} B ( e^{-t}, -\alpha, 0) + e^{\alpha t} B( e^{-t}, \alpha, 0 ) + e^{\alpha t} B(e^t, - \alpha, 0) + e^{- \alpha t} B( e^t, \alpha,0))\nonumber\\
&+ \frac{1}{\alpha^2}+ \cosh(\alpha t) \left ( - \frac{1}{\alpha^2} + 6 H_{-\alpha} + 6H_\alpha + 6 i \pi -5 \right) + 12 \log \left( 2 \sinh \left ( \frac{t}{2} \right) \right ) + 5\Big \rbrack \nonumber\\
&-t \frac{(13 \alpha^2 -1 ) \coth \left (\frac{\alpha t}{2}\right ) }{2\alpha} + 12 \log \left ( \frac{2 \sinh \left(\frac{\alpha t}{2}\right) }{\alpha} \right), \nonumber\\
&f_{01} = 6  \Big({\text{csch}}^2 \left ( \frac{\alpha t}{2} \right) \bigg \lbrack \frac{B(e^{-t},-\alpha,0)+B(e^t, - \alpha, 0)+B(e^{-t}, \alpha, 0)+B(e^t, \alpha,0)}{2} \nonumber\\
& + H_{- \alpha} + H_\alpha + 2\log \left ( 2 \sinh \left( \frac{t}{2} \right) \right) + i \pi  \bigg \rbrack + 2  \left( \log  \left ( \alpha \sinh \left(\frac{t}{2}\right) \text{csch} \left(\frac{\alpha t}{2}\right)\right ) +1 \right )\Big),
\end{flalign}
where $B(x, \beta, 0) = \frac{x^\beta  {}_2F_1(1, \beta, 1+\beta, x) }{\beta} $ is the incomplete Beta function, $H_n$ is the harmonic function, and $\alpha = \sqrt{1- 24\eta_H}$.

Also, defining $z \equiv 1- e^{-t}$, 
\begin{flalign}
f_{200} &= \frac{1728 (z^2-1)( \zeta(3) - \text{Li}_3 (1-z)) }{z^2} + \frac{288 \text{Li}_2 (z) (7(z-2)z -12(z-1) 
\log(1-z))}{z^2}\nonumber\\
&- \frac{1728(z-2) \text{Li}_3(z) }{z} - \frac{144(z-1) \log^2(1-z) ( 6 (z+1) \log (z) - 7z + 7 ) }{z^2}+ 1128 \nonumber\\
& + \frac{12 \left (24 \pi^2 (z^2 -1) + (z-2) z\right) \log (1-z) }{z^2} + \frac{288 (z-2) (z-1)^2 \log^3 (1-z)}{z^3}.
\end{flalign}

\begin{flalign}
f_{201} &=  \frac{432 (( z-1) z ( 15 z -46) + 4 \pi^2 ( z (( z02) z -10) + 12 )) \log^2 (1-z)}{z^3} \nonumber\\
& + \frac{864 ( z ( z ( z ( 5z -44) + 103) -96) + 33) \log^4 (1-z) }{z^4} + \frac{10368 ( z-2)^2 \text{Li}_2 (z)^2}{z^2}\nonumber\\
&+ \frac{864 (( 9z-46) ( z-1)^2 + ( 4z ( 15 -2(z-2) z) -72) \log(z)) \log^3 (1-z)}{z^3} \nonumber\\
& + \frac{5184 \text{Li}_2 (z) \log (1-z) (z ( z (7z-32) + 2(z-9) (z-2) (z-1) \log(1-z))}{z^3} \nonumber\\
& + \frac{5184 \text{Li}_3 (1-z) (z(z( 5z-14) + 16) -4 (z-3) (z-1) (z+2) \log (1-z)) }{z^3}\nonumber\\
& +  \frac{10368 \text{Li}_3 (z) (( z-2)z + 2((z-8) z +8) \log(1-z)) }{z^2} + \frac{20736(z-2) \text{Li}_4 (1-z)}{z} \nonumber\\
& + \frac{20736 (( z-6) z +6 ) \left( \text{Li}_4 \left(\frac{z}{z-1} \right) + \text{Li}_4(z) \right)}{z^2} + \frac{12960 (z_2) \text{Li}_2(z)}{z} \nonumber\\
& + \frac{216 ((z-2) z^2 + 96 ( 6-5z) \zeta(3) - 4\pi^2 ( z (5z-14) + 16) z) \log(1-z) }{z^3} \nonumber\\
& - \frac{144( -525z^2 + 180 ( z ( 5z-14) + 16 ) \zeta (3) + 8 \pi^4 ( z-2) z) }{5z^2} \nonumber\\
& + \frac{2592 ( z ( 5z-14) + 16 ) \log (z) \log^2 (1-z)}{z^2},
\end{flalign}

\begin{flalign}
f_{111}& = \frac{864 ( 3z ( z^2 - 8z + 7 ) + 8 \pi^2 (2z^2-9z +8)) \log^2 (1-z)}{z^3} + \frac{5184(z-2) \text{Li}_2 (z) }{z} \nonumber\\
& + \frac{1728 ( z(z(z(3z-44) + 127) -136) + 51) \log^4 ( 1-z)}{z^4} - \frac{41472(z-1) \text{Li}_2 (z)^2}{z^2} \nonumber\\
& + \frac{3456 ((z-1) (( z-17) z +21) - 6(z(2z-9) + 8 ) \log(z)) \log^3 (1-z) }{z^3} \nonumber\\
& + \frac{10368 (( z-7) z + 7 ) \log(z) \log^2 (1-z) }{z^2} - \frac{41472 ((z-6) z +6) \text{Li}_3 (z) \log(1-z)}{z^2} \nonumber\\
& + \frac{20736 \text{Li}_2 (z) \log (1-z) (z ((z-9)z + 9) + ( z (( z-14) z + 34) - 22 ) \log (1-z))}{z^3} \nonumber\\
& + \frac{20737 \text{Li}_3 ( 1-z) ( z (( z-7) z +7) - 2 ( z ( 2z-9 + 8 ) \log (1-z))}{z^3} \nonumber\\
& + \frac{41472 (( z -6 ) z + 6  \left (\text{Li}_4 ( z) + \text{Li}_4 \left(\frac{z}{z-1} \right) \right) }{z^2} + 20736 \left ( 2 - \frac{((z-7) z + 7 ) \zeta (3)}{z^2} \right) \nonumber\\
& \frac{432 ( 3 (z-2) z^2 + 96 ( z ( 2z-9 ) + 8 ) \zeta (3) - 8 \pi^2 ((z-7) z + 7) z) \log(1-z)}{z^3}, 
\end{flalign}
and 
\begin{flalign}
f_{300} &= \frac{864 ( 2z ( z (z^2 - 8z +17) -14) + 9) \log^4(1-z)}{z^4} - \frac{20736 ( z-1) \text{Li}_2(z)^2}{z^2}\nonumber\\
& + \frac{216 \log^2 ( 1-z) ( 8 \pi^2 (z-2) ( z^2-2) -108z(z^2-1) \log(z) + 73z(z-1)^2 )}{z^3} \nonumber\\
& - \frac{864(z-2) \log^3 (1-z) ( 4 (2z^2-3) \log(z) - 9(z-1)^2 )}{z^3} \nonumber\\
& + \frac{432 \text{Li}_2 (z) (73(z-2) z^2 + 24 (z-1) \log(1-z) ((6-4z) \log (1-z) -9z)) }{z^3}\nonumber\\
& + \frac {5184(z^2-1) \text{Li}_3 (1-z) ( 9z+4(z-2) \log(1-z)) }{z^3} - \frac{20736(2z-3) \text{Li}_4 \left(\frac{z}{z-1}  \right) }{z^2}\nonumber\\
& - \frac{5184(z-2) \text{Li}_3(z)(9z+4(z-2) \log(1-z)) }{z^2} + \frac{20736(z-3)(z-1) \text{Li}_4(z) }{z^2} \nonumber\\
& + \frac{20736(z-2) \text{Li}_4 (1-z)}{z} + \frac{192(1215 (z^2-1) \zeta(3) + 320 z^2 - 6\pi^4(z-2) z)}{5z^2}\nonumber\\
& + \frac{12((z-2) ( z^2-1728 \zeta(3)) + 648 \pi^2 z ( z^2-1)) \log(1-z)}{z^3}.
\end{flalign}

 \medskip

\bibliography{Qcorrect.bib}

@article{Iliesiu:2020qvm,
    author = "Iliesiu, Luca V. and Turiaci, Gustavo J.",
    title = "{The statistical mechanics of near-extremal black holes}",
    eprint = "2003.02860",
    archivePrefix = "arXiv",
    primaryClass = "hep-th",
    doi = "10.1007/JHEP05(2021)145",
    journal = "JHEP",
    volume = "05",
    pages = "145",
    year = "2021"
}

@article{Brown:2024ajk,
    author = "Brown, Adam R. and Iliesiu, Luca V. and Penington, Geoff and Usatyuk, Mykhaylo",
    title = "{The evaporation of charged black holes}",
    eprint = "2411.03447",
    archivePrefix = "arXiv",
    primaryClass = "hep-th",
    doi = "10.1007/JHEP01(2026)109",
    journal = "JHEP",
    volume = "01",
    pages = "109",
    year = "2026"
}

@article{Heydeman:2020hhw,
    author = "Heydeman, Matthew and Iliesiu, Luca V. and Turiaci, Gustavo J. and Zhao, Wenli",
    title = "{The statistical mechanics of near-BPS black holes}",
    eprint = "2011.01953",
    archivePrefix = "arXiv",
    primaryClass = "hep-th",
    reportNumber = "PUPT-2621",
    doi = "10.1088/1751-8121/ac3be9",
    journal = "J. Phys. A",
    volume = "55",
    number = "1",
    pages = "014004",
    year = "2022"
}

@article{Kapec:2023ruw,
    author = "Kapec, Daniel and Sheta, Ahmed and Strominger, Andrew and Toldo, Chiara",
    title = "{Logarithmic Corrections to Kerr Thermodynamics}",
    eprint = "2310.00848",
    archivePrefix = "arXiv",
    primaryClass = "hep-th",
    doi = "10.1103/PhysRevLett.133.021601",
    journal = "Phys. Rev. Lett.",
    volume = "133",
    number = "2",
    pages = "021601",
    year = "2024"
}

@article{Maulik:2024dwq,
    author = "Maulik, Sabyasachi and Pando Zayas, Leopoldo A. and Ray, Augniva and Zhang, Jingchao",
    title = "{Universality in logarithmic temperature corrections to near-extremal rotating black hole thermodynamics in various dimensions}",
    eprint = "2401.16507",
    archivePrefix = "arXiv",
    primaryClass = "hep-th",
    reportNumber = "LCTP-24-01",
    doi = "10.1007/JHEP06(2024)034",
    journal = "JHEP",
    volume = "06",
    pages = "034",
    year = "2024"
}

@article{Jensen:2016pah,
    author = "Jensen, Kristan",
    title = "{Chaos in AdS$_2$ Holography}",
    eprint = "1605.06098",
    archivePrefix = "arXiv",
    primaryClass = "hep-th",
    doi = "10.1103/PhysRevLett.117.111601",
    journal = "Phys. Rev. Lett.",
    volume = "117",
    number = "11",
    pages = "111601",
    year = "2016"
}

@article{Iliesiu:2022onk,
    author = "Iliesiu, Luca V. and Murthy, Sameer and Turiaci, Gustavo J.",
    title = "{Revisiting the logarithmic corrections to the black hole entropy}",
    eprint = "2209.13608",
    archivePrefix = "arXiv",
    primaryClass = "hep-th",
    doi = "10.1007/JHEP07(2025)058",
    journal = "JHEP",
    volume = "07",
    pages = "058",
    year = "2025"
}

@article{Fitzpatrick:2015zha,
    author = "Fitzpatrick, A. Liam and Kaplan, Jared and Walters, Matthew T.",
    title = "{Virasoro Conformal Blocks and Thermality from Classical Background Fields}",
    eprint = "1501.05315",
    archivePrefix = "arXiv",
    primaryClass = "hep-th",
    doi = "10.1007/JHEP11(2015)200",
    journal = "JHEP",
    volume = "11",
    pages = "200",
    year = "2015"
}

@article{Fitzpatrick:2014vua,
    author = "Fitzpatrick, A. Liam and Kaplan, Jared and Walters, Matthew T.",
    title = "{Universality of Long-Distance AdS Physics from the CFT Bootstrap}",
    eprint = "1403.6829",
    archivePrefix = "arXiv",
    primaryClass = "hep-th",
    doi = "10.1007/JHEP08(2014)145",
    journal = "JHEP",
    volume = "08",
    pages = "145",
    year = "2014"
}

@article{Chen:2017dnl,
    author = "Chen, Hongbin and Fitzpatrick, A. Liam and Kaplan, Jared and Li, Daliang",
    title = "{The AdS$_{3}$ propagator and the fate of locality}",
    eprint = "1712.02351",
    archivePrefix = "arXiv",
    primaryClass = "hep-th",
    doi = "10.1007/JHEP04(2018)075",
    journal = "JHEP",
    volume = "04",
    pages = "075",
    year = "2018"
}

@article{Fitzpatrick:2016mjq,
    author = "Fitzpatrick, A. Liam and Kaplan, Jared",
    title = "{On the Late-Time Behavior of Virasoro Blocks and a Classification of Semiclassical Saddles}",
    eprint = "1609.07153",
    archivePrefix = "arXiv",
    primaryClass = "hep-th",
    doi = "10.1007/JHEP04(2017)072",
    journal = "JHEP",
    volume = "04",
    pages = "072",
    year = "2017"
}

@article{Ghodrati:2022hbb,
    author = "Ghodrati, Mahdis",
    title = "{Encoded information of mixed correlations: the views from one dimension higher}",
    eprint = "2209.04548",
    archivePrefix = "arXiv",
    primaryClass = "hep-th",
    doi = "10.1007/JHEP08(2023)059",
    journal = "JHEP",
    volume = "08",
    pages = "059",
    year = "2023"
}

@article{Ghodrati:2026xky,
    author = "Ghodrati, Mahdis",
    title = "{Some universalities in the partition functions of low-dimensional gravity models}",
    eprint = "2605.25251",
    archivePrefix = "arXiv",
    primaryClass = "hep-th",
    month = "5",
    year = "2026"
}

@article{Ghodrati:2025fah,
    author = "Ghodrati, Mahdis",
    title = "{Photonic Exceptional Points in Holography and QCD}",
    eprint = "2510.15518",
    archivePrefix = "arXiv",
    primaryClass = "hep-th",
    month = "10",
    year = "2025"
}

@article{Ghodrati:2023uef,
    author = "Ghodrati, Mahdis",
    title = "{String scattering amplitudes and mutual information in confining backgrounds: The partonic behavior}",
    eprint = "2307.13454",
    archivePrefix = "arXiv",
    primaryClass = "hep-th",
    reportNumber = "Phys. Rev. D 112, 046011",
    doi = "10.1103/mv9d-kmq4",
    journal = "Phys. Rev. D",
    volume = "112",
    number = "4",
    pages = "046011",
    year = "2025"
}

@article{Lindblad:1975ef,
    author = "Lindblad, Goran",
    title = "{On the Generators of Quantum Dynamical Semigroups}",
    reportNumber = "TRITA-TFY-75-1",
    doi = "10.1007/BF01608499",
    journal = "Commun. Math. Phys.",
    volume = "48",
    pages = "119",
    year = "1976"
}

@article{Gielen:2009qya,
    author = "Gielen, Steffen",
    title = "{Does Black Hole Evaporation Imply That Physics is Nonunitary, and if so, What Must the Laws of Physics Look Like?}",
    eprint = "0905.1497",
    archivePrefix = "arXiv",
    primaryClass = "quant-ph",
    doi = "10.1142/s0219749909005596",
    journal = "Int. J. Quant. Inf.",
    volume = "07",
    number = "05",
    pages = "847--878",
    year = "2009"
}

@article{Zheng:2025apa,
    author = "Zheng, Jie-ping and Dukelsky, Jorge and Molina, Rafael A. and Garc{\'\i}a-Garc{\'\i}a, Antonio M.",
    title = "{Role of exceptional points in the dynamics of the Lindblad Sachdev-Ye-Kitaev model}",
    eprint = "2510.15793",
    archivePrefix = "arXiv",
    primaryClass = "quant-ph",
    doi = "10.1103/m3ww-1fnj",
    journal = "Phys. Rev. D",
    volume = "113",
    number = "6",
    pages = "066018",
    year = "2026"
}

@article{Kulkarni:2021gtt,
    author = "Kulkarni, Anish and Numasawa, Tokiro and Ryu, Shinsei",
    title = "{Lindbladian dynamics of the Sachdev-Ye-Kitaev model}",
    eprint = "2112.13489",
    archivePrefix = "arXiv",
    primaryClass = "cond-mat.stat-mech",
    doi = "10.1103/PhysRevB.106.075138",
    journal = "Phys. Rev. B",
    volume = "106",
    number = "7",
    pages = "075138",
    year = "2022"
}

@article{Liu:2024gxr,
    author = "Liu, Xiao-Long and Nian, Jun and Pando Zayas, Leopoldo A.",
    title = "{Quantum Corrections to Holographic Strange Metal at Low Temperature}",
    eprint = "2410.11487",
    archivePrefix = "arXiv",
    primaryClass = "hep-th",
    reportNumber = "PCFT-24-32, LCTP-24-18",
    month = "10",
    year = "2024"
}

@article{Daguerre:2023cyx,
    author = "Daguerre, Lucas",
    title = "{Boundary correlators and the Schwarzian mode}",
    eprint = "2310.19885",
    archivePrefix = "arXiv",
    primaryClass = "hep-th",
    doi = "10.1007/JHEP01(2024)118",
    journal = "JHEP",
    volume = "01",
    pages = "118",
    year = "2024"
}

@article{Castro:2025pst,
    author = "Castro, Alejandra and Hollander, Jildou and Martinez, Pedro J. and Verheijden, Evita",
    title = "{The effects of near-AdS$_{2}$ backreaction on matter fields}",
    eprint = "2509.08046",
    archivePrefix = "arXiv",
    primaryClass = "hep-th",
    doi = "10.1007/JHEP05(2026)282",
    journal = "JHEP",
    volume = "05",
    pages = "282",
    year = "2026"
}

@article{Ghosh:2019rcj,
    author = "Ghosh, Animik and Maxfield, Henry and Turiaci, Gustavo J.",
    title = "{A universal Schwarzian sector in two-dimensional conformal field theories}",
    eprint = "1912.07654",
    archivePrefix = "arXiv",
    primaryClass = "hep-th",
    doi = "10.1007/JHEP05(2020)104",
    journal = "JHEP",
    volume = "05",
    pages = "104",
    year = "2020"
}

@article{Chen:2025qdj,
    author = "Chen, Wenjing and Ma, Yixuan and Han, Si-Wei and Wang, Zihao and Feng, Jun",
    title = "{Quantum thermodynamics in a rotating BTZ black hole spacetime}",
    eprint = "2507.16787",
    archivePrefix = "arXiv",
    primaryClass = "hep-th",
    month = "7",
    year = "2025"
}

@article{Fitzpatrick:2015foa,
    author = "Fitzpatrick, A. Liam and Kaplan, Jared and Walters, Matthew T. and Wang, Junpu",
    title = "{Hawking from Catalan}",
    eprint = "1510.00014",
    archivePrefix = "arXiv",
    primaryClass = "hep-th",
    doi = "10.1007/JHEP05(2016)069",
    journal = "JHEP",
    volume = "05",
    pages = "069",
    year = "2016"
}

@article{Kulkarni_2022,
   title={Lindbladian dynamics of the Sachdev-Ye-Kitaev model},
   volume={106},
   ISSN={2469-9969},
   url={http://dx.doi.org/10.1103/PhysRevB.106.075138},
   DOI={10.1103/physrevb.106.075138},
   number={7},
   journal={Physical Review B},
   publisher={American Physical Society (APS)},
   author={Kulkarni, Anish and Numasawa, Tokiro and Ryu, Shinsei},
   year={2022},
   month=aug }

@article{Maulik:2025hax,
    author = "Maulik, Sabyasachi and Meng, Xin and Pando Zayas, Leopoldo A.",
    title = "{Quantum-corrected Hawking radiation from near-extremal Kerr-Newman black holes}",
    eprint = "2501.08252",
    archivePrefix = "arXiv",
    primaryClass = "hep-th",
    reportNumber = "LCTP-24-22",
    doi = "10.1007/JHEP02(2026)205",
    journal = "JHEP",
    volume = "02",
    pages = "205",
    year = "2026"
}

@article{PandoZayas:2025snm,
    author = "Pando Zayas, Leopoldo A. and Zhang, Jingchao",
    title = "{One-loop Corrected Holographic Shear Viscosity to Entropy Density Ratio at Low Temperatures}",
    eprint = "2510.16100",
    archivePrefix = "arXiv",
    primaryClass = "hep-th",
    reportNumber = "LITP-25-12",
    month = "10",
    year = "2025"
}

@article{Pathak:2026yyb,
    author = "Pathak, Anamika Avinash and Bhattacharya, Swastik",
    title = "{Temperature Fluctuations and quantum corrections near Black Hole Horizon}",
    eprint = "2603.11386",
    archivePrefix = "arXiv",
    primaryClass = "gr-qc",
    month = "3",
    year = "2026"
}

@article{Nian:2025oei,
    author = "Nian, Jun and Pando Zayas, Leopoldo A. and Yue, Cong-Yuan",
    title = "{Quantum Corrections in the Low-Temperature Fluid/Gravity Correspondence}",
    eprint = "2510.15411",
    archivePrefix = "arXiv",
    primaryClass = "hep-th",
    reportNumber = "PCFT-25-43, LITP-25-01",
    month = "10",
    year = "2025"
}

@article{Lv:2025nww,
    author = "Lv, Zhiwei and Shaukat, Sulaman and Donmez, Orhan and Javed, Faisal and Waseem, Arfa",
    title = "{Quantum effects on greybody factor via quantum Oppenheimer{\textendash}Snyder-dS spacetime}",
    doi = "10.1140/epjc/s10052-025-14434-0",
    journal = "Eur. Phys. J. C",
    volume = "85",
    number = "7",
    pages = "719",
    year = "2025"
}

@article{Cremonini:2025yqe,
    author = "Cremonini, Sera and Li, Li and Liu, Xiao-Long and Nian, Jun",
    title = "{Quantum Corrections to eta/s from JT Gravity}",
    eprint = "2510.21602",
    archivePrefix = "arXiv",
    primaryClass = "hep-th",
    month = "10",
    year = "2025"
}

@article{Law:2022zdq,
    author = "Law, Y. T. Albert and Parmentier, Klaas",
    title = "{Black hole scattering and partition functions}",
    eprint = "2207.07024",
    archivePrefix = "arXiv",
    primaryClass = "hep-th",
    doi = "10.1007/JHEP10(2022)039",
    journal = "JHEP",
    volume = "10",
    pages = "039",
    year = "2022"
}

@article{Jiang:2025cyl,
    author = "Jiang, Zheng and Nian, Jun and Shao, Caiying and Tian, Yu and Zhang, Hongbao",
    title = {{Quantum Gravity Corrections to the Scalar Quasi-Normal Modes in Near-Extremal Reissener-Nordstrom Black Holes}},
    eprint = "2506.22945",
    archivePrefix = "arXiv",
    primaryClass = "hep-th",
    reportNumber = "PCFT-25-25",
    doi = "10.1103/s27b-58gw",
    month = "6",
    year = "2025"
}

@article{David:2021qaa,
    author = "David, Marina and Lezcano Gonzalez, Alfredo and Nian, Jun and Pando Zayas, Leopoldo A.",
    title = "{Logarithmic corrections to the entropy of rotating black holes and black strings in AdS$_{5}$}",
    eprint = "2106.09730",
    archivePrefix = "arXiv",
    primaryClass = "hep-th",
    reportNumber = "LCTP-21-14",
    doi = "10.1007/JHEP04(2022)160",
    journal = "JHEP",
    volume = "04",
    pages = "160",
    year = "2022"
}

@article{Liu:2024qnh,
    author = "Liu, Xiao-Long and Yue, Cong-Yuan and Nian, Jun and Zheng, Wenni",
    title = "{Quantum-Corrected Holographic Wilson Loop Expectation Values and Super-Yang-Mills Confinement}",
    eprint = "2412.11107",
    archivePrefix = "arXiv",
    primaryClass = "hep-th",
    month = "12",
    year = "2024"
}

@article{Cardoso:2001hn,
    author = "Cardoso, Vitor and Lemos, Jose P. S.",
    title = "{Scalar, electromagnetic and Weyl perturbations of BTZ black holes: Quasinormal modes}",
    eprint = "gr-qc/0101052",
    archivePrefix = "arXiv",
    doi = "10.1103/PhysRevD.63.124015",
    journal = "Phys. Rev. D",
    volume = "63",
    pages = "124015",
    year = "2001"
}

@article{Denef:2009kn,
    author = "Denef, Frederik and Hartnoll, Sean A. and Sachdev, Subir",
    title = "{Black hole determinants and quasinormal modes}",
    eprint = "0908.2657",
    archivePrefix = "arXiv",
    primaryClass = "hep-th",
    reportNumber = "NSF-KITP-09-145",
    doi = "10.1088/0264-9381/27/12/125001",
    journal = "Class. Quant. Grav.",
    volume = "27",
    pages = "125001",
    year = "2010"
}

@article{Liu:2025iei,
    author = "Liu, Yu-Qi and Yu, Hao-Wei and Cheng, Peng",
    title = "{Quantum-corrected black hole thermodynamics from the gravitational path integral}",
    eprint = "2506.15261",
    archivePrefix = "arXiv",
    primaryClass = "hep-th",
    month = "6",
    year = "2025"
}

@article{Castro:2021csm,
    author = "Castro, Alejandra and Godet, Victor and Sim\'on, Joan and Song, Wei and Yu, Boyang",
    title = "{Gravitational perturbations from NHEK to Kerr}",
    eprint = "2102.08060",
    archivePrefix = "arXiv",
    primaryClass = "hep-th",
    doi = "10.1007/JHEP07(2021)218",
    journal = "JHEP",
    volume = "07",
    pages = "218",
    year = "2021"
}

@article{Cvetic:1997uw,
    author = "Cvetic, Mirjam and Larsen, Finn",
    title = "{General rotating black holes in string theory: Grey body factors and event horizons}",
    eprint = "hep-th/9705192",
    archivePrefix = "arXiv",
    reportNumber = "UPR-0752-T",
    doi = "10.1103/PhysRevD.56.4994",
    journal = "Phys. Rev. D",
    volume = "56",
    pages = "4994--5007",
    year = "1997"
}

@article{Setare:2006hq,
    author = "Setare, M. R.",
    title = "{Gauge and gravitational anomalies and Hawking radiation of rotating BTZ black holes}",
    eprint = "hep-th/0608080",
    archivePrefix = "arXiv",
    doi = "10.1140/epjc/s10052-006-0148-8",
    journal = "Eur. Phys. J. C",
    volume = "49",
    pages = "865--868",
    year = "2007"
}

@article{Jiang:2007pn,
    author = "Jiang, Qing-Quan and Wu, Shuang-Qing and Cai, Xu",
    title = "{Hawking radiation from the (2+1)-dimensional BTZ black holes}",
    eprint = "hep-th/0701048",
    archivePrefix = "arXiv",
    doi = "10.1016/j.physletb.2007.05.058",
    journal = "Phys. Lett. B",
    volume = "651",
    pages = "58--64",
    year = "2007"
}

@article{Coussaert:1995zp,
    author = "Coussaert, Oliver and Henneaux, Marc and van Driel, Peter",
    title = "{The Asymptotic dynamics of three-dimensional Einstein gravity with a negative cosmological constant}",
    eprint = "gr-qc/9506019",
    archivePrefix = "arXiv",
    reportNumber = "ULB-TH-95-08",
    doi = "10.1088/0264-9381/12/12/012",
    journal = "Class. Quant. Grav.",
    volume = "12",
    pages = "2961--2966",
    year = "1995"
}

@article{Gaberdiel:2007ve,
    author = "Gaberdiel, Matthias R.",
    title = "{Constraints on extremal self-dual CFTs}",
    eprint = "0707.4073",
    archivePrefix = "arXiv",
    primaryClass = "hep-th",
    doi = "10.1088/1126-6708/2007/11/087",
    journal = "JHEP",
    volume = "11",
    pages = "087",
    year = "2007"
}

@article{Geng:2024xpj,
    author = "Geng, Hao",
    title = "{Replica wormholes and entanglement islands in the Karch-Randall braneworld}",
    eprint = "2405.14872",
    archivePrefix = "arXiv",
    primaryClass = "hep-th",
    doi = "10.1007/JHEP01(2025)063",
    journal = "JHEP",
    volume = "01",
    pages = "063",
    year = "2025"
}

@article{Geng:2025rov,
    author = "Geng, Hao",
    title = "{The Mechanism behind the Information Encoding for Islands}",
    eprint = "2502.08703",
    archivePrefix = "arXiv",
    primaryClass = "hep-th",
    month = "2",
    year = "2025"
}

@article{Czech:2025jnw,
    author = "Czech, Bartlomiej and Shuai, Sirui and Wang, Yixu",
    title = "{Entropy Inequalities Constrain Holographic Erasure Correction}",
    eprint = "2502.12246",
    archivePrefix = "arXiv",
    primaryClass = "hep-th",
    month = "2",
    year = "2025"
}

@article{Maloney:2007ud,
    author = "Maloney, Alexander and Witten, Edward",
    title = "{Quantum Gravity Partition Functions in Three Dimensions}",
    eprint = "0712.0155",
    archivePrefix = "arXiv",
    primaryClass = "hep-th",
    doi = "10.1007/JHEP02(2010)029",
    journal = "JHEP",
    volume = "02",
    pages = "029",
    year = "2010"
}

@article{Cvetic:2009jn,
    author = "Cvetic, Mirjam and Larsen, Finn",
    title = "{Greybody Factors and Charges in Kerr/CFT}",
    eprint = "0908.1136",
    archivePrefix = "arXiv",
    primaryClass = "hep-th",
    reportNumber = "UPR-1210-T",
    doi = "10.1088/1126-6708/2009/09/088",
    journal = "JHEP",
    volume = "09",
    pages = "088",
    year = "2009"
}

@article{Nian:2020bzf,
    author = "Nian, Jun and Zayas, Leopoldo A. Pando",
    title = "{Retarded Green\textquoteright{}s functions from rotating AdS black holes: Emergent CFT2 and viscosity}",
    eprint = "2012.02797",
    archivePrefix = "arXiv",
    primaryClass = "hep-th",
    reportNumber = "LCTP-20-29, LCTP-20-29",
    doi = "10.1103/PhysRevD.104.066020",
    journal = "Phys. Rev. D",
    volume = "104",
    number = "6",
    pages = "066020",
    year = "2021"
}

@article{Nian:2019buz,
    author = "Nian, Jun",
    title = "{Kerr black hole evaporation and Page curve}",
    eprint = "1912.13474",
    archivePrefix = "arXiv",
    primaryClass = "hep-th",
    reportNumber = "LCTP-19-38",
    doi = "10.1142/S0218271824500305",
    journal = "Int. J. Mod. Phys. D",
    volume = "33",
    number = "07n08",
    pages = "2450030",
    year = "2024"
}

@article{Nian:2019pxj,
    author = "Nian, Jun and Pando Zayas, Leopoldo A.",
    title = "{Microscopic entropy of rotating electrically charged AdS$_{4}$ black holes from field theory localization}",
    eprint = "1909.07943",
    archivePrefix = "arXiv",
    primaryClass = "hep-th",
    reportNumber = "LCTP-19-22",
    doi = "10.1007/JHEP03(2020)081",
    journal = "JHEP",
    volume = "03",
    pages = "081",
    year = "2020"
}

@article{Maldacena:2016upp,
    author = "Maldacena, Juan and Stanford, Douglas and Yang, Zhenbin",
    title = "{Conformal symmetry and its breaking in two dimensional Nearly Anti-de-Sitter space}",
    eprint = "1606.01857",
    archivePrefix = "arXiv",
    primaryClass = "hep-th",
    doi = "10.1093/ptep/ptw124",
    journal = "PTEP",
    volume = "2016",
    number = "12",
    pages = "12C104",
    year = "2016"
}

@article{Banados:1992wn,
    author = "Banados, Maximo and Teitelboim, Claudio and Zanelli, Jorge",
    title = "{The Black hole in three-dimensional space-time}",
    eprint = "hep-th/9204099",
    archivePrefix = "arXiv",
    reportNumber = "PRINT-92-0151 (CHILE), IASSNS-HEP-92-29",
    doi = "10.1103/PhysRevLett.69.1849",
    journal = "Phys. Rev. Lett.",
    volume = "69",
    pages = "1849--1851",
    year = "1992"
}

@article{Banados:1992gq,
    author = "Banados, Maximo and Henneaux, Marc and Teitelboim, Claudio and Zanelli, Jorge",
    title = "{Geometry of the (2+1) black hole}",
    eprint = "gr-qc/9302012",
    archivePrefix = "arXiv",
    reportNumber = "IASSNS-HEP-92-81",
    doi = "10.1103/PhysRevD.48.1506",
    journal = "Phys. Rev. D",
    volume = "48",
    pages = "1506--1525",
    year = "1993",
    note = "[Erratum: Phys.Rev.D 88, 069902 (2013)]"
}

@article{Robinson:2005pd,
    author = "Robinson, Sean P. and Wilczek, Frank",
    title = "{A Relationship between Hawking radiation and gravitational anomalies}",
    eprint = "gr-qc/0502074",
    archivePrefix = "arXiv",
    reportNumber = "MIT-CTP-3561",
    doi = "10.1103/PhysRevLett.95.011303",
    journal = "Phys. Rev. Lett.",
    volume = "95",
    pages = "011303",
    year = "2005"
}

@article{BERTLMANN2001137,
title = {Two-Dimensional Gravitational Anomalies, Schwinger Terms, and Dispersion Relations},
journal = {Annals of Physics},
volume = {288},
number = {1},
pages = {137-163},
year = {2001},
issn = {0003-4916},
doi = {https://doi.org/10.1006/aphy.2000.6110},
url = {https://www.sciencedirect.com/science/article/pii/S0003491600961104},
author = {R.A. Bertlmann and E. Kohlprath}
}

@article{Sfetsos:1997xs,
    author = "Sfetsos, Konstadinos and Skenderis, Kostas",
    title = "{Microscopic derivation of the Bekenstein-Hawking entropy formula for nonextremal black holes}",
    eprint = "hep-th/9711138",
    archivePrefix = "arXiv",
    reportNumber = "CERN-TH-97-330, KUL-TF-97-33",
    doi = "10.1016/S0550-3213(98)00023-6",
    journal = "Nucl. Phys. B",
    volume = "517",
    pages = "179--204",
    year = "1998"
}

@article{Hyun:1997jv,
    author = "Hyun, Seungjoon",
    title = "{U duality between three-dimensional and higher dimensional black holes}",
    eprint = "hep-th/9704005",
    archivePrefix = "arXiv",
    journal = "J. Korean Phys. Soc.",
    volume = "33",
    pages = "S532--S536",
    year = "1998"
}

@article{Lee:1998pd,
    author = "Lee, H. W. and Myung, Y. S.",
    title = "{Greybody factor for the BTZ black hole and a 5-D black hole}",
    eprint = "hep-th/9804095",
    archivePrefix = "arXiv",
    reportNumber = "INJE-TP-98-4",
    doi = "10.1103/PhysRevD.58.104013",
    journal = "Phys. Rev. D",
    volume = "58",
    pages = "104013",
    year = "1998"
}

@article{Murata:2006pt,
    author = "Murata, Keiju and Soda, Jiro",
    title = "{Hawking radiation from rotating black holes and gravitational anomalies}",
    eprint = "hep-th/0606069",
    archivePrefix = "arXiv",
    reportNumber = "KUNS-2029",
    doi = "10.1103/PhysRevD.74.044018",
    journal = "Phys. Rev. D",
    volume = "74",
    pages = "044018",
    year = "2006"
}

@article{Gao:2022uop,
    author = "Gao, Xin and Hebecker, Arthur and Schreyer, Simon and Venken, Victoria",
    title = "{Loops, local corrections and warping in the LVS and other type IIB models}",
    eprint = "2204.06009",
    archivePrefix = "arXiv",
    primaryClass = "hep-th",
    doi = "10.1007/JHEP09(2022)091",
    journal = "JHEP",
    volume = "09",
    pages = "091",
    year = "2022"
}

@article{Vagenas:2006qb,
    author = "Vagenas, Elias C. and Das, Saurya",
    title = "{Gravitational anomalies, Hawking radiation, and spherically symmetric black holes}",
    eprint = "hep-th/0606077",
    archivePrefix = "arXiv",
    doi = "10.1088/1126-6708/2006/10/025",
    journal = "JHEP",
    volume = "10",
    pages = "025",
    year = "2006"
}

@article{Stanford:2017thb,
    author = "Stanford, Douglas and Witten, Edward",
    title = "{Fermionic Localization of the Schwarzian Theory}",
    eprint = "1703.04612",
    archivePrefix = "arXiv",
    primaryClass = "hep-th",
    doi = "10.1007/JHEP10(2017)008",
    journal = "JHEP",
    volume = "10",
    pages = "008",
    year = "2017"
}

@article{Chen:2016cms,
    author = "Chen, Hongbin and Fitzpatrick, A. Liam and Kaplan, Jared and Li, Daliang and Wang, Junpu",
    title = "{Degenerate Operators and the $1/c$ Expansion: Lorentzian Resummations, High Order Computations, and Super-Virasoro Blocks}",
    eprint = "1606.02659",
    archivePrefix = "arXiv",
    primaryClass = "hep-th",
    doi = "10.1007/JHEP03(2017)167",
    journal = "JHEP",
    volume = "03",
    pages = "167",
    year = "2017"
}

@article{Simon:2024dwm,
    author = "Sim\'on, Joan and Yu, Boyang",
    title = "{BMS$_{3}$ fermionic localization}",
    eprint = "2412.05038",
    archivePrefix = "arXiv",
    primaryClass = "hep-th",
    doi = "10.1007/JHEP04(2025)137",
    journal = "JHEP",
    volume = "04",
    pages = "137",
    year = "2025"
}

@article{Qi:2019gny,
    author = "Qi, Yong-Hui and Sin, Sang-Jin and Yoon, Junggi",
    title = "{Quantum Correction to Chaos in Schwarzian Theory}",
    eprint = "1906.00996",
    archivePrefix = "arXiv",
    primaryClass = "hep-th",
    reportNumber = "KIAS-P19031",
    doi = "10.1007/JHEP11(2019)035",
    journal = "JHEP",
    volume = "11",
    pages = "035",
    year = "2019"
}

@article{Fitzpatrick:2016mtp,
    author = "Fitzpatrick, A. Liam and Kaplan, Jared and Li, Daliang and Wang, Junpu",
    title = "{Exact Virasoro Blocks from Wilson Lines and Background-Independent Operators}",
    eprint = "1612.06385",
    archivePrefix = "arXiv",
    primaryClass = "hep-th",
    doi = "10.1007/JHEP07(2017)092",
    journal = "JHEP",
    volume = "07",
    pages = "092",
    year = "2017"
}

@article{Chen:2025rcc,
    author = "Chen, Ying-Jian and Nian, Jun",
    title = "{Quantum Corrections to Randall-Sundrum Model from JT Gravity}",
    eprint = "2512.06686",
    archivePrefix = "arXiv",
    primaryClass = "hep-th",
    month = "12",
    year = "2025"
}

@misc{zhang2024brownianspinlockingeffect,
      title={Brownian spin-locking effect}, 
      author={Xiao Zhang and Peiyang Chen and Mei Li and Yuzhi Shi and Erez Hasman and Bo Wang and Xianfeng Chen},
      year={2024},
      eprint={2412.00879},
      archivePrefix={arXiv},
      primaryClass={physics.optics},
      url={https://arxiv.org/abs/2412.00879}, 
}

@article{Anninos:2010sq,
    author = "Anninos, Dionysios and Hartnoll, Sean A. and Iqbal, Nabil",
    title = "{Holography and the Coleman-Mermin-Wagner theorem}",
    eprint = "1005.1973",
    archivePrefix = "arXiv",
    primaryClass = "hep-th",
    doi = "10.1103/PhysRevD.82.066008",
    journal = "Phys. Rev. D",
    volume = "82",
    pages = "066008",
    year = "2010"
}

@article{Blommaert:2018oro,
    author = "Blommaert, Andreas and Mertens, Thomas G. and Verschelde, Henri",
    title = "{The Schwarzian Theory - A Wilson Line Perspective}",
    eprint = "1806.07765",
    archivePrefix = "arXiv",
    primaryClass = "hep-th",
    doi = "10.1007/JHEP12(2018)022",
    journal = "JHEP",
    volume = "12",
    pages = "022",
    year = "2018"
}

@article{Cotler:2018zff,
    author = "Cotler, Jordan and Jensen, Kristan",
    title = "{A theory of reparameterizations for AdS$_3$ gravity}",
    eprint = "1808.03263",
    archivePrefix = "arXiv",
    primaryClass = "hep-th",
    doi = "10.1007/JHEP02(2019)079",
    journal = "JHEP",
    volume = "02",
    pages = "079",
    year = "2019"
}

@article{Iso:2006wa,
    author = "Iso, Satoshi and Umetsu, Hiroshi and Wilczek, Frank",
    title = "{Hawking radiation from charged black holes via gauge and gravitational anomalies}",
    eprint = "hep-th/0602146",
    archivePrefix = "arXiv",
    reportNumber = "MIT-CTP-3714, KEK-TH-1062, OIQP-05-23",
    doi = "10.1103/PhysRevLett.96.151302",
    journal = "Phys. Rev. Lett.",
    volume = "96",
    pages = "151302",
    year = "2006"
}

@article{Iso:2006ut,
    author = "Iso, Satoshi and Umetsu, Hiroshi and Wilczek, Frank",
    title = "{Anomalies, Hawking radiations and regularity in rotating black holes}",
    eprint = "hep-th/0606018",
    archivePrefix = "arXiv",
    reportNumber = "MIT-CTP-3730, KEK-TH-1077, OIQP-06-02",
    doi = "10.1103/PhysRevD.74.044017",
    journal = "Phys. Rev. D",
    volume = "74",
    pages = "044017",
    year = "2006"
}
\bibliographystyle{JHEP}
\end{document}